\documentclass[a4paper,11pt]{article}
\pdfoutput=1 

\usepackage{jheppub} 
\usepackage[T1]{fontenc} 
\usepackage{slashed}
\usepackage{subcaption}

\title{\boldmath Collider Constraints and Prospects of a Scalar
Singlet Extension to Higgs Portal Dark Matter}

\author{Grace Dupuis}
\affiliation{Department of Physics, McGill University, \\
Rue University, Montr\'{e}al, Qu\'{e}bec, Canada}
\emailAdd{dupuisg@physics.mcgill.ca}

\abstract{This work considers an extension of the Standard Model (SM) Higgs sector by a real, scalar singlet field, including
applicability to a dark matter (DM) model with the addition of a Yukawa coupling to a fermionic dark matter candidate. 
The collider signatures and constraints on the mixed two-Higgs scenario are determined, including limits
from Higgs production signals and exclusion searches, as well as constraints arising from the Higgs total and invisible
widths. As there is overwhelming Higgs data which is consistent with a SM scenario, the case in which an additional
scalar has evaded detection is further explored in the context of Higgs precision measurement. The discovery
reach and prospective signatures of the model at a proposed linear collider are investigated, with particular
focus on the Higgs triple coupling, and di-Higgs production processes.}

\begin{document} 
\maketitle
\flushbottom

\section{Introduction}
\label{scn:intro}

The discovery of the Higgs boson at the Large Hadron Collider (LHC) is one of the 
most significant scientific achievements of late. By finally providing long-awaited 
evidence of the previously undiscovered scalar sector of the Standard
Model (SM), this groundbreaking result has sparked much excitement. Not only does this
discovery verify the well-established fundamental theory of particle physics, it also
presents new possibilities to probe physics beyond the Standard Model.
Although this result is a major achievement in the field of particle physics, 
the Higgs sector still remains largely unexplored. Specifically, it has yet to be determined 
whether the discovered scalar is indeed the Higgs boson of the Standard Model, or a piece of some extended 
theory. Such a question lies under the realm of Higgs precision measurements. Future experiments, 
reaching higher energies and employing advancing techniques will shed light on precision aspects
such as the Higgs CP nature, its self-couplings and couplings to electroweak 
vector bosons, and possible deviations from the Standard Model that may indicate a larger Higgs sector.

As successful as the Standard Model has proven to be, there are still missing
pieces, one of which is the failure to account for the identity
of dark matter (DM) as a particle species.
Extensions of the SM scalar sector lend themselves to models of DM,
through potential DM candidates and/or new mediators bridging the dark and visible sectors. 
The study presented here considers one of the simplest such extensions --- the
addition of a scalar singlet field, which couples to a fermionic DM candidate. The 
scalar singlet extension is attractive in its simplicity, hence, similar
models have been studied extensively in the literature. The implications on the Higgs
sector of a pure scalar singlet extension have been investigated \cite{Robens:2015gla, 
Pruna:2013bma, Godunov:2015nea, Martin-Lozano:2015dja, Falkowski:2015iwa},
as well as the viability of such a model in a 
dark matter context \cite{Berlin:2015wwa,Kouvaris:2014uoa,Buckley:2014fba,Ghorbani:2014qpa},
with subtle distinctions. 
There is a wide range of models which contain a new scalar singlet, including similar studies
of the implications of an additional Higgs-like scalar, with varying 
instances of the scalar potential and Higgs-scalar interaction terms \cite{Basso:2013nza, Basso:2012nh}.

Often such models, which are typically referred to as Higgs portal, couple the SM field
directly to a DM candidate \cite{Fedderke:2014wda,Baek:2011aa}. Studies of scalar singlet extensions
also exist which take the additional scalar as the DM candidate, as it is protected under 
a $Z_{2}$ symmetry \cite{Queiroz:2014pra,Feng:2014vea,Martin:2014sxa,Cline:2013gha}.
The new scalar here acquires a vacuum expectation value (vev)
and mixes with the SM Higgs field, giving two scalar mediators that act as the portal to
the dark fermion. The case of a dark fermion with its SM interactions mediated by a new scalar
or pseudoscalar has also been considered, but in the context of a bottom-up, effective field 
theory (EFT) approach, without considering the mixing of the scalar and SM Higgs \cite{Izaguirre:2014vva},
or embedded in a more complex model with additional extensions \cite{Varzielas:2015sno} .

In the following study, a scalar singlet extension is investigated, with applicability to a 
model of dark matter. Focus is placed on the collider implications of the mixing in the
Higgs sector, particularly the prospective influence of the additional scalar in Higgs precision
measurements at the proposed International Linear Collider (ILC).
The applicability of the model in a dark matter context is addressed in associated
invisible signatures, and compatibility with direct detection (DD) limits.
As current LHC data is consistent with a SM Higgs, it is vital to further consider
the scenario in the context of future experiments, in the case where an additional
Higgs-like scalar has evaded detection. Under this assumption,
the discovery potential at a proposed linear collider is explored. A lepton collider
is advantageous in possessing greater sensitivity to Higgs precision measurements, to
which the LHC may not be sensitive. It is discussed how a precision environment 
may probe the scalar parameters of the model, and what distinguishing features
may be observable. In particular, the study focuses on the Higgs self-coupling through 
double Higgs production processes.

This work is structured as follows. The model and its parametrization are presented in 
section \ref{scn:model}. As the implications of new decay modes are examined, expressions
for the new decay widths are given in section \ref{scn:widths}. The LHC constraints are presented
in section \ref{scn:LHC}, including both a review of Higgs exclusion searches and measured signals, 
as well as specific invisible searches. The topic of discovery reach and projected limits at a future
linear collider is explored in section \ref{scn:ILC}. 
The compatibility with direct detection data is addressed in section \ref{scn:DD}. 
A concluding discussion is then given on the viability and prospects of the model.

\section{Model}
\label{scn:model}

The scalar sector of the SM is supplemented with an additional real scalar field, which 
is a singlet under the SM gauge group and couples to a fermionic dark matter candidate 
via a Yukawa term. The additional scalar acquires a vev in the same way as the familiar 
SM Higgs field, thereby inducing mixing between the new scalar and the Higgs. The model
is parametrized as follows.
\\
\\
The scalar potential is modified, according to
\begin{equation}
V(\Phi_1, \phi_2) = \lambda_1 (\Phi_1^{\dagger}\Phi_1 - v_1^2/2)^2 + 
\frac{\lambda_2}{4} (\phi_2^2 - v_2^2)^2 + \frac{\lambda_{12}}{2} (\Phi_1^{\dagger}\Phi_1 - v_1^2/2)
(\phi_2^2 - v_2^2),
\label{eqn:potl}
\end{equation} 
where $\Phi_1$ is the SM Higgs doublet and $v_1$ is equivalent to the SM Higgs vev,
specifically $v_1 = 246 $ GeV; $\phi_2$ is the additional field. As the new field
is a singlet under the SM gauge, the familiar symmetry breaking mechanism, with respect to
the electroweak sector, is unchanged. Hence, the SM Higgs vev and other electroweak parameters
are the same as in the SM case. 

A Yukawa interaction is introduced, coupling a dark, vector-like fermion to the scalar singlet 
field, as described by the following interaction Lagrangian:
\begin{equation}
\mathcal{L}_{\mathrm{dark}} \supset  -\frac{1}{2} \phi_2 \left( g_{L} \bar{\psi}_{L}\psi_{L}^{c} + 
g_{R}\bar{\psi}_{R} \psi_{R}^{c} \right) + \mathrm{h.c.}
\label{eqn:Ldark}
\end{equation}
There is a discrete symmetry in this case by which the fields transform as
\begin{align*}
\psi_{L}  &\rightarrow i \psi_{L} \\
\psi_{R}  &\rightarrow -i \psi_{R} \\
\phi_2 &\rightarrow -\phi_2.
\end{align*}
This symmetry forbids both a bare Dirac mass term, as well as any 
terms of odd order in $\phi_2$ in the scalar potential. Eq. \ref{eqn:potl}
represents the most general scalar potential consistent with the symmetry. 

Both scalar fields are allowed to acquire vevs,
\begin{equation}
\Phi_1 \rightarrow \frac{1}{\sqrt{2}} {0 \choose v_1 + h_1(x)}, 
\quad \phi_2 \rightarrow  v_2 + h_2(x).
\label{eqn:vevs}
\end{equation}
Beginning with the scalar sector,
after diagonalizing the mass terms, one finds the following mass eigenvalues:
\begin{subequations}
\begin{equation}
m_{H,S}^2 = (\lambda_1 v_1^2 + \lambda_2 v_2^2) \pm \sqrt{(\lambda_1 v_1^2 - 
\lambda_2 v_2^2)^2 +(\lambda_{12} v_1 v_2)^2},
\label{eqn:massvals1}
\end{equation}
or
\begin{equation}
m_{S}^2 = m_{H}^2 \pm \delta m^2; \quad \delta m^2 = 2 \sqrt{(\lambda_1 v_1^2 - 
\lambda_2 v_2^2)^2 +(\lambda_{12} v_1 v_2)^2}.
\label{eqn:massvals2}
\end{equation}
\label{eqn:massvals}
\end{subequations}
with mass eigenstates given by
\begin{equation}
\left( \begin{array}{c} H \\ S \end{array} \right) = \begin{pmatrix} c_{\theta_h} & -
s_{\theta_h} \\ s_{\theta_h} & c_{\theta_h} \end{pmatrix} 
\left( \begin{array}{c} h_1 \\ h_2\end{array} \right).
\label{eqn:mixing}
\end{equation}
Here $c_{\theta_h}$ and $s_{\theta_h}$ denote $\cos\theta_h$ and $\sin\theta_h$ respectively. 
The mixing angle is given by the following expression:
\begin{equation}
\tan 2\theta_h = \frac{-\lambda_{12} v_1 v_2}{\lambda_1v_1^2 - \lambda_2v_2^2}
\end{equation}
with $\theta_h \in \left[ -\pi/2, \pi/2 \right]$.
$H$ is taken to be the recently discovered Higgs boson, with mass fixed at
$m_H = 125.09$ GeV \cite{Agashe:2014kda}, and the mass of the additional scalar is allowed to be 
either lighter or heavier than $H$. 

The scalar sector is described by three additional parameters --- the vev of the singlet field, 
the mass of the new scalar, and the scalar-Higgs mixing angle, $\{ v_2, m_S, \theta_h \}$. 
In terms of these parameters, the potential coefficients are 

\begin{subequations}
\begin{align}
\lambda_1 &= \frac{1}{2v_1^2} \left( c_{\theta_h}^2 m_H^2 + s_{\theta_h}^2 m_S^2 \right) 
\label{eqn:lambda1} \\
\lambda_2 &= \frac{1}{2 v_2^2} \left( s_{\theta_h}^2 m_H^2 + c_{\theta_h}^2 m_S^2 \right)
\label{eqn:lambda2} \\
\lambda_{12} & = \frac{(m_S^2 - m_H^2)}{2 v_1 v_2}  \sin 2\theta_h 
\label{eqn:lambda12}.
\end{align}
\label{eqn:lambdas}
\end{subequations}
The couplings of $H$ and $S$ to SM particles are simply
given by those of the SM Higgs, scaled respectively by $\cos\theta_h$ and $\sin\theta_h$. 
The extension of the scalar potential results in modified expressions for the scalar
self-couplings. The triple scalar self-coupling coefficients
for $H$ and $S$, denoted respectively by $g_H$ and $g_S$, are given by
\begin{subequations}
\begin{align}
g_H = \frac{m_H^2}{2 v_1 v_2}\left( v_2 c_{\theta_h}^3 - v_1 s_{\theta_h}^3 \right) \label{eqn:gH} \\
g_S = \frac{m_S^2}{2 v_1 v_2}\left( v_1 c_{\theta_h}^3 + v_2 s_{\theta_h}^3 \right) \label{eqn:gS}.
\end{align}
\label{eqn:tricouplings}
\end{subequations}
Furthermore, scalar mixing gives rise to additional interaction vertices between the two scalars,
 $H$-$H$-$S$ and $H$-$S$-$S$, with respective coupling strengths 
\begin{align}
\mu &= \frac{\sin 2\theta_h }{2 v_1 v_2} \left( v_1 s_{\theta_h}  + 
v_2 c_{\theta_h} \right) \left( m_H^2 + \frac{m_S^2}{2} \right)  \label{eqn:mucoup} \\
\eta &= \frac{\sin 2\theta_h }{2 v_1 v_2} \left( -v_1 c_{\theta_h}  + 
v_2 s_{\theta_h}  \right) \left( m_S^2 + \frac{m_H^2}{2} \right). \label{eqn:etacoup}
\end{align}

The triple scalar interaction terms, both self-interactions and mixed scalar interactions
are summarized in the following Lagrangian:
\begin{equation}
-\mathcal{L} \supset g_H H^3 + g_S S^3 + \mu H^2 S + \eta H S^2.
\end{equation}

Returning to the interaction term which describes the dark sector
of the model, as given in eq. \ref{eqn:Ldark}, as $S$ acquires a vev,
one generically obtains two Majorana mass states, 
\begin{equation*}
\chi_1 = \begin{pmatrix} \psi_L \\ \psi_L^c \end{pmatrix}, \qquad 
\chi_2 = \begin{pmatrix} \psi_R^c \\ \psi_R \end{pmatrix}
\end{equation*}
with masses $m_1 = g_L v_2$ and $m_2 = g_R v_2$. 
In this analysis the degenerate
case is considered, in which the left and right couplings are related by $g_L = -g_R \equiv g$.
This represents the simplest extension of this scalar extension to a dark
matter scenario, giving a single dark matter candidate; the more general
model described by an extended parameter space is left for future study.
In the case of two mass-degenerate Majorana states, one can equivalently
describe the picture in terms of a single Dirac fermion, 
$\chi = \left( \xi \quad \eta^c \right)^{T}$, where
\begin{equation*}
\xi = \frac{1}{\sqrt{2}} \left( \psi_L + i \psi_R^c \right) \quad 
\eta= \frac{1}{\sqrt{2}} \left( \psi_L - i \psi_R^c \right),
\end{equation*}
with mass $m_{\chi} = g v_2$. The dark sector is then described by
\begin{equation}
\bar{\chi} ( i\slashed{\partial} - m_{\chi}) \chi - \frac{m_{\chi}}{v_2} h_2 \bar{\chi} \chi.
\end{equation}
With the addition of the dark matter mass, the model is then completely
described by the set of four parameters,
\begin{equation*}
\{ v_2, m_S, \theta_h, m_{\chi} \}.
\end{equation*}

\section{New Decay Channels}
\label{scn:widths}

\subsection{Partial Widths}
A dark matter candidate that couples to the Higgs introduces a new contribution to the 
Higgs invisible width, $\Gamma_{\mathrm{inv}}$. The associated experimental signature is a
channel resulting in missing energy, denoted $\slashed{E}_{T}$.
For $m_{\chi} < m_H/2$, the channel $H \rightarrow \chi \overline{\chi}$ gives a contribution
\begin{equation}
\Gamma_{\mathrm{inv}} = \frac{s_{\theta_{h}}^2 m_H}{8\pi} \left( \frac{m_{\chi}}{v_2} \right)^2
\left (1 - \frac{4m_{\chi}^2}{m_H^2}\right)^{3/2}.
\label{eqn:inviswidth}
\end{equation} 
Moreover, additional non-SM decay modes arise from the scalar mixing terms. 
The total Higgs width receives new contributions from the decay to an $S$ pair. If $S$ is 
sufficiently light, the decay $H \rightarrow S S$ is kinematically allowed, with a partial width
given by
\begin{equation}
\Gamma_{HSS} = \frac{\eta^2}{8 \pi m_H} \left ( 1 - \frac{4m_S^2}{m_H^2} \right )^{1/2}. 
\label{eqn:WdthHSS}
\end{equation}

\begin{table}
\begin{center}
\begin{tabular}{ | l | l | }
\hline
Scalar mass hierarchy & Accessible decay modes \\
\hline
\hline 
$m_S < m_H/2$ & $H \rightarrow SS$ \\
\hline
$m_H/2 < m_S < m_H$ & $H \rightarrow Sf\bar{f}$ \\
\hline
$m_H < m_S < 2m_H$ & $S \rightarrow Hf\bar{f}$ \\
\hline
$2m_H < m_S$ & $S \rightarrow HH$ \\
\hline 
\end{tabular}
\end{center}
\caption{Summary of the available scalar-to-scalar decay modes for various regions of the 
additional scalar mass, $m_S$, relative to the SM Higgs mass. It is implicit that for each case
the invisible decay to $\chi\bar{\chi}$ is included for each scalar if its mass is greater
than $2m_{\chi}$. For the intermediate mass ranges, $f$ may denote either
a SM fermion, or $\chi$, if kinematically permitted.}
\label{tab:wdthtab}
\end{table}

In light of experimental exclusion limits on a light Higgs, perhaps the other mass 
regions are the more interesting cases --- that is either the case in which 
$m_S \sim m_H$, or the scenario in which the new scalar is much heavier and has not yet
been discovered at the energies currently probed by colliders. If $S$ is very heavy, 
the dominant new scalar decay mode is its decay to an $H$ pair; the expression for the
width is analogous to eq. \ref{eqn:WdthHSS} upon the exchanges $m_H \leftrightarrow m_S$ 
and $\eta \rightarrow \mu$. For the intermediate regions, one may consider the three-body 
decays to one real and one virtual scalar, with the latter decaying to a fermion pair.
More specifically, for the case in which $m_H/2 < m_S < m_H$, the accessible decay is 
$H \rightarrow SS^{*} \rightarrow Sf\bar{f}$, where $f$ may be any SM fermion (excluding the 
top quark), or $\chi$, if the DM is sufficiently light. The expressions for the 
relevant three-body widths (distinguishing between $f_{SM}$ and $\chi$)
 are given by the following:
\begin{subequations}
\begin{align}
\Gamma(H \rightarrow Sf\overline{f}) =& \frac{s_{\theta_h}^2 \eta^2}{8\pi^3} 
\frac{m_S^2}{m_H^3} \left( \frac{m_f}{v_1}\right)^2  I(2m_f/m_S, m_H/m_S) ,
\\
\Gamma(H \rightarrow S\chi\overline{\chi}) =& \frac{c_{\theta_h}^2 \eta^2}{8\pi^3}
\frac{m_S^2}{m_H^3}\left( \frac{m_{\chi}}{v_2}\right)^2 I(2m_{\chi}/m_S, m_H/m_S).
\end{align}
Here
\begin{equation}
I(y,z) = \int_{1}^{x_{max}} \mathrm{d}x \frac{\sqrt{x^2 - 1}}{(z - 2x)^2} 
\frac{(1 + z^2 - y^2 - 2zx)^{3/2}}{(1 + z^2 - 2zx)^{1/2}},
\end{equation}
\end{subequations}
where $x_{max} = (1 + z^2 - y^2)/(2z)$. The $y$ term is neglected for $y = 2m_f/m_S$. 
As would be expected, the expressions for the intermediate case with $S$ heavier than
$H$ are obtained by exchanging $m_H \leftrightarrow m_S$, $\eta \rightarrow \mu$, and 
$s_{\theta_h} \leftrightarrow c_{\theta_h}$.
The mass regions and accessible decay channels are summarized in table \ref{tab:wdthtab}.

\subsection{Scalar Couplings}
\label{sscn:couplings}

\begin{figure}
\begin{center}
\begin{subfigure}[b]{0.45\textwidth}
\includegraphics[width=\linewidth]{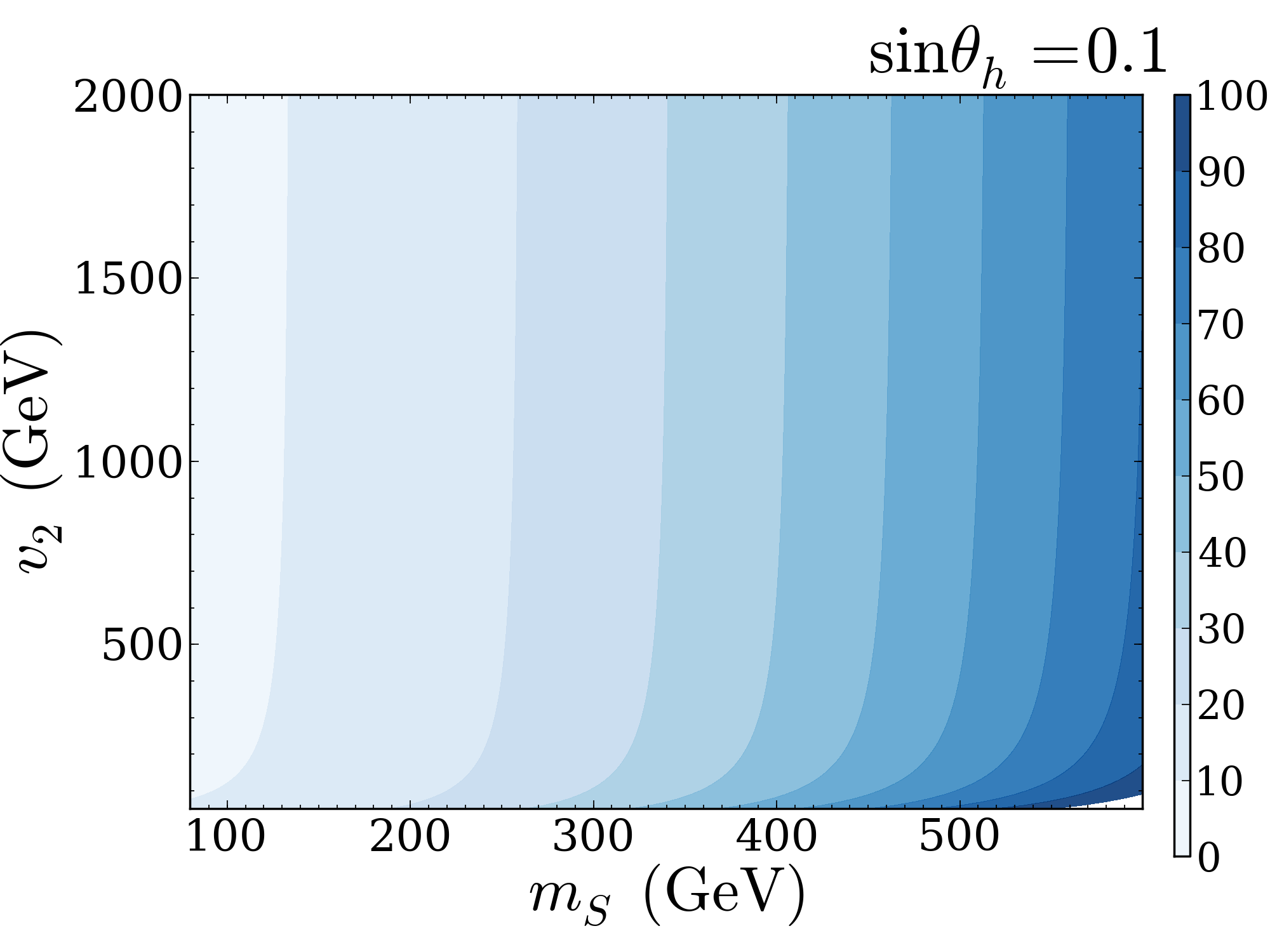}
\end{subfigure}
\begin{subfigure}[b]{0.45\textwidth}
\includegraphics[width=\linewidth]{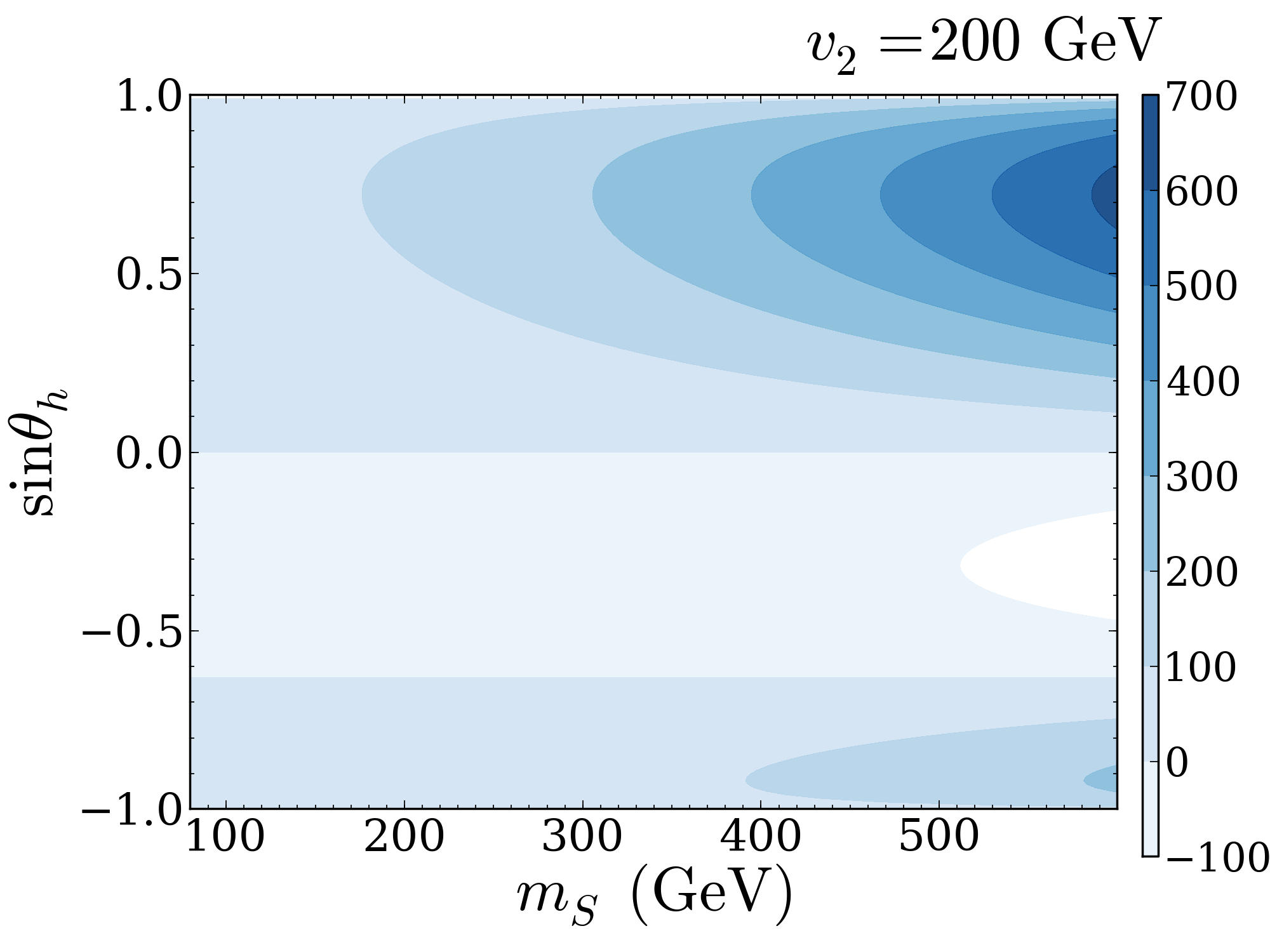}
\end{subfigure}

\end{center}
\caption{Value of the $H$-$H$-$S$ coupling constant, $\mu$, over various regions of scalar
parameter space.The left figure gives the dependence in 
$(m_S, v_2)$ space for  a discrete value of the mixing angle, while the right
shows the dependence over $(m_S, s_{\theta_h})$ space for chosen value of
$v_2$.}
\label{fig:mucoup}
\end{figure}

\begin{figure}
\begin{center}
\begin{subfigure}[b]{0.45\textwidth}
\includegraphics[width=\linewidth]{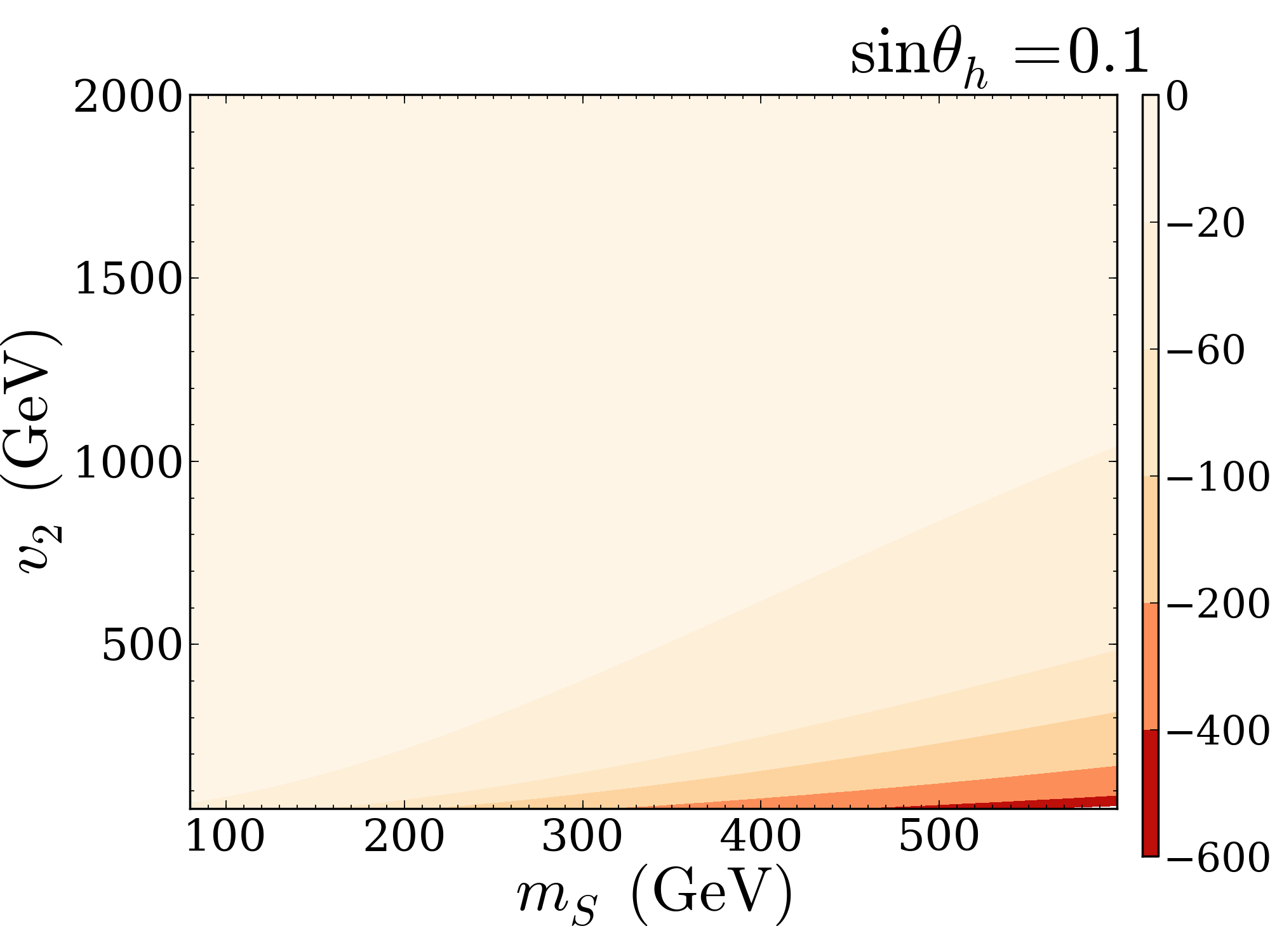}
\end{subfigure}
\begin{subfigure}[b]{0.45\textwidth}
\includegraphics[width=\linewidth]{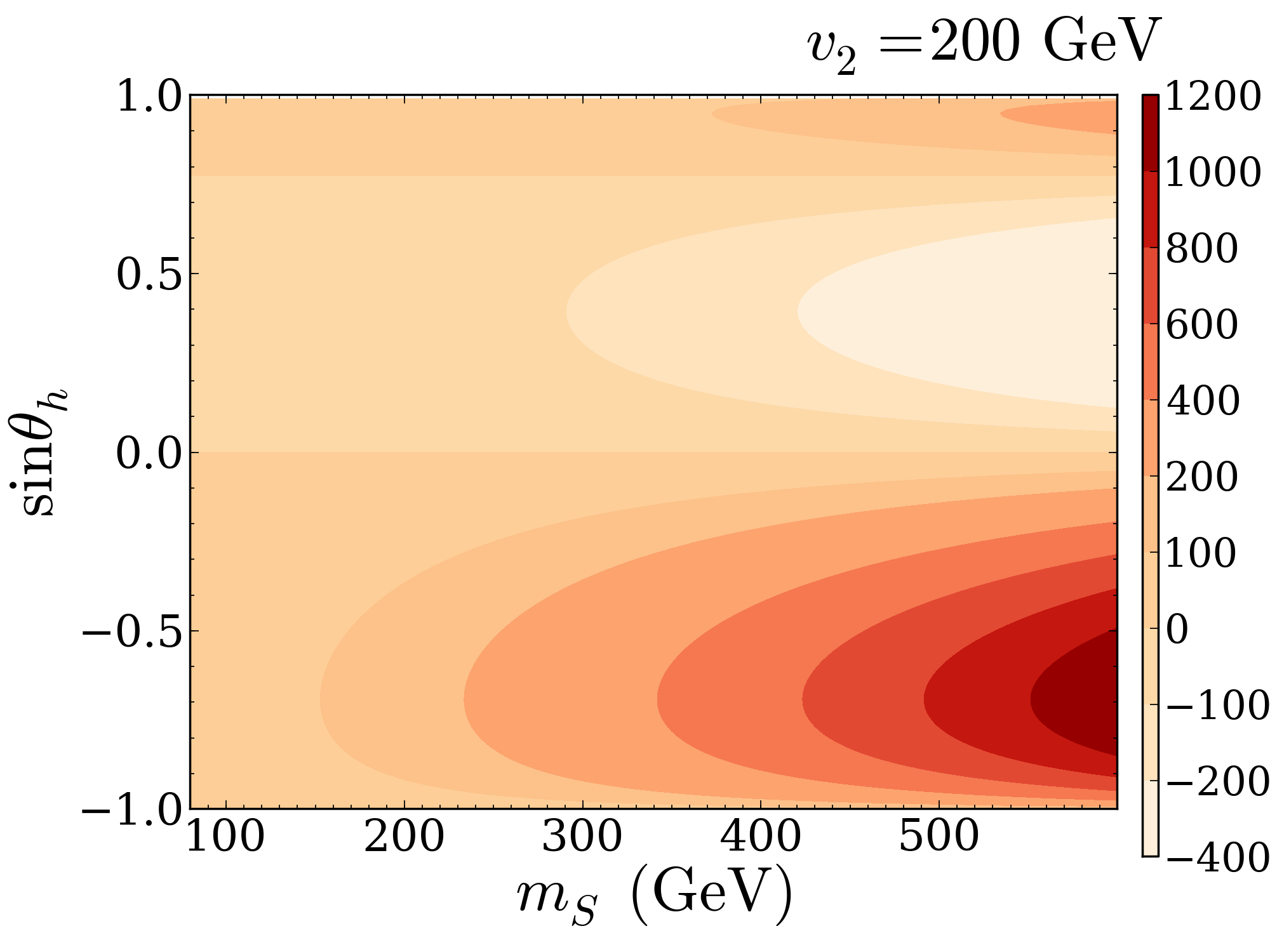}
\end{subfigure}
\end{center}
\caption{Dependence of $H$-$S$-$S$ coupling, $\eta$, on the scalar parameter
space. The description of subfigures is as given in figure \ref{fig:mucoup}.}
\label{fig:etacoup}
\end{figure}

The interaction vertices between $H$ and $S$, arising from the mixed cubic terms in
the scalar potential, provide possibly significant new contributions to decay
widths and production processes. \footnote{The contribution of this vertex to Higgs production at a linear collider will be further discussed in section \ref{sscn:diHiggs}} 
It can be seen
from eqs. \ref{eqn:mucoup} and \ref{eqn:etacoup} that these couplings may become
large for certain values of the mass and vev of the additional scalar. 
Physical quantities which would otherwise be suppressed ---  either by a small mixing 
angle or by nature of being a higher order effect --- may be non-negligible. Figures
\ref{fig:mucoup}  and \ref{fig:etacoup} show the dependence of the scalar couplings,
$\mu$ and $\eta$ respectively, on the scalar parameters.  

\subsection{Three-Body Decays}
\label{sscn:3wdths}

\begin{figure}
\begin{center}
\begin{subfigure}[b]{0.49\textwidth}
\includegraphics[width=\linewidth]{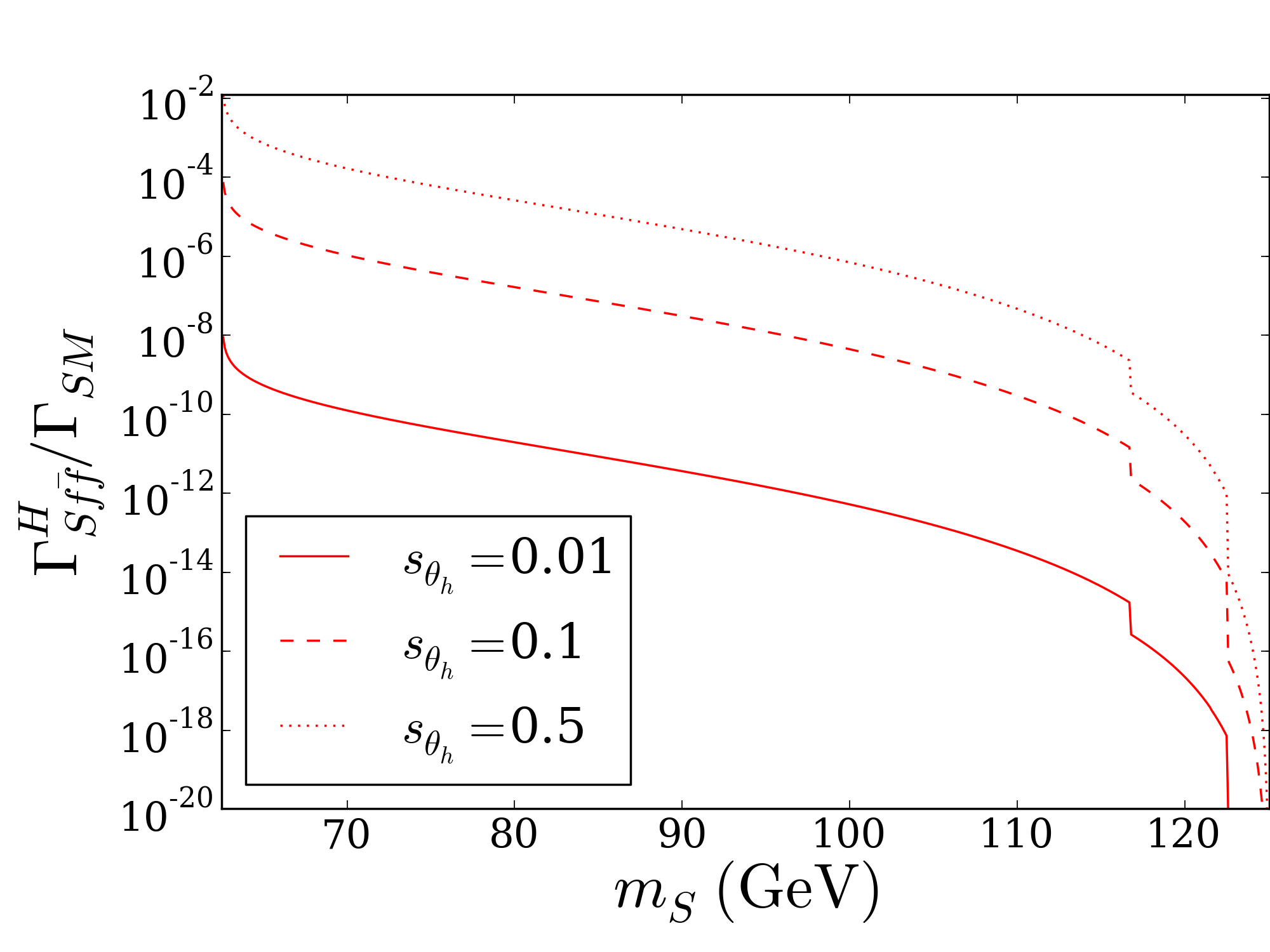}
\end{subfigure}
\begin{subfigure}[b]{0.49\textwidth}
\includegraphics[width=\linewidth]{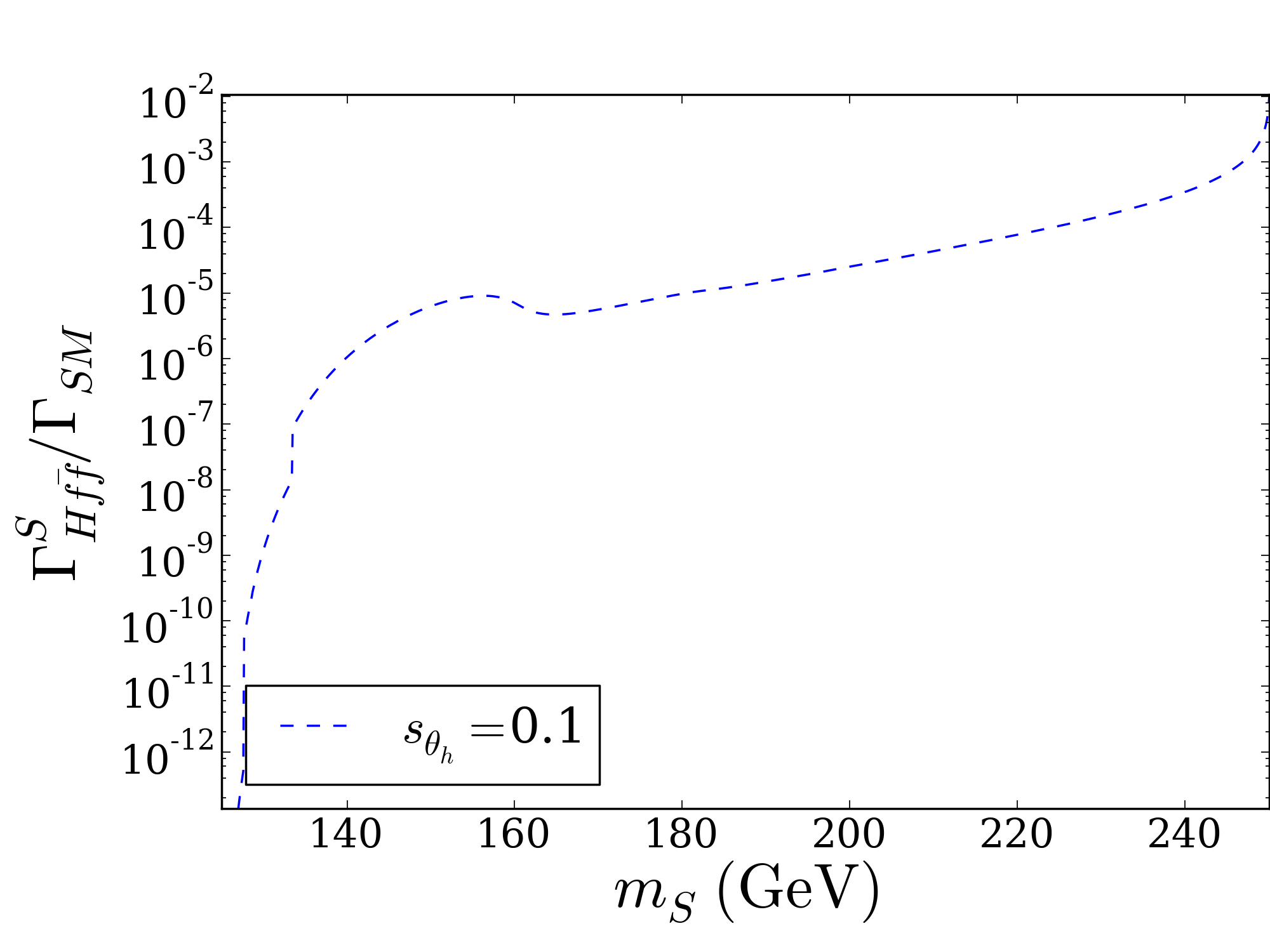}
\end{subfigure}
\end{center}
\caption{Relative magnitude of the three-body decay widths to a scalar and SM fermion pair.
A simplified branching ratio is shown, taken as the ratio of the three-body width to the
total visible width. The width $H \rightarrow S f \bar{f}$ is shown in the left figure, and
$S \rightarrow H f \bar{f}$ on the right. In both cases, the scalar vev is fixed to be
$v_2 = 200$ GeV, and the mixing angle is varied discretely.}
\label{fig:3wdthsff}
\end{figure}

\begin{figure}
\begin{center}
\begin{subfigure}[b]{0.45\textwidth}
\includegraphics[width=\linewidth]{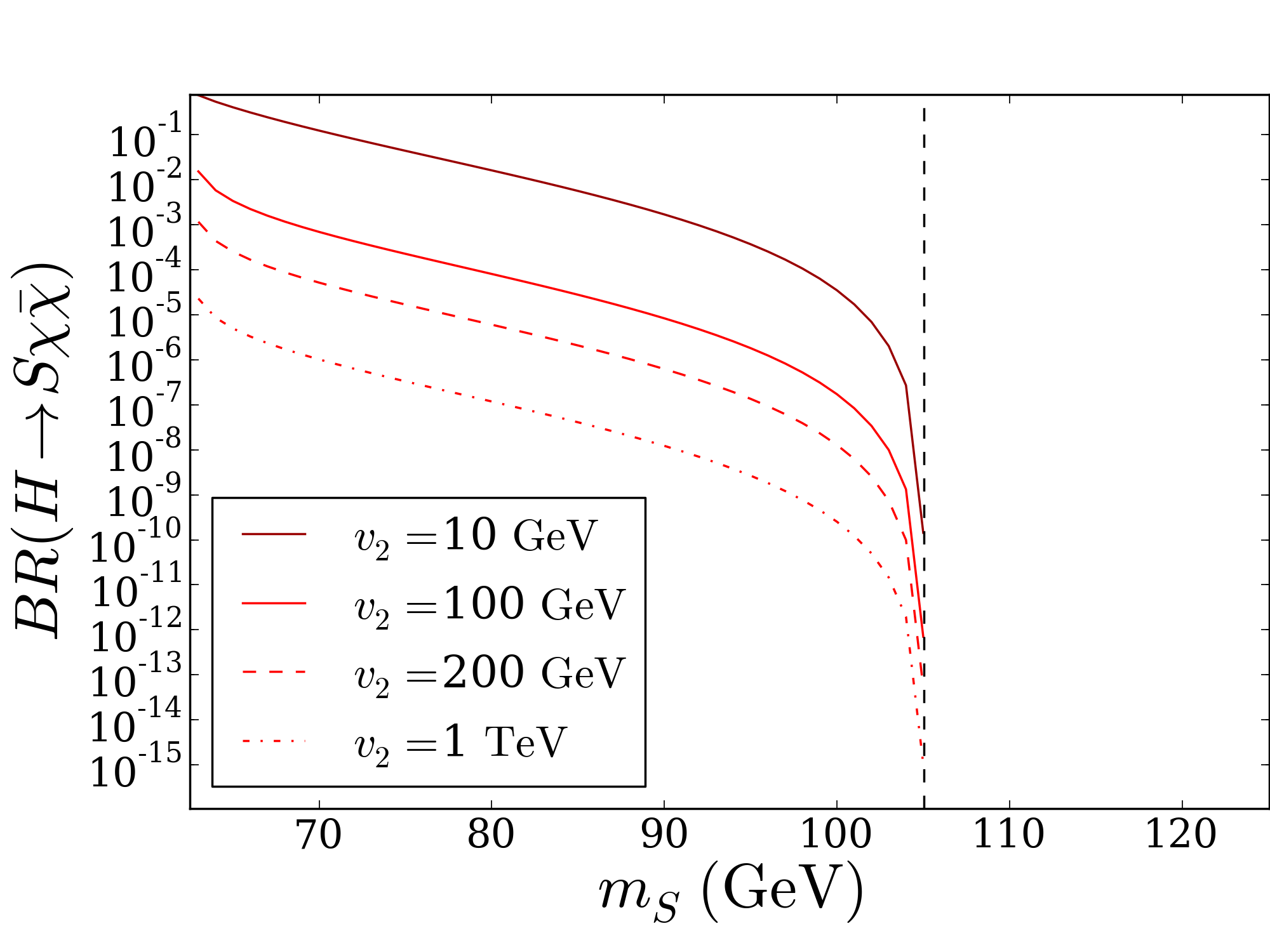}
\end{subfigure}
\begin{subfigure}[b]{0.45\textwidth}
\includegraphics[width=\linewidth]{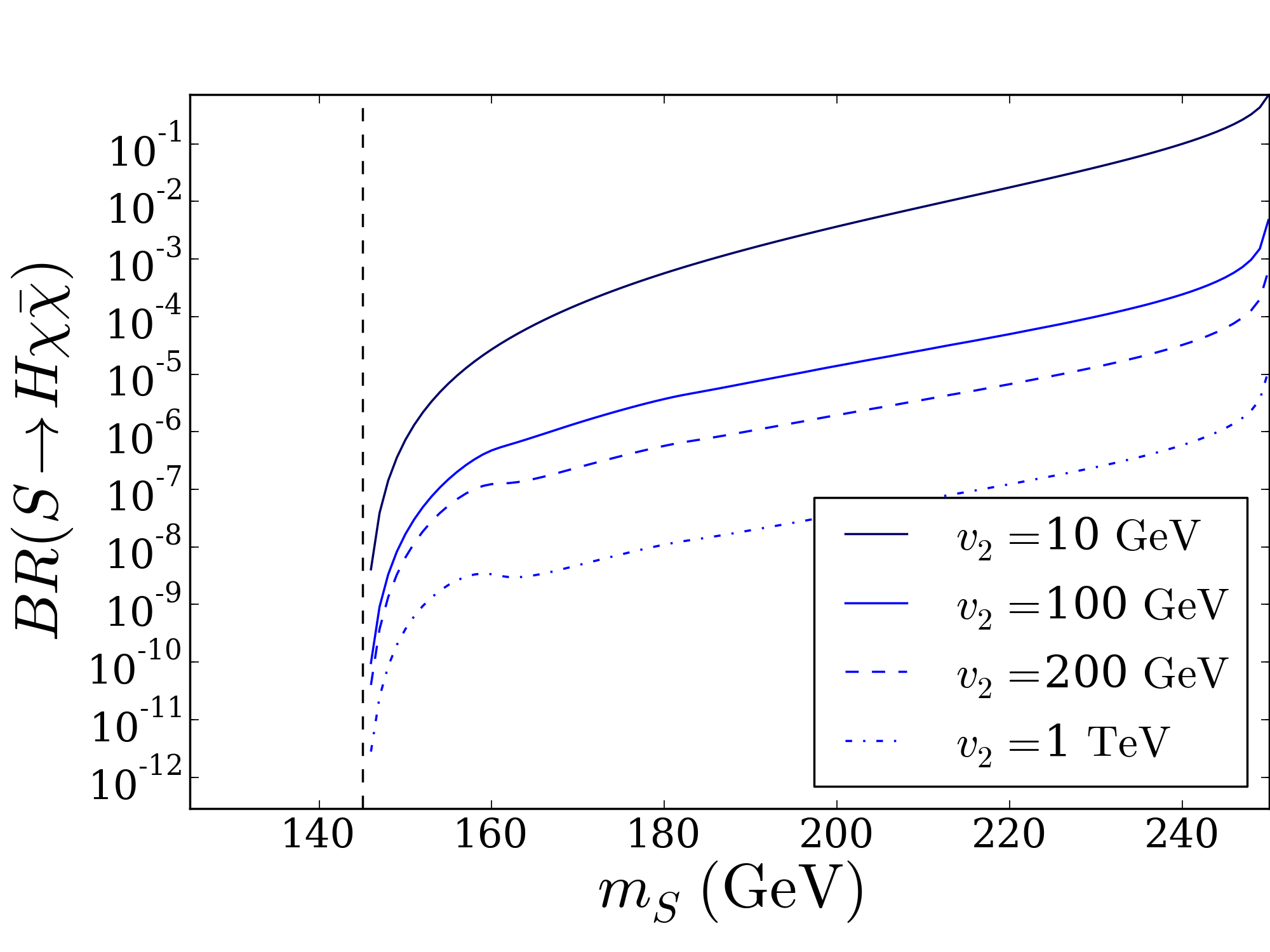}
\end{subfigure}
\end{center}
\caption{Branching ratio for the three-body decays to a scalar and dark matter pair. The
dark matter mass is fixed at $m_{\chi} = 10$ GeV, and the scalar mixing angle
is set to $\sin\theta_h = 0.5$. Branching ratios for $H \rightarrow S \chi \bar{\chi}$ and 
$S \rightarrow H \chi \bar{\chi}$ are shown in the left and right figures respectively.}
\label{fig:3wdthsXX}
\end{figure}

One typically expects the contribution of the three-body widths to be
subdominant, due to suppression by the phase space integration. As previously noted,
for certain regions of parameter space, the tri-scalar couplings may be large
enough to partially offset these effects. In the intermediate mass regions, where the 
three-body decays are kinematically relevant, the magnitudes of these couplings grow
very large only for extremal values of the mixing angle or the scalar vev, however it warrants 
further consideration to verify that the three-body decay widths remain subdominant,
and may be neglected. A more detailed exploration of the relative size of the three-body 
contribution is presented in the following. The partial width of the three-body 
decay is calculated and compared with the more dominant contributions. 

The mass regions of interest relevant for the three-body decays of $H$ and $S$
are the two intermediate ranges --- that is,
$m_H/2 < m_S < m_H$ and $m_H < m_S < 2 m_H$. The different choices of final-state fermion
pair -- either a SM fermion, or $\chi$ --- are presented separately. The decay $H(S)$ to $S(H)$
and a SM fermion pair is shown in figure \ref{fig:3wdthsff}. In this case, the reference
point for comparison is chosen to be the total width for decay to SM particles,
\begin{align}
\Gamma_{SM}^{H} &= c_{\theta_h}^2 \Gamma_{h}^{SM}(m_H) \\
\Gamma_{SM}^{S} &= s_{\theta_h}^2 \Gamma_{h}^{SM}(m_S),
\label{eqn:viswidths}
\end{align}
i.e., the total Higgs width under the SM, scaled by the appropriate factor of the mixing angle. 
This effectively gives the branching ratio under a few simplifications. The other new 
contributions to the total width are the two body decays to a scalar pair, which is kinematically
forbidden in the intermediate mass regions, and to $\chi\bar{\chi}$, which is
neglected for simplicity; the inclusion of the invisible decay to the total width would 
simply result in a further overall reduction of the branching ratio. The result is 
presented as a function of $m_S$, discretely varying the mixing angle.
It can be seen in figures \ref{fig:mucoup} and \ref{fig:etacoup}, that over the 
intermediate mass intervals, both couplings $\mu$ and $\eta$ show little dependence 
on the scalar vev, except for a small increase towards very small values of $v_2$.
Hence, its value is fixed to $v_2 = 200$ GeV.  Several values of
$\sin\theta_h$ are shown for the $H$ decay width, given in the left figure. Despite the non-trivial
functional dependence of the coupling, the primary effect of varying the mixing angle
is an overall scaling of the width, $\propto \sin^2\theta_h$. This dependence vanishes
 in the case of the corresponding $S$ decay, due to the factor of $\sin^2\theta_h$ 
in the normalizing visible width, as seen in eq. \ref{eqn:viswidths}. 
A very minimal effect was seen upon varying the mixing angle, and so only one value is shown. 

When considering the analogous decay to $\chi$ rather than a SM fermion, the dark matter
mass enters into the three-body width; the invisible decay is thus included in the
analysis as the parameter space is already extended. The normalizing quantity is chosen
as the total width, giving the branching ratio for $H(S) \rightarrow S(H) \chi\bar{\chi}$ . The
results are shown in figure \ref{fig:3wdthsXX}. In the given result, the value of $m_{\chi}$
was chosen to be $10$ GeV. The primary influence of the $\chi$ mass on the magnitude 
of the three-body width is through its contribution to the coupling, through the ratio
$m_{\chi}/v_2$, and results primarily in an overall scaling. For most of the scalar 
mass range of interest however, the values of the
dark matter mass for which the relevant decays are kinematically permitted, are at most 
of order $\sim 10$ GeV, i.e. $m_{\chi} \lesssim 50$ GeV. Hence, as the order of $m_{\chi}$
does not change drastically in comparison to the variation of the scalar vev, its
value is fixed such that it is near the larger end of the allowed range, and the
three-body decays of both $H$ and $S$ are allowed over most of the $m_S$ range considered.
In keeping with the aim of investigating the possible extremal values of these partial widths,
a further simplification is made with respect to the mixing angle. The most apparent effect
of the mixing angle on the overall magnitude is an overall scaling by $\sim \sin\theta_h$;
therefore it is set to roughly what will be later shown to be the maximum value allowed by
experimental constraints, with $\sin\theta_h=0.5$.

Over most of parameter space, the three-body contributions to the $H$ and $S$ widths
are considerably small; except for large values of the mixing angle or small values of the vev,
and values of $m_S \simeq m_H/2$ or $\simeq 2 m_H$, in most cases $BR \lesssim 10^{-4}$. 
For most parameter values, one is justified in neglecting these contributions in calculating
the total width. 
In the case of the decay involving a $\chi\bar{\chi}$ pair, the branching ratio does become more
significant for very small values of the new scalar vev, i.e. $\mathcal{O}(10)$ GeV, or more specifically,
when $v_2 \sim m_{\chi}$. However, as will be discussed later, such small values of $v_2$ are
disfavoured by other experimental and theoretical limits, and the remaining analysis
focuses on vevs of order $100$ GeV and larger, neglecting these three-body contributions. 
The couplings $\mu$ and $\eta$ do assume larger values for regions of parameter space over
which the three-body decays are relevant, but they only become very large (i.e. orders of
magnitude larger) in regions where these decays are kinematically forbidden, and the
quantity of interest is not the total widths. This may have relevance towards off-shell scalar
production processes, a topic which will be further discussed in later
sections.

\section{LHC Limits}
\label{scn:LHC}

Current LHC Higgs data have been found to be consistent with the SM Higgs sector. 
Such a result places limits on a scalar extension, both 
with respect to the modification of production cross sections of the SM-like Higgs,
as well as the fact that a second scalar has evaded discovery as yet.
This scenario, in which a new scalar has escaped detection by hadron colliders at
the currently achieved energies and luminosities, generally constrains either the mixing
parameter to be small or the scalar mass to be heavy. The limits posed by 
LHC Higgs data are reviewed here, and limits resulting from new contributions
to the Higgs widths are determined.

\subsection{Higgs Signal Strength and Exclusion Searches}
\label{sscn:HBHS}

\begin{figure}
\begin{center}
   \begin{subfigure}{0.46\textwidth}
   \includegraphics[width=\linewidth]{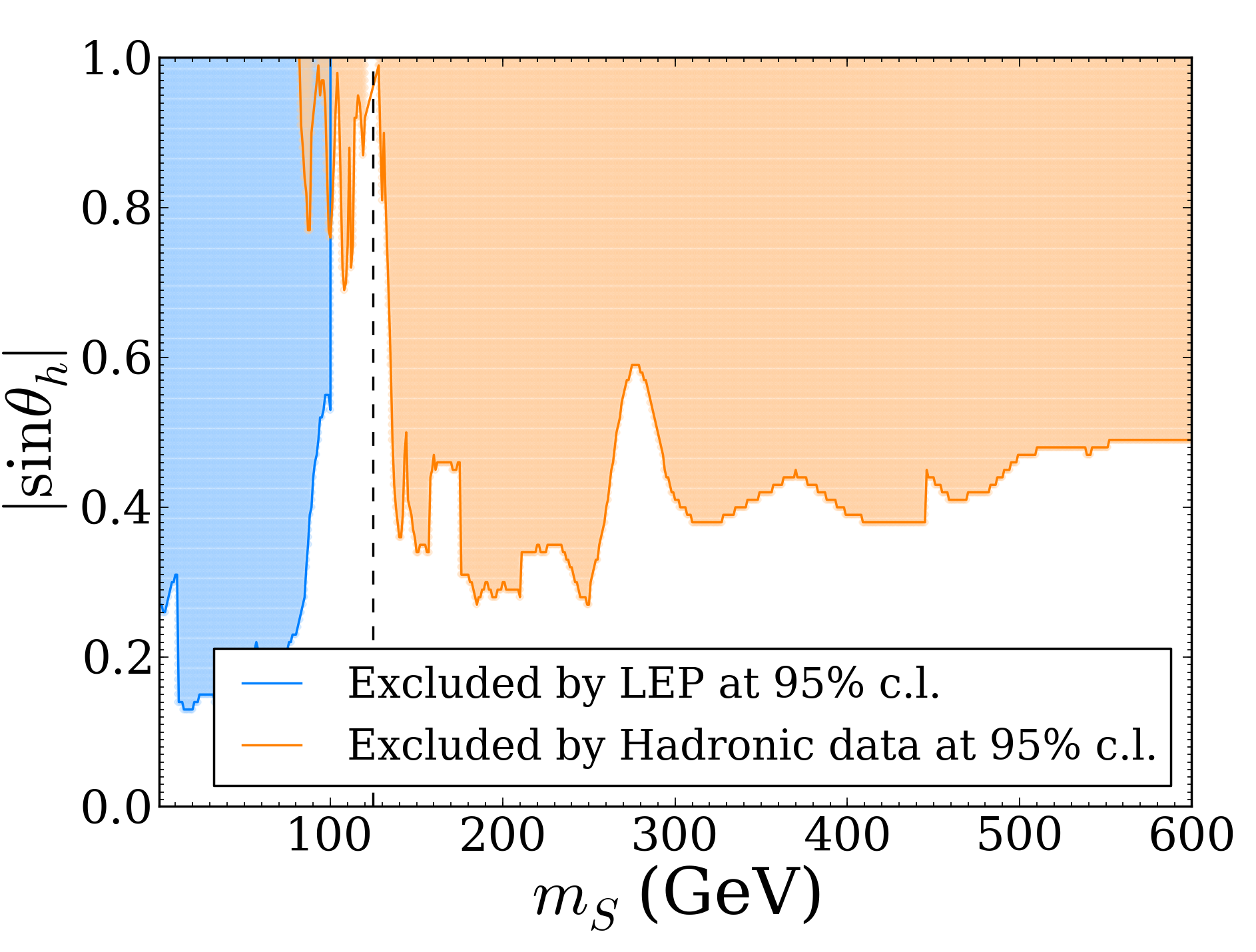}
   \end{subfigure}
   \begin{subfigure}{0.47\textwidth}
   \includegraphics[width=\linewidth]{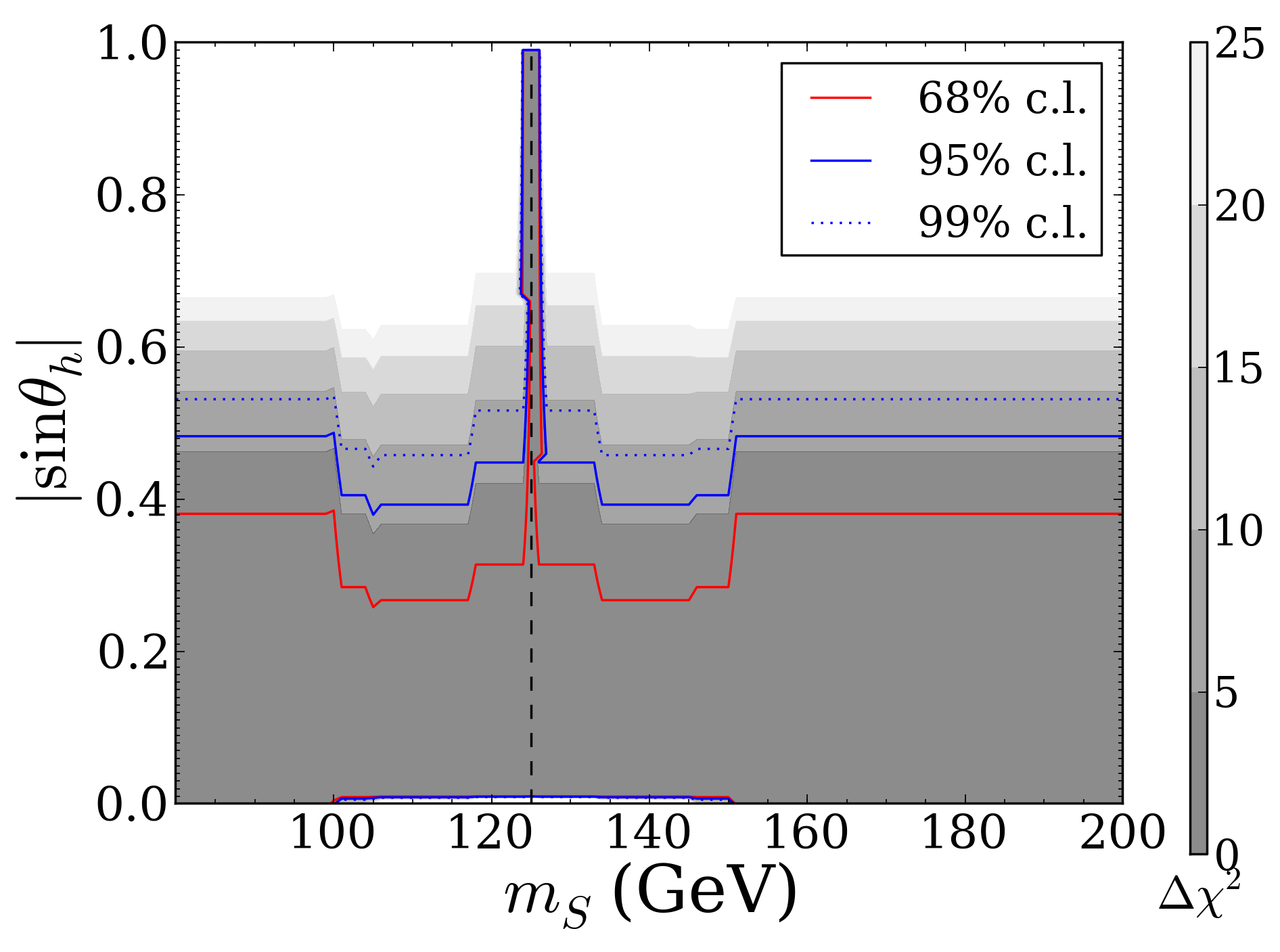}
   \end{subfigure}
\end{center}
\caption{Limits on scalar parameter space, $(m_S, |\sin \theta_h|)$ from current collider 
data. Left: Regions excluded by LEP and Tevatron/LHC Higgs exclusion searches, at 
$95\%$ c.l. Right: Preferred regions compatible with measured LHC Higgs signal strength. 
The regions allowed at $68\%$, $95\%$, and $99\%$ c.l. are shown.}
\label{fig:HBHSlim}
\end{figure}

In the following section, the compatibility of a mixed two-Higgs scenario with current 
hadronic collider data is reviewed. The scalar parameter space is strictly constrained
by Higgs searches, with respect to both past exclusion searches, and measured signals,
particularly in light of the recent Higgs discovery.
The precise limits on a scalar singlet extension posed by LHC Higgs data have been 
determined by ref. \cite{Robens:2015gla}.  
Since publication of this work, updated LHC Higgs results
have been released, as well as a version of the publicly available code used throughout the analysis, which incorporates these new data. 
There is also a slight distinction from the present analysis,  in the convention 
chosen to define the mass eigenstates, as well as the inclusion here of the nonzero 
$H \rightarrow SS$ and $S \rightarrow HH$ branching ratios.
\footnote{Ref. \cite{Robens:2015gla} takes the scalar defined by
$c_{\theta_h} h_1- s_{\theta_h}h_2$ to be the lighter of the two, rather than fixing it to be the $125$ GeV boson, as is done here,
effectively switching $\sin\theta_h \rightarrow \cos\theta_h$, for $m_S < 125$ GeV.}
In light of this, the limits are reproduced here for completion, under these
modifications, following the procedure of \cite{Robens:2015gla}.
\footnote{Ref. \cite{Martin-Lozano:2015dja} conducts a similar analysis, 
using an alternative approach to the signal strength limit.}
The publicly available code 
\textsc{HiggsBoundsv4.3.1} \cite{Bechtle:2015pma,arXiv:1311.0055,arXiv:1301.2345,
arXiv:1102.1898,arXiv:0811.4169} is used to determine the limiting region in the space
defined by the scalar mass and mixing angle, by exclusion from LEP, Tevatron and LHC
data. In particular, LEP searches from refs.
\cite{Searches:2001aa,Abbiendi:2002qp,Searches:2001ab,Abdallah:2003ry,Abbiendi:2007ac,Searches:2001ac,
Abbiendi:2001kp,Achard:2004cf,Schael:2006cr,Abdallah:2003wd,Abdallah:2004wy} are used, 
and experimental LHC/Tevatron results are those in refs. 
\cite{ATLASnotes, LHWGnotes, Aad:2014xva, Chatrchyan:2012sn,Chatrchyan:2014tja, Aad:2012tfa,
Aad:2012tj, Chatrchyan:2012tx,Aad:2011ec,ATLAS:2012ac,ATLAS:2011af, Khachatryan:2014wca, 
ATLAS:2011ae, Chatrchyan:2012dg,Aad:2014fia,Aad:2014yja,Aad:2014iia, Aad:2011rv,ATLAS:2011aa, 
ATLAS:2012ad, Chatrchyan:2013vaa, ATLAS:2012ae,Aad:2014ioa, Chatrchyan:2012ft, CDFnotes, D0notes, 
Abazov:2010ci, Aaltonen:2009vf, Abazov:2010zk, Aaltonen:2011nh, Aaltonen:2010cm,Benjamin:2011sv,
Group:2012zca,Aaltonen:2009ke,Abazov:2011qz,Abazov:2008wg,Aaltonen:2012qt, Abazov:2009yi,
Benjamin:2010xb,Abazov:2011jh,Abazov:2011ed, TEVNPHWorking:2011aa,Abazov:2010hn,Aaltonen:2008ec,
Aaltonen:2009dh,Abazov:2009aa,Abazov:2010ct,Abazov:2009kq}, with more recent LHC
results from Refs. \cite{Aad:2014vgg, arXiv:1509.04670,arXiv:1506.02301,arXiv:1504.00936,
arXiv:1507.05930,arXiv:1509.00389} .
The region allowed at $95\%$ c.l. is shown in figure \ref{fig:HBHSlim}, with regions excluded
by LEP and Tevatron/LHC presented separately.     

While the limits posed by exclusion searches constrain the model in the absence of a 
signal, the recent Higgs discovery necessitates the complementary limits from
compatibility with this observed signal. Results are determined using 
\textsc{HiggsSignalsv1.4.0} \cite{Stal:2013hwa, Bechtle:2013xfa}. 
The experimental results used to obtain the constraints are given in refs.
\cite{Aad:2015gba, Aad:2014eha, Aad:2015vsa, ATLAS:2014aga, ATLAS-CONF-2015-005,
Aad:2014eva, Aad:2014xzb, Khachatryan:2014ira, Chatrchyan:2014vua, Chatrchyan:2013iaa,
Chatrchyan:2013mxa, CMS-PAS-HIG-13-005}.
For light scalars, $m_S < 80$ GeV, the strictest limit is found by LEP exclusion, giving an 
upper bound of $|\sin\theta_h| \lesssim 0.15$, while the upper limit from LHC exclusion searches
is between $|\sin\theta_h | \lesssim 0.3$ at its most stringent, in the region $m_S \sim 170$--$250$ GeV,
and is as large as 0.4--0.6 for higher masses. 
The constraint posed by the measured signal strength in Higgs
production at LHC, which places an upper bound (at $95\%$ c.l.) on the mixing angle of 
$| \sin\theta_h | \lesssim 0.5$, for scalar masses $m_S < 100$ GeV or $m_S > 150$ GeV, 
and $| \sin\theta_h | \lesssim 0.4$ for scalar masses within a $25$ GeV window of $125$ GeV.
When $m_S$ approaches $m_H$, within the detector mass resolution, this represents the 
degenerate limit which simply reproduces the SM signal --- in other words, the 
signal strength for a $125$ GeV scalar with SM couplings scaled by $\cos^2 \theta_h + \sin^2 \theta_h = 1$ ---
hence the mixing angle is unconstrained in this narrow region. Figure \ref{fig:HBHSlim} 
shows the regions preferred by LHC Higgs signal compatibility.  

Additional limits arise from electroweak precision data (EWPD), and theoretical constraints
such as perturbativity and renormalization group (RG) evolution of the couplings, which
may tighten the mixing angle upper bound, to $|\sin\theta_h | \lesssim 0.3$ in the heavier $S$ region
 --- $m_S \gtrsim 600$ GeV \cite{Robens:2015gla}.
Such considerations also place further lower bounds on the $S$ vev in some cases, again disfavouring
small $v_2$; the restrictions were not in conflict with any of the present results.
Similar analyses in refs. \cite{ Martin-Lozano:2015dja, Falkowski:2015iwa} also 
include such theoretical considerations, obtaining similar bounds.

\subsection{Invisible Signals}
\label{sscn:invis}

\begin{figure}
\begin{center}
  \begin{subfigure}{0.36\textwidth}
    \includegraphics[width=\linewidth]{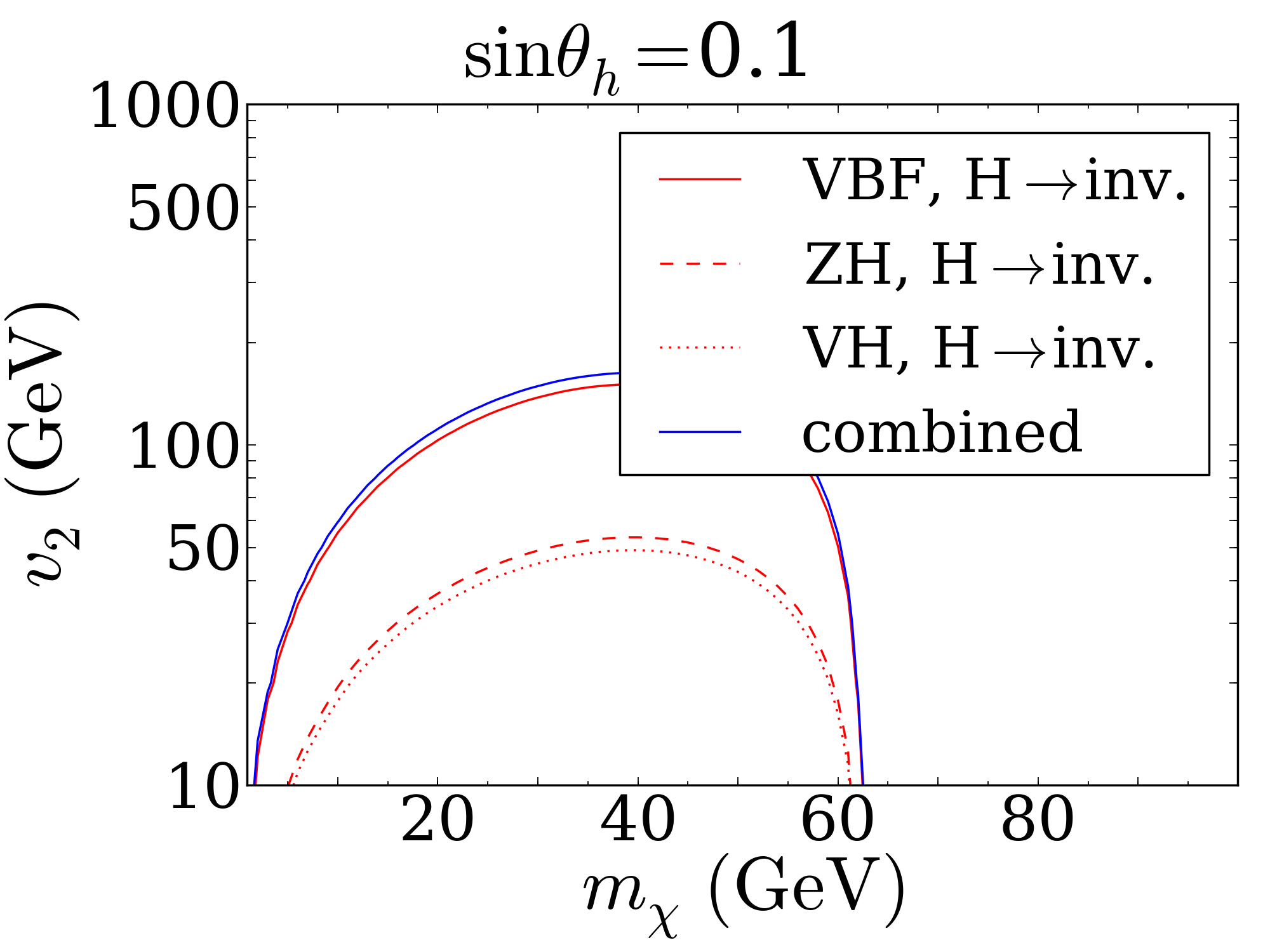}
  \end{subfigure}
  \begin{subfigure}{0.3\textwidth}
    \includegraphics[width=\linewidth]{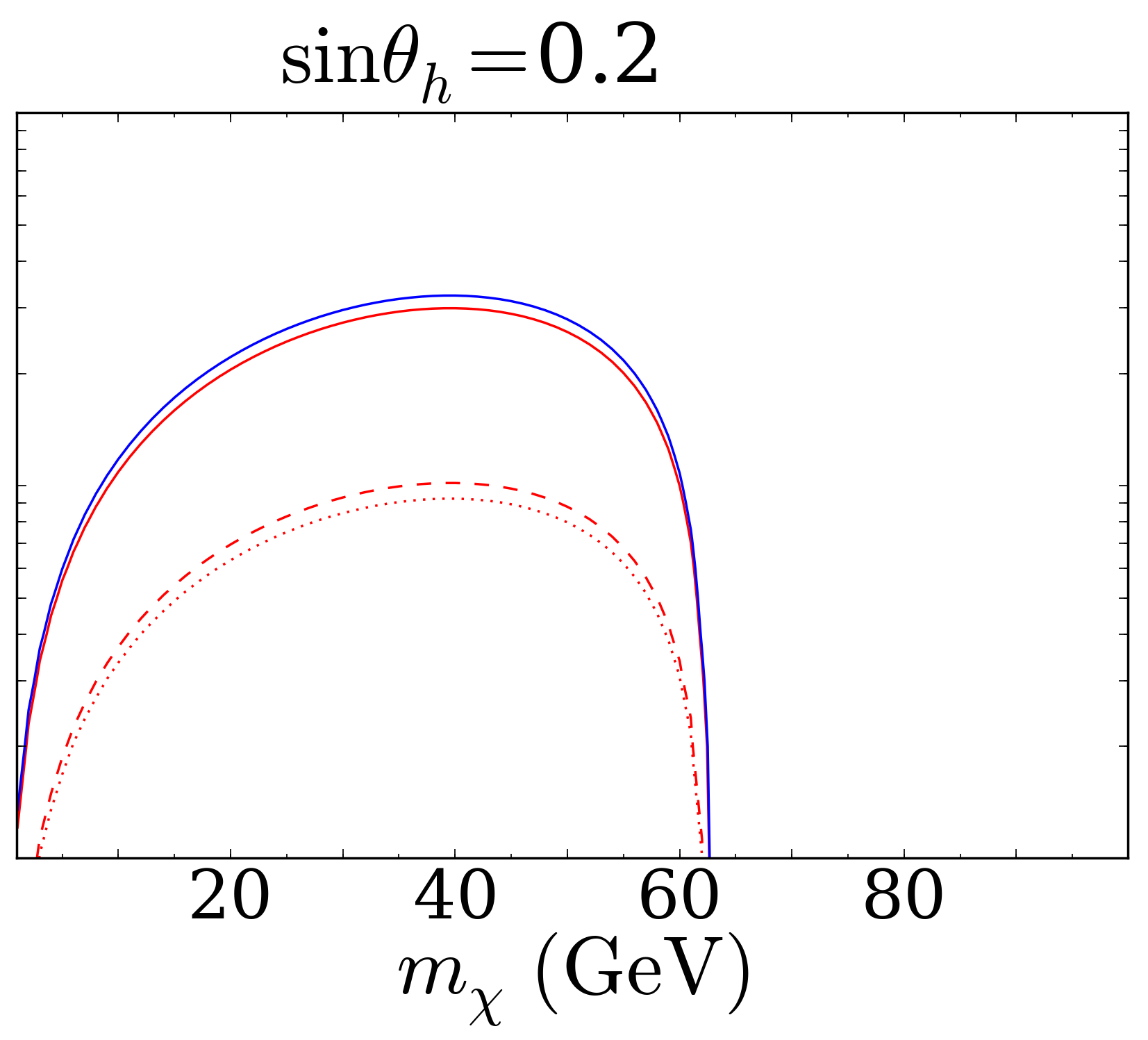}
  \end{subfigure}
  \begin{subfigure}{0.3\textwidth}
    \includegraphics[width=\linewidth]{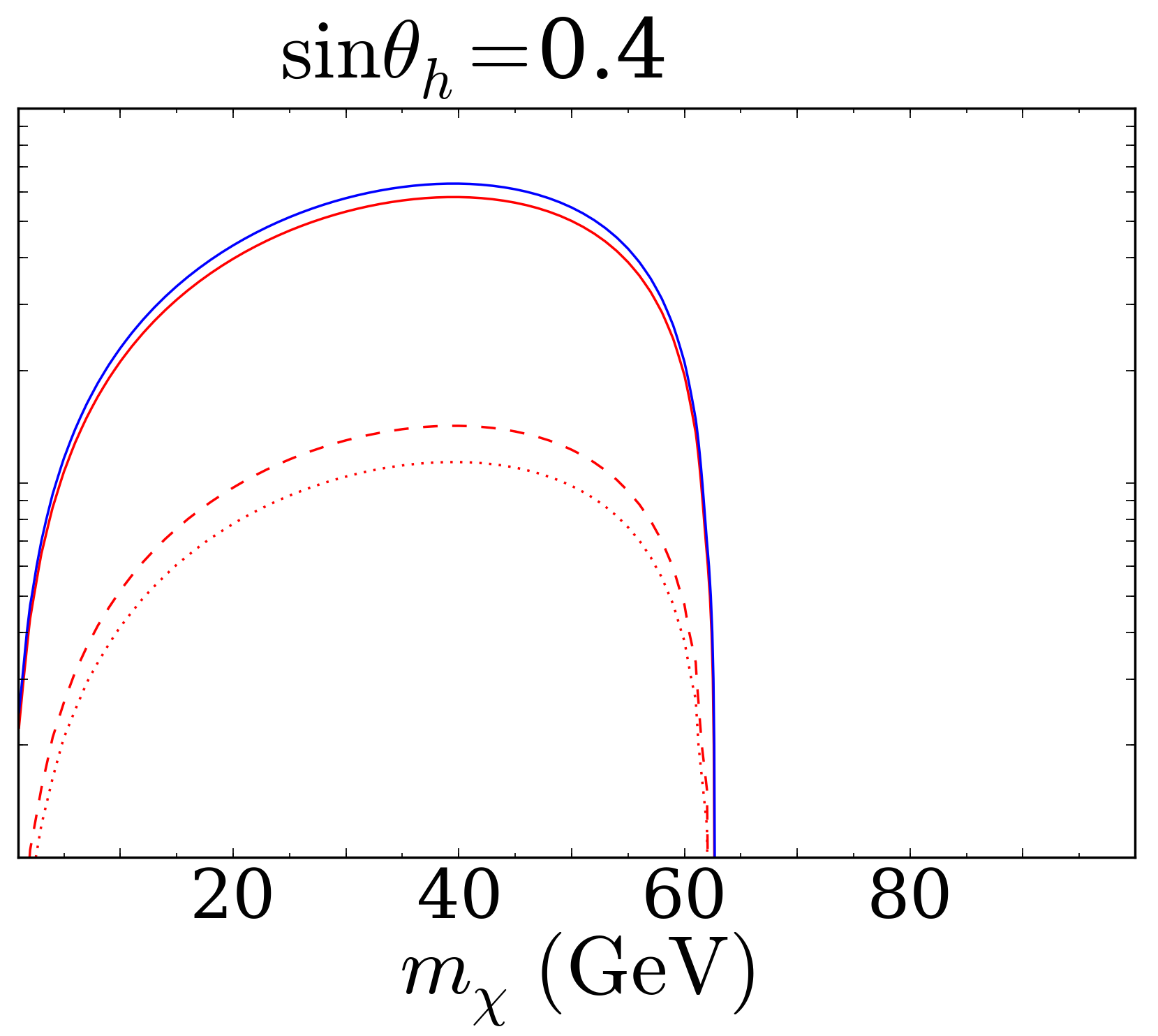}
  \end{subfigure}
\end{center}
\caption{Limits on $(m_{\chi}, v_2)$ parameter space, from ATLAS upper bounds on signal strength
in Higgs invisible decays, for discrete values of the mixing angle, below the 
upper bound set in section \ref{sscn:HBHS}. The regions below the curves are 
excluded at $95\%$ c.l.}
\label{fig:invisWdth}
\end{figure}

The extension to the SM scalar sector results in a strict constraint 
posed by Higgs production searches and measurements, due to scaling of Higgs couplings
 to SM particles and the allowed mass of a new Higgs-like scalar; the allowed values
of the scalar parameters are described in the previous subsection. 
The remaining parameter --- the dark matter mass --- becomes significant 
when focusing on searches with invisible signatures. A direct limit on the dark matter
mass, as well as the additional scalar vev, is found in Higgs production with subsequent invisible
decay channels.

A recent ATLAS search \cite{Aad:2015pla} for invisible Higgs decays, in both vector
boson fusion (VBF), and associated vector boson production modes, gives an upper
limit on the Higgs invisible branching ratio. ATLAS obtains limits from three
production channels --- vector boson fusion, associated Z production with subsequent
leptonic Z decays, and associated vector boson, $V$ ($W$ or $Z$), with hadronic $W$/$Z$
decay, as well the combined limit from the three. The search corresponds to $4.7$ fb$^{-1}$
of data at centre of mass energy $7$ TeV, and $20.3$ fb$^{-1}$ at $8$ TeV. Upper
limits are given at $95\%$ c.l. on the production cross section times invisible branching
ratio signal strength. This signal strength, denoted here by $\zeta$ is simply the 
production cross section times branching ratio, normalized by the SM production cross section
\begin{equation}
\zeta = \frac{\sigma}{\sigma_{SM}} \times \mathcal{B}(h \rightarrow \mathrm{inv}).
\label{eqn:zetadefn}
\end{equation}
More specifically, for $H$, this becomes
\begin{equation}
\zeta_{H} = c_{\theta_h}^2 \frac{\Gamma_{\mathrm{inv}}}{\Gamma_{\mathrm{tot}}}
= \frac{c_{\theta_h}^2 s_{\theta_h}^2 \Gamma_{\mathrm{inv}}}{c_{\theta_h}^2 \Gamma_{SM} 
+ s_{\theta_h}^2 \Gamma_{\mathrm{inv}}}. 
\label{eqn:zeta}
\end{equation}

Here, $\Gamma_{\mathrm{inv}}(m_{\chi}, v_2)$ is as given in eq. \ref{eqn:inviswidth}, 
with the factor $s_{\theta_h}$ written explicitly. The corresponding observable for 
$S$ production and decay is obtained under exchange $m_H \rightarrow m_{S}$ and 
$s_{\theta_h} \leftrightarrow c_{\theta_h}$.

ATLAS sets upper limits on the normalized signal strength of $0.28$, $0.75$, and $0.78$ in the
VBF, leptonic $ZH$, and hadronic $VH$ channels, with a combined limit of $0.25$.
The resulting bound on $(m_{\chi}, v_{2})$ is given in figure \ref{fig:invisWdth}.
The $95\%$ c.l. exclusion is shown, using the individual channels, as well the combined 
results. The mixing angle is varied discretely, with values below the upper 
limit obtained in section \ref{sscn:HBHS}. The constraint represents the  
experimental bound to the invisible $H$ width, at mass $125$ GeV. A similar search by 
CMS \cite{Chatrchyan:2014tja} gives an upper bound on the invisible branching ratio as
a function of Higgs mass, although the limit at $m_H=125$ GeV is looser than that given 
by ref. \cite{Aad:2015pla}. As this gives a bound for a general scalar mass, this is 
applied to the invisible $S$ width. For mixing angle values which remain consistent with
the Higgs production signal strength however, the corresponding limit applied to the invisible
branching ratio of $S$ does not significantly constrain the parameter space.

\subsection{Total Higgs Width}
\label{sscn:totalwidth}

An analysis by CMS obtains a limit on the width of the Higgs boson, by a method which uses 
the ratio between on-shell and off-shell cross section measurements \cite{Khachatryan:2015mma}.
CMS obtains an upper limit on the total Higgs width of $26$ MeV, at $95 \%$ c.l.
\begin{figure}
\begin{center}
  \begin{subfigure}{0.36\textwidth}
    \includegraphics[width=\linewidth]{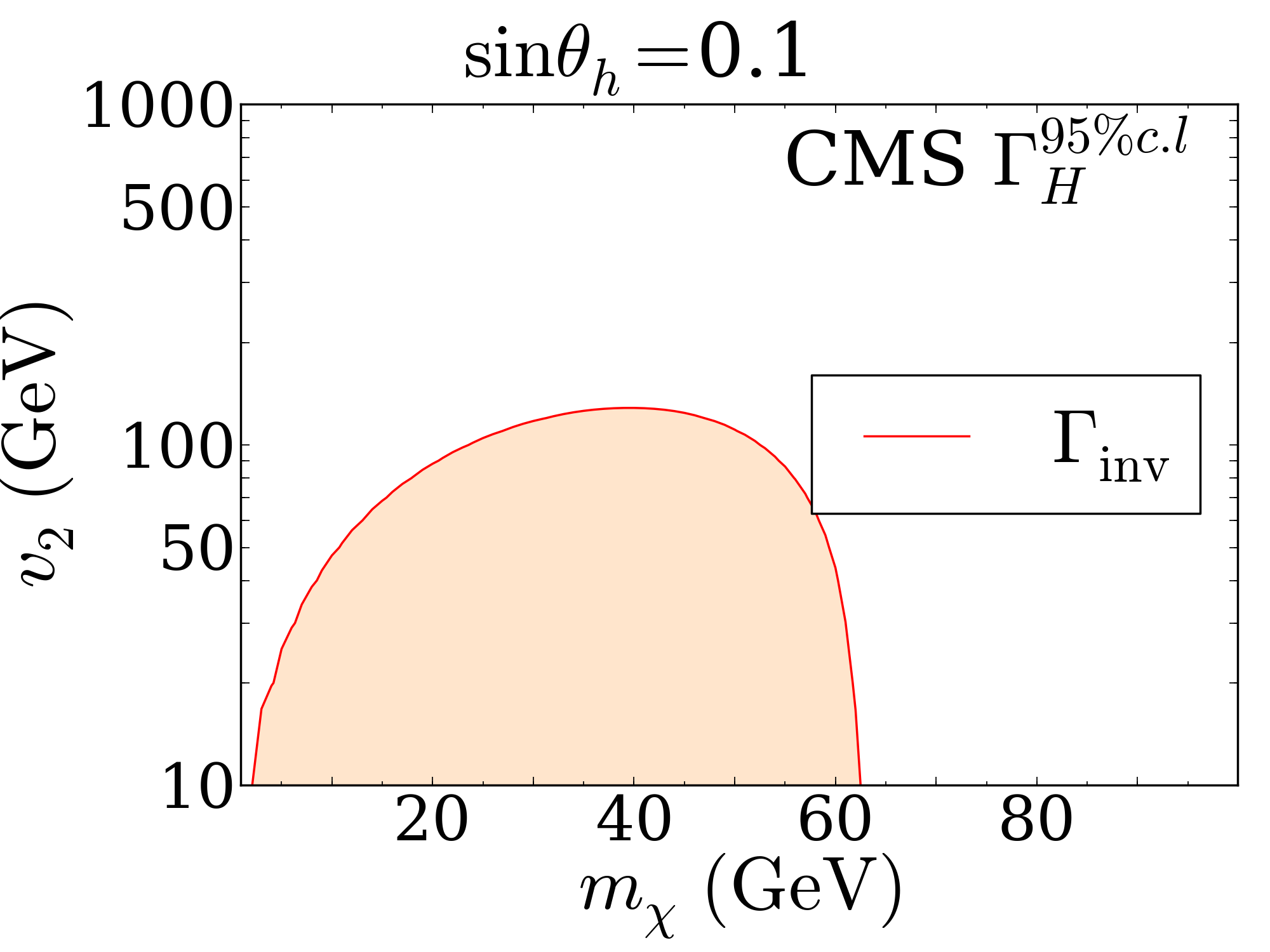}
  \end{subfigure}
  \begin{subfigure}{0.3\textwidth}
    \includegraphics[width=\linewidth]{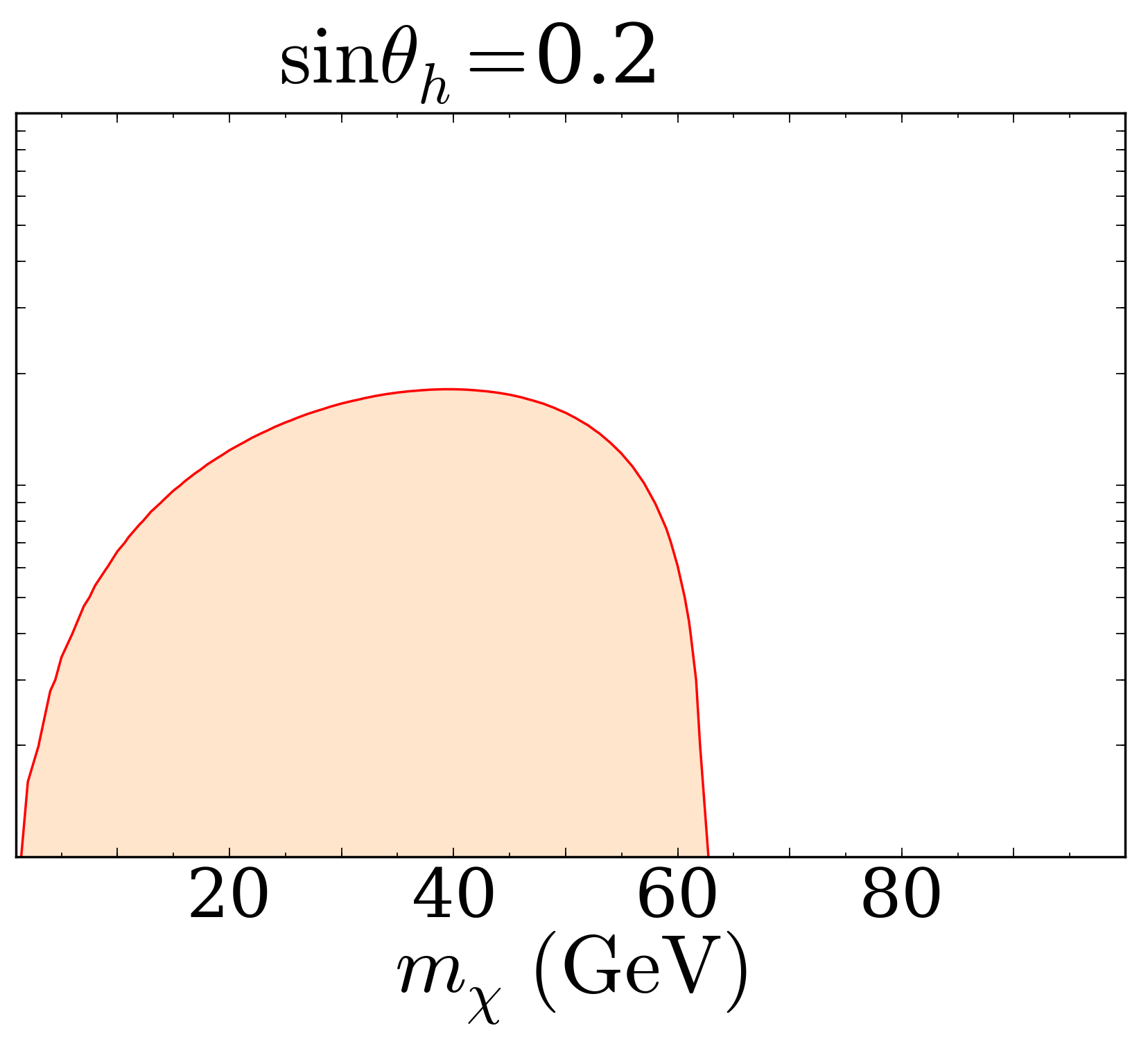}
  \end{subfigure}
  \begin{subfigure}{0.3\textwidth}
    \includegraphics[width=\linewidth]{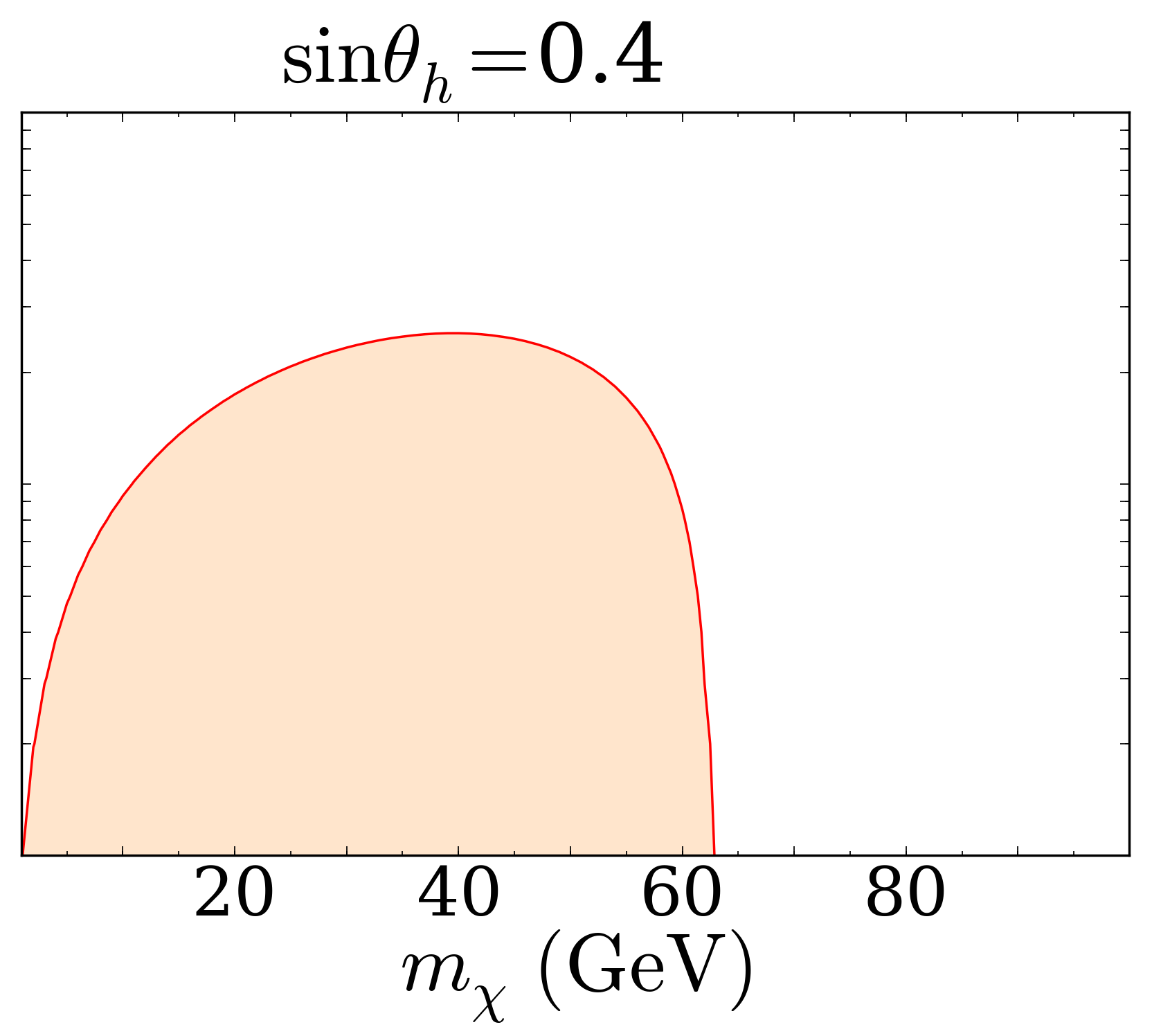}
  \end{subfigure}
\end{center}
\caption{Limit on dark matter parameters resulting from invisible contribution to total 
Higgs width. Results are obtained using the CMS upper bound on the total Higgs width,
assuming the absence of additional two-body scalar decays, i.e. in the heavy $S$ regime.
The shaded regions are excluded at $95\%$ c.l.}
\label{fig:totwidthinv}
\end{figure}
For most of the scalar mass range, this provides an additional constraint on the
dark matter mass and coupling; if $m_{S} > m_{H}/2 \sim 63$ GeV, the decay of $H$ to an $S$ pair is 
kinematically forbidden, and as seen in section \ref{sscn:3wdths}, the three-body 
contributions are negligible.
Assuming that the only non-SM contribution to the Higgs width is the invisible
decay $H \rightarrow \chi \bar{\chi}$, the resulting limit on the dark matter
mass and scalar vev is given in figure \ref{fig:totwidthinv}. The constraint
posed by measured invisible decays however, gives the stricter limit. 

\begin{figure}
\begin{center}
  \begin{subfigure}{0.37\textwidth}
    \includegraphics[width=\linewidth]{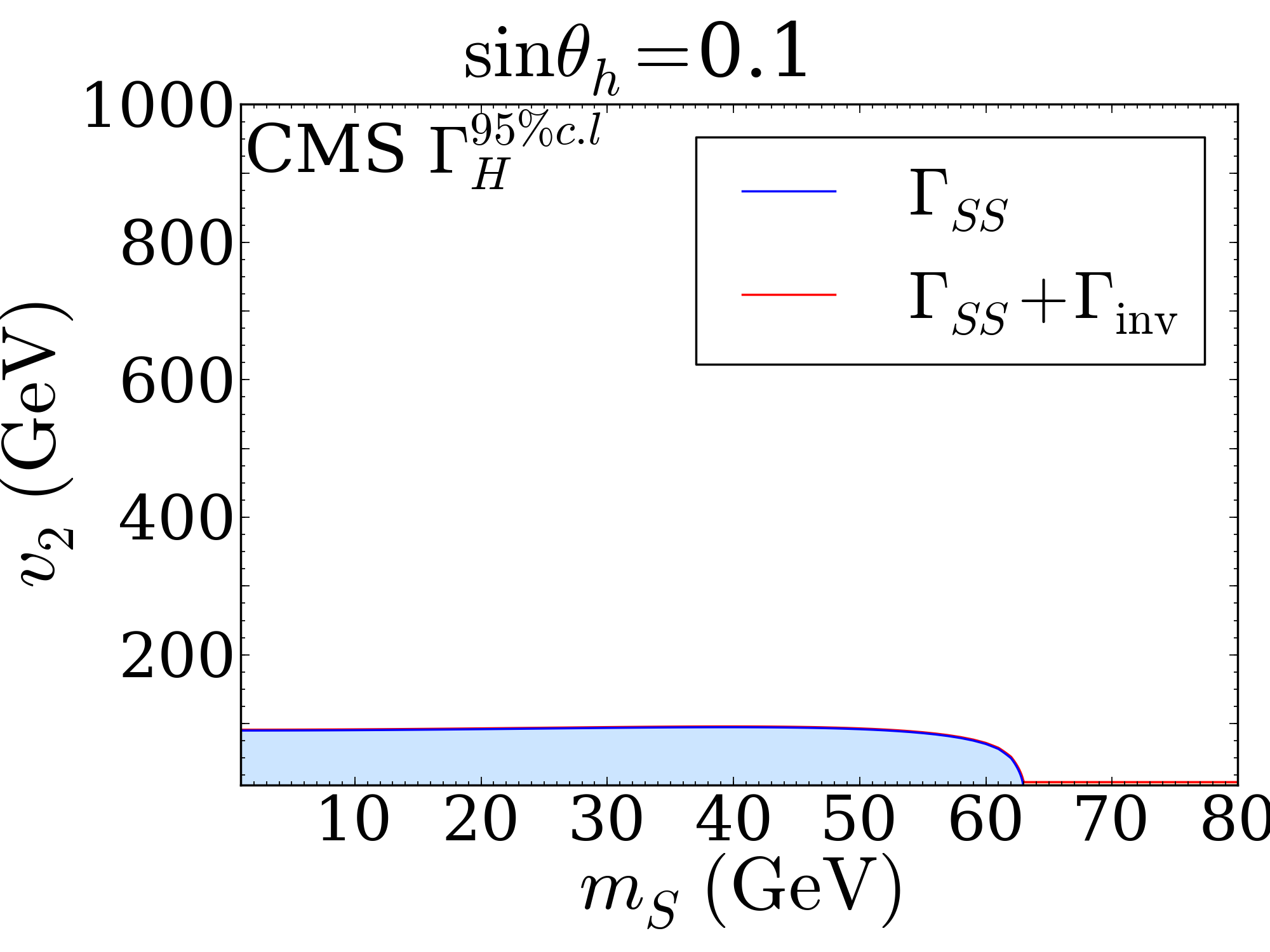}
  \end{subfigure}
  \begin{subfigure}{0.3\textwidth}
    \includegraphics[width=\linewidth]{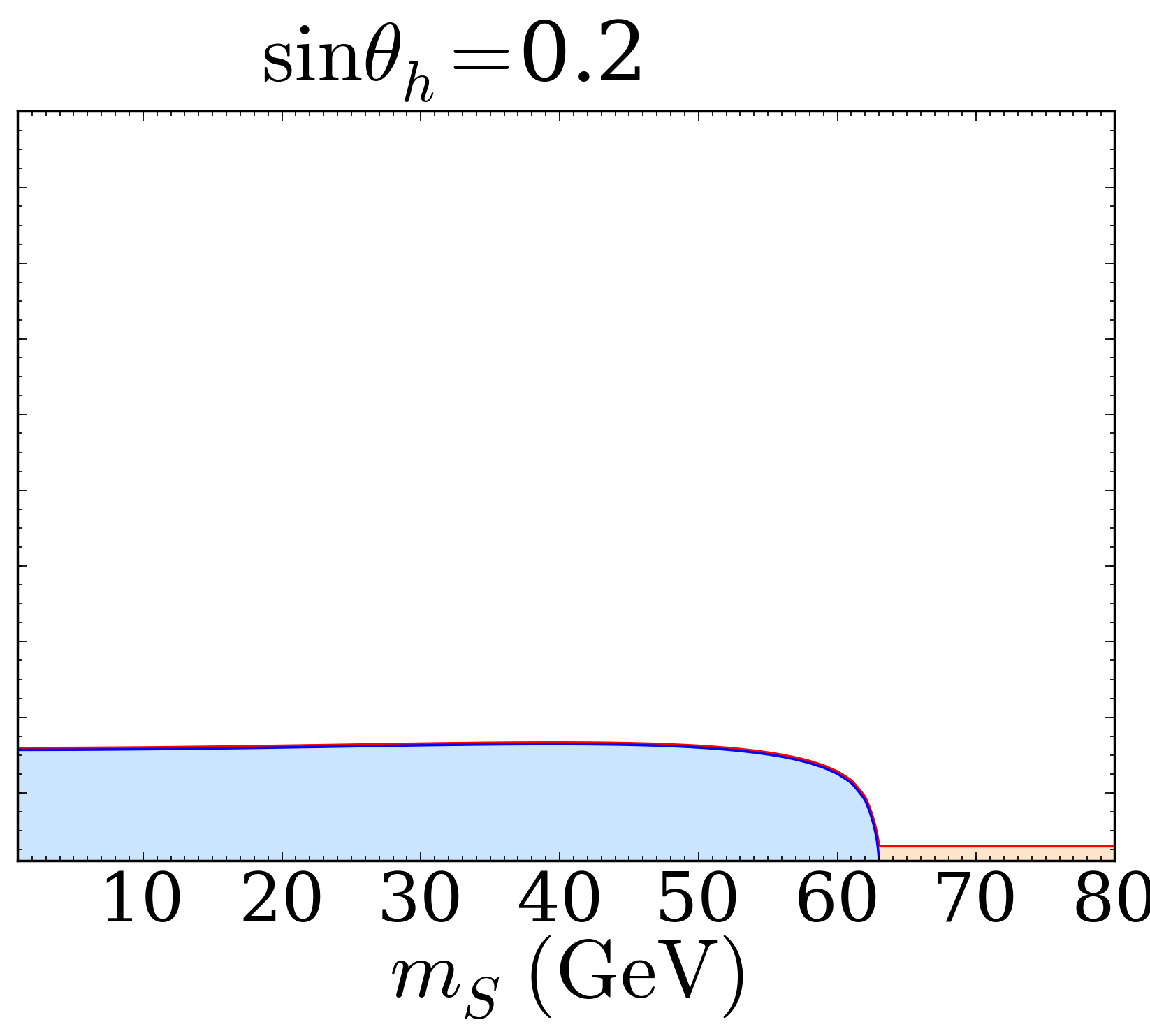}
  \end{subfigure}
  \begin{subfigure}{0.3\textwidth}
    \includegraphics[width=\linewidth]{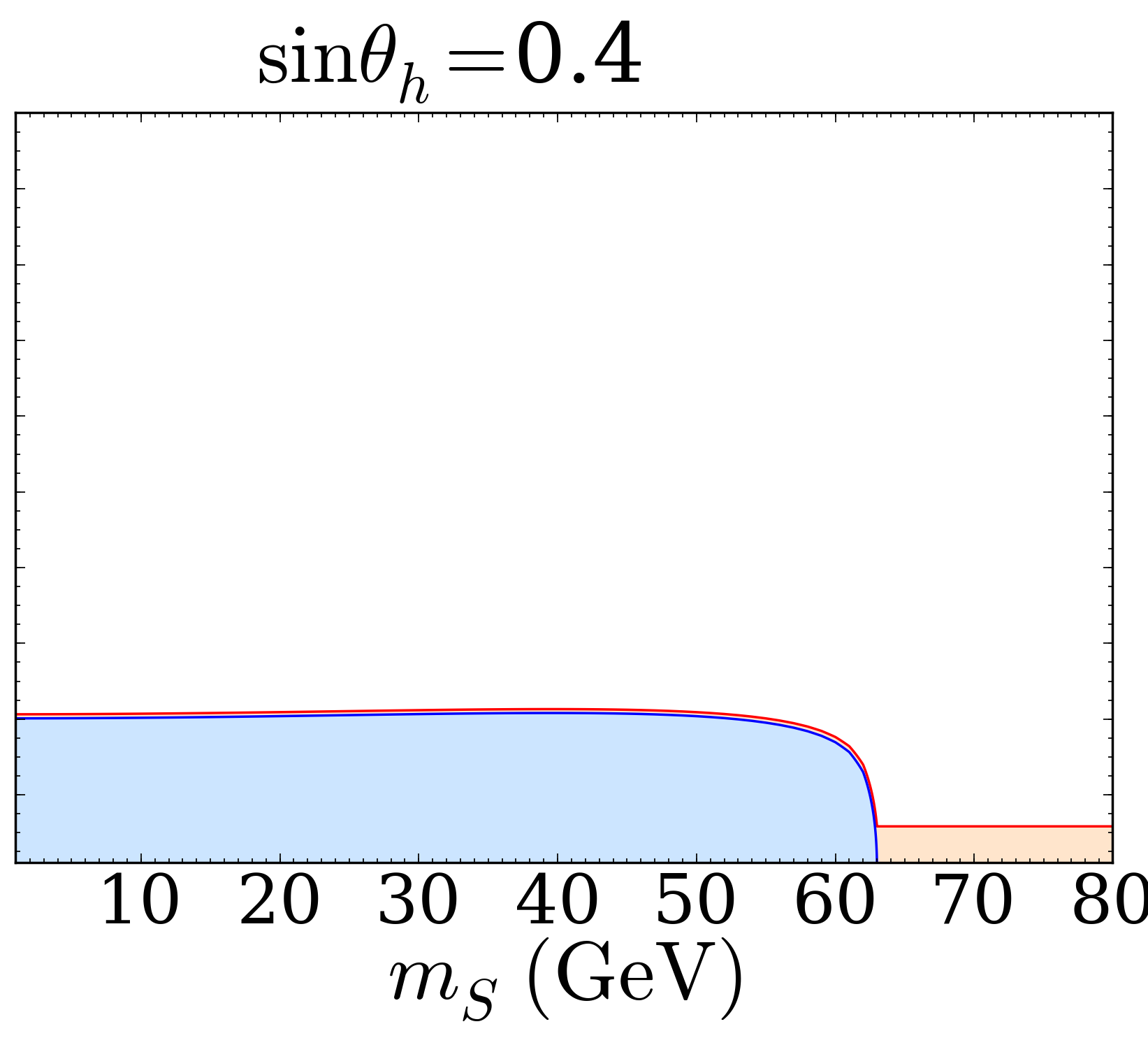}
  \end{subfigure}

  \begin{subfigure}{0.37\textwidth}
    \includegraphics[width=\linewidth]{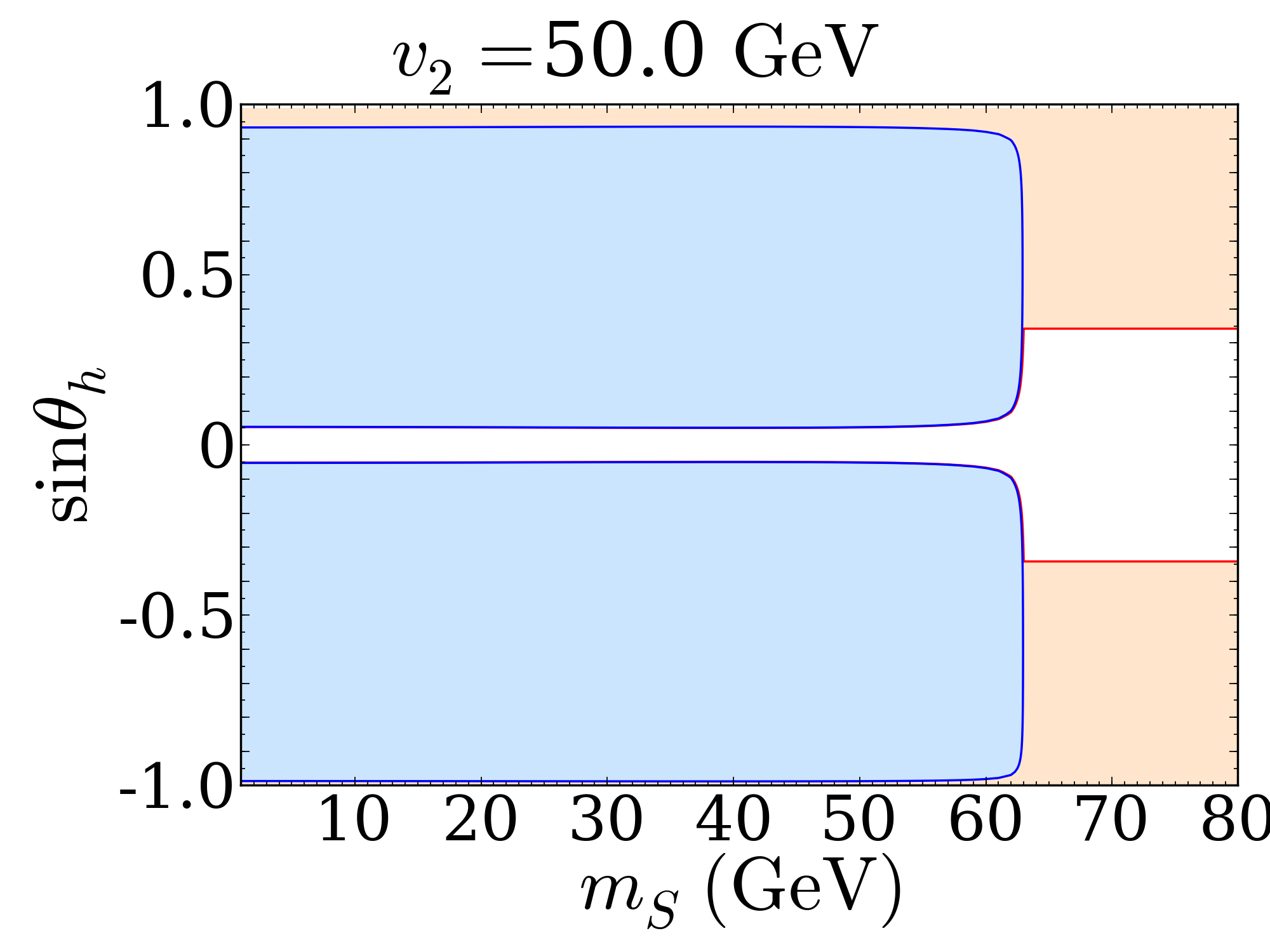}
  \end{subfigure} 
  \begin{subfigure}{0.3\textwidth}
    \includegraphics[width=\linewidth]{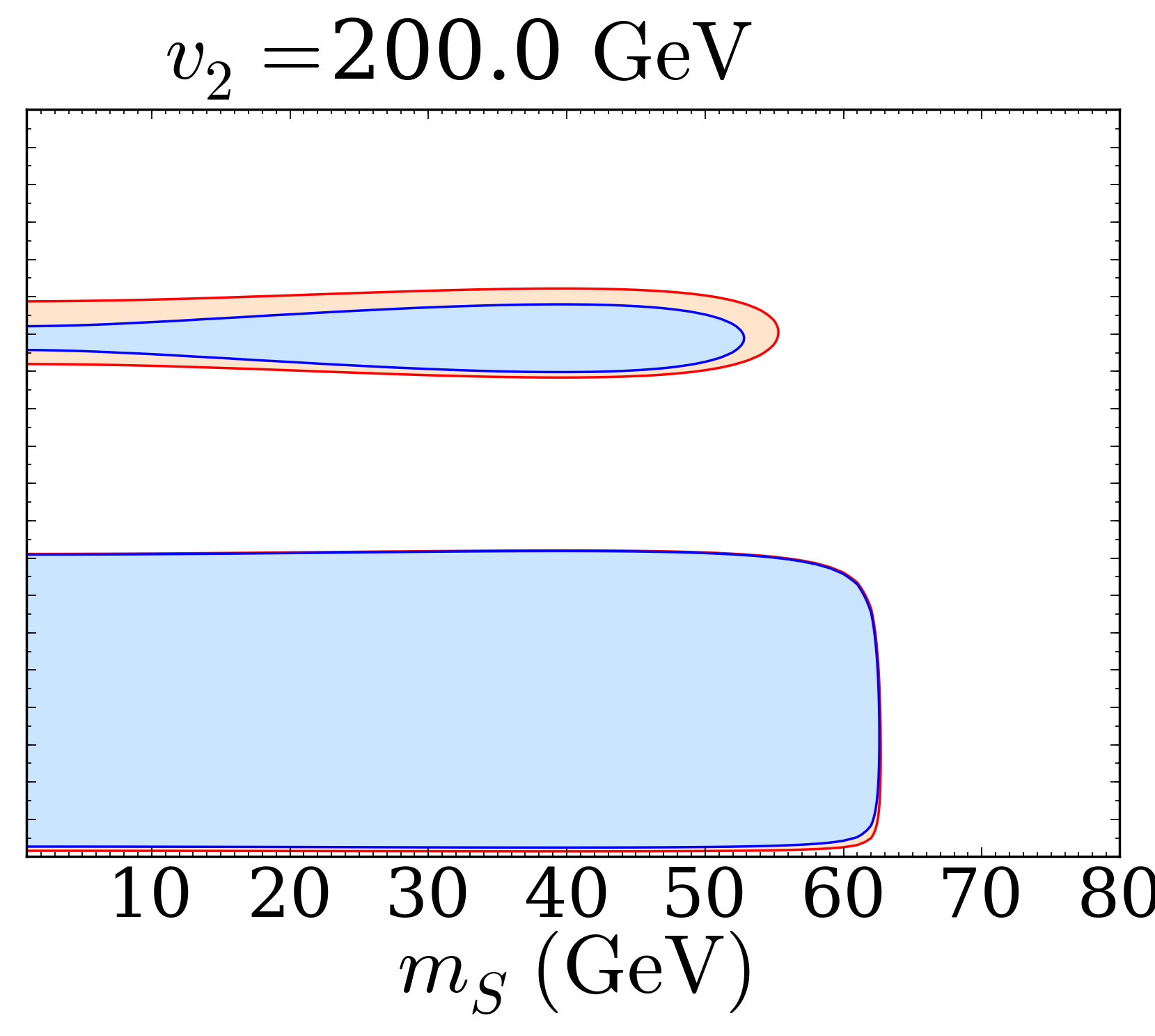}
  \end{subfigure}
  \begin{subfigure}{0.3\textwidth}
    \includegraphics[width=\linewidth]{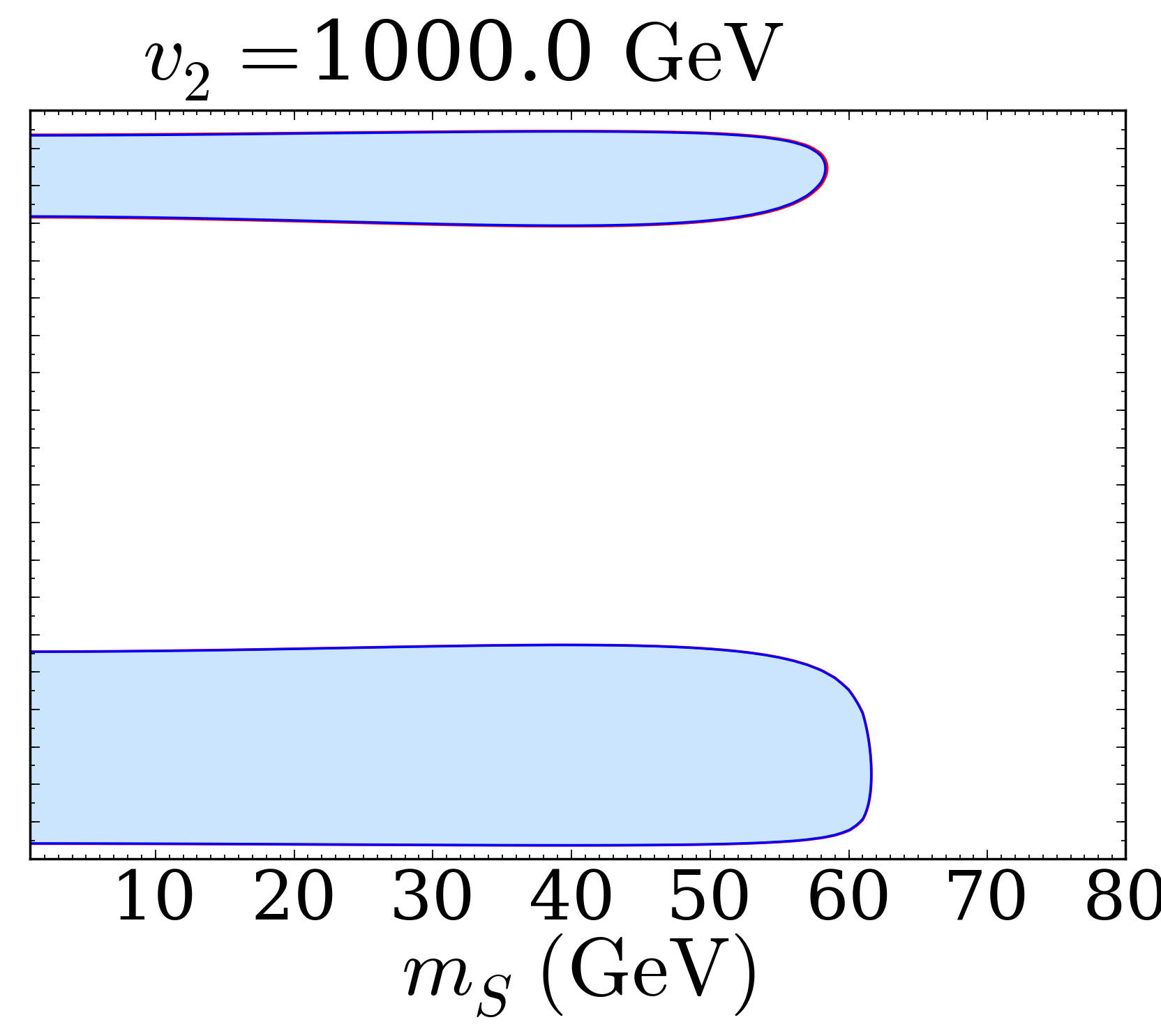}
  \end{subfigure}

  \begin{subfigure}{0.37\textwidth}
    \includegraphics[width=\linewidth]{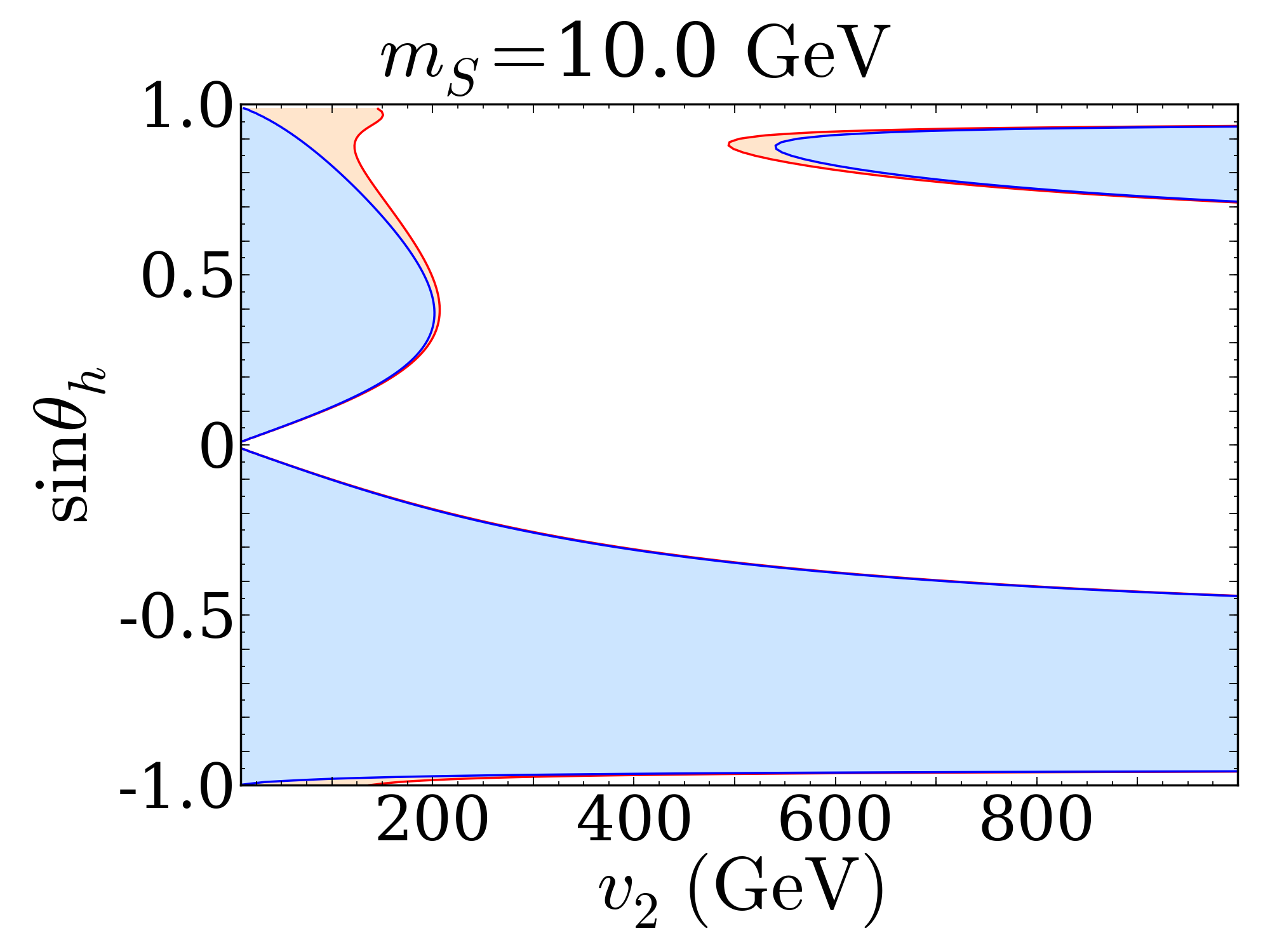}
  \end{subfigure} 
  \begin{subfigure}{0.3\textwidth}
    \includegraphics[width=\linewidth]{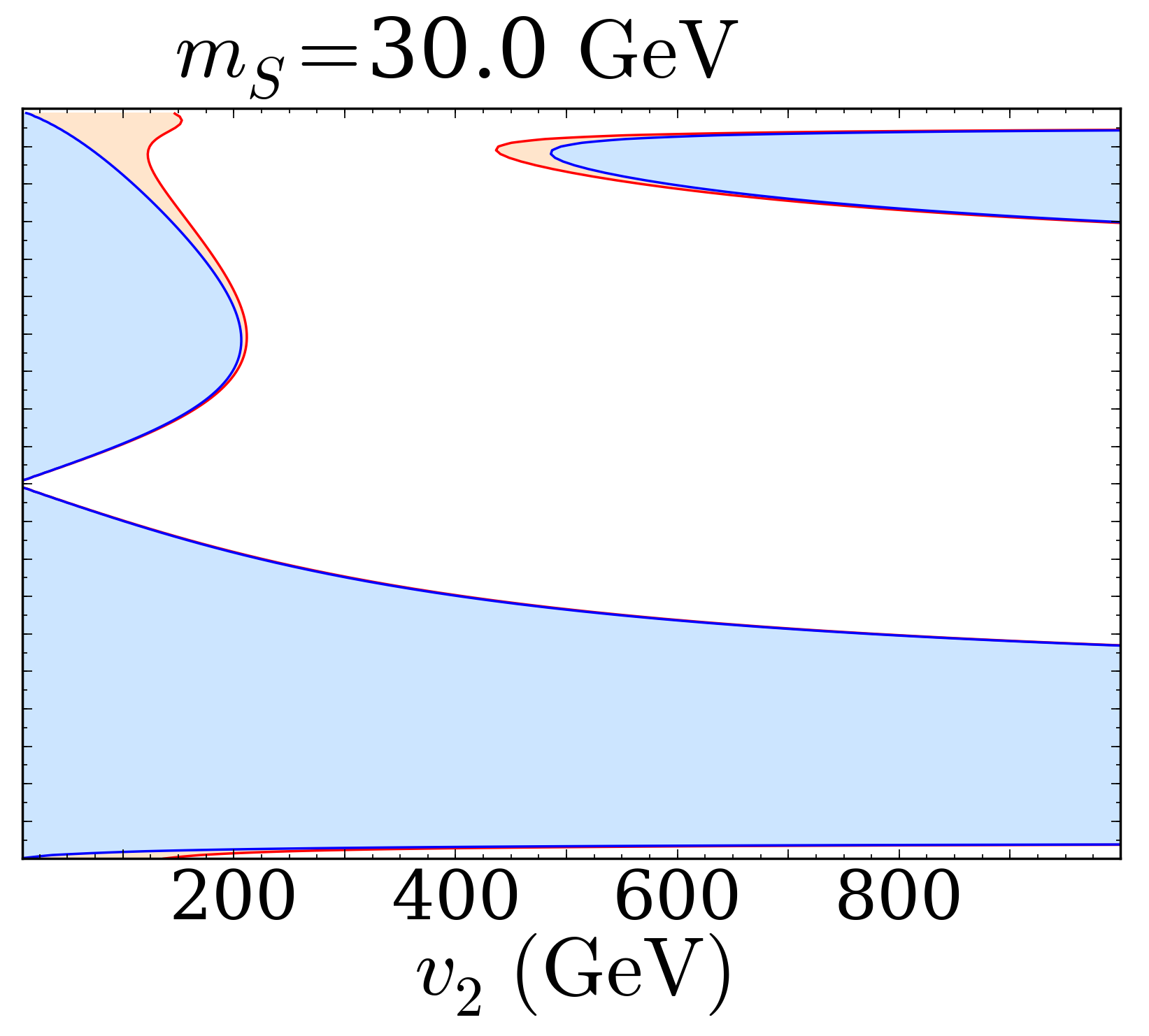}
  \end{subfigure}
  \begin{subfigure}{0.3\textwidth}
    \includegraphics[width=\linewidth]{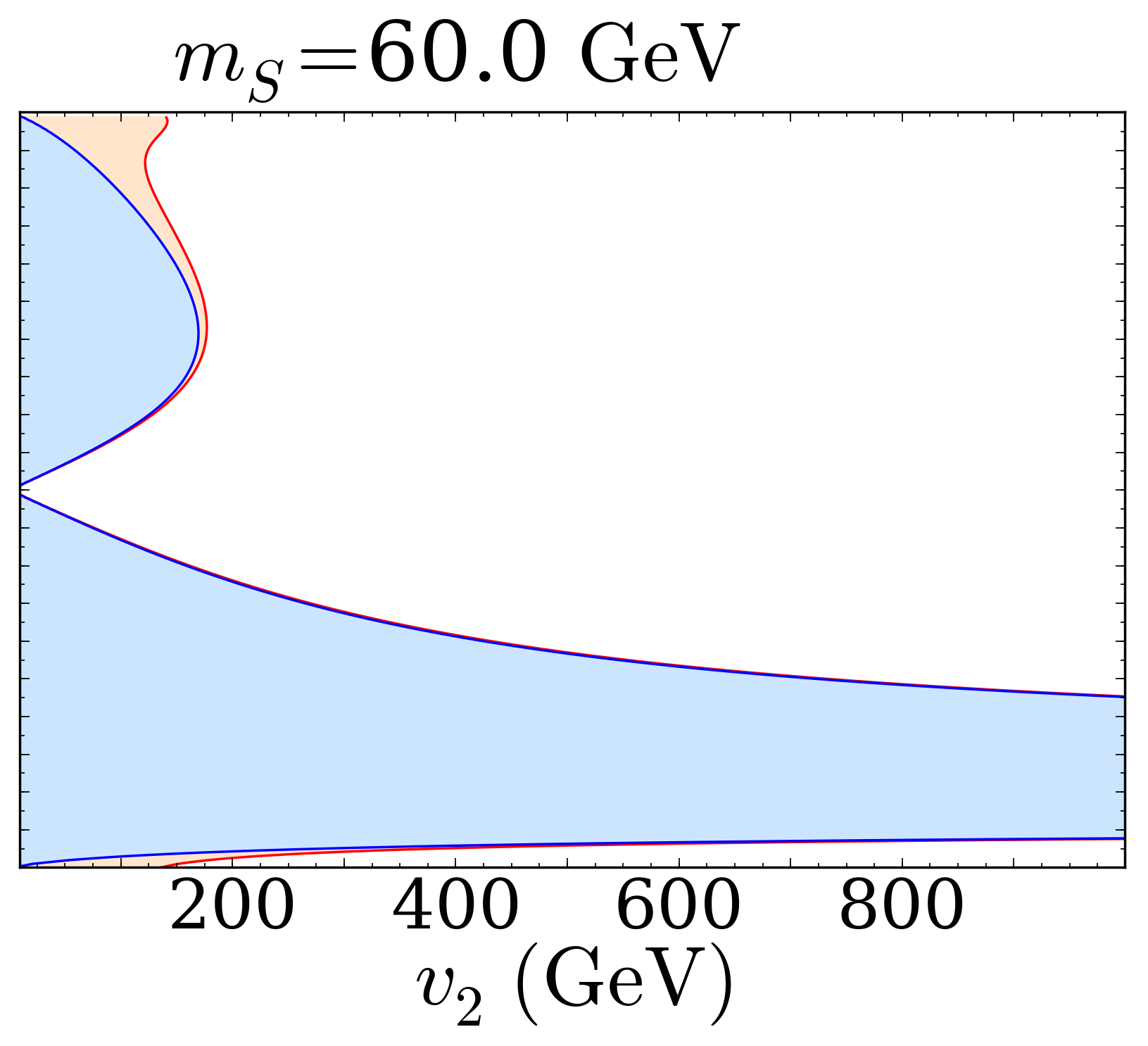}
  \end{subfigure}
\end{center}
\caption{Limit resulting from contribution of $H \rightarrow S S$ decay to total 
Higgs width. The shaded regions represent the region excluded at $95\%$ c.l. by the CMS
limit. From the top row down, $\sin \theta_h$, $v_2$, $m_S$ are each varied discretely in turn.
The blue curve/region corresponds to the case for which the decay to $SS$ is the only non-SM
contribution, and red to the case which also includes the invisible channel.}
\label{fig:totwidth}
\end{figure}

In the light $S$ regime --- that is, in the subregion of parameter space for which
$m_S < m_H/2$ --- the total width also receives a contribution from the decay channel
$H \rightarrow S S$. This provides additional direct experimental sensitivity on the allowed values 
of the scalar vev, as it enters into the mixed scalar coupling, which
depends additionally on the mixing angle and scalar mass. Each of these three
parameters, $(\sin\theta_h, v_2, m_S)$, is varied discretely in turn, giving the
exclusion region in the other two. The results are given in figure \ref{fig:totwidth}.
Both the case for which the scalar decay channel is the only new contribution (which is
the case for heavy dark matter), as well as that for which both scalar and invisible 
decays are present. In the latter case, $m_{\chi}$ is chosen such 
that the invisible contribution is below the upper bound over most of the $v_2$ range
considered, as shown in figures 
\ref{fig:invisWdth} and \ref{fig:totwidthinv}; the value is taken to be $m_{\chi} = 10$ GeV.

The resulting exclusion regions in figure \ref{fig:totwidth} show the extremal  allowed
values of the mixing angle and scalar vev, which are more or less proportionate as each
is varied. The figures in the top row show a lower bound on $v_2$, which increases with
the mixing angle, and remains constant with $m_S$, in the kinematically allowed range.
Varying $v_2$ discretely (shown in the middle row), one sees an upper bound on 
$|\sin \theta_h|$ which relaxes with increasing $v_2$. The regions in $(v_2, s_h)$
(bottom row) illustrate this behaviour. Smaller values of the scalar vev, $v_2 \lesssim 100$
GeV are disfavoured, except for vary small values of the mixing angle; the allowed range of 
$\sin \theta_h$ is relaxed with increasing $v_2$, and for $v_2 > 200$ GeV the upper bound
on $\sin \theta_h$ is well above that imposed by the LHC signal strength.

\section{ILC Prospects}
\label{scn:ILC}

The current experimental results impose that an additional Higgs-like scalar, 
or any general extended Higgs sector, has evaded detection at the current collider 
energies. A possible next step which suggests itself, is to move toward a precision 
environment and search for signatures of a scalar extension, through its influence 
on Higgs precision measurements. In the following discussion, the potential discovery 
reach, and possible signatures of a mixed Higgs scenario are considered at 
the ILC. 

By nature of being a lepton collider the ILC will offer substantially cleaner 
signals, and is therefore much better suited to precision measurements; such a 
task is not possible at the LHC due to the significant QCD background. The cleaner 
environment, as well as detectors and instrumentation associated with a lepton 
collider, allow for the possibility to reconstruct Higgses more cleanly and in 
additional decay channels. Furthermore, the capability for polarized beams adds
greater sensitivity to spin effects, by allowing greater control 
on the initial helicity states. This is advantageous in that beam  polarization 
provides the ability to fully characterize a process with respect to the differences in 
interactions and couplings between left and right-handed particles.
Beam polarization is particularly important in electroweak processes, which
are sensitive to spin.
There are several key motivations for using polarized beams at ILC, which vary
with the particular type of process under study. Firstly, oppositely polarized
beams enhance luminosity in electron-positron annihilation processes,
as an electron annihilates a positron of opposite helicity.
Beam polarization asymmetry is also an informative variable in precision
electroweak measurements, via the $e^+ e^- \rightarrow f \bar{f}$ process. 
Another advantage is the ability to increase
the signal to background ratio, in processes that occur predominantly through a 
specific initial helicity configuration, thereby
optimizing either certain SM signals or new physics searches.
In light of its proposed features and capabilities,
Higgs parameters such as the top Yukawa coupling, Higgs branching ratios, couplings 
to vector bosons, and the Higgs self coupling (of particular focus here), 
are perhaps well within the reach of the ILC.

\begin{figure}
\begin{center}
\begin{subfigure}[c]{0.4\textwidth}
\includegraphics[width=0.9\linewidth]{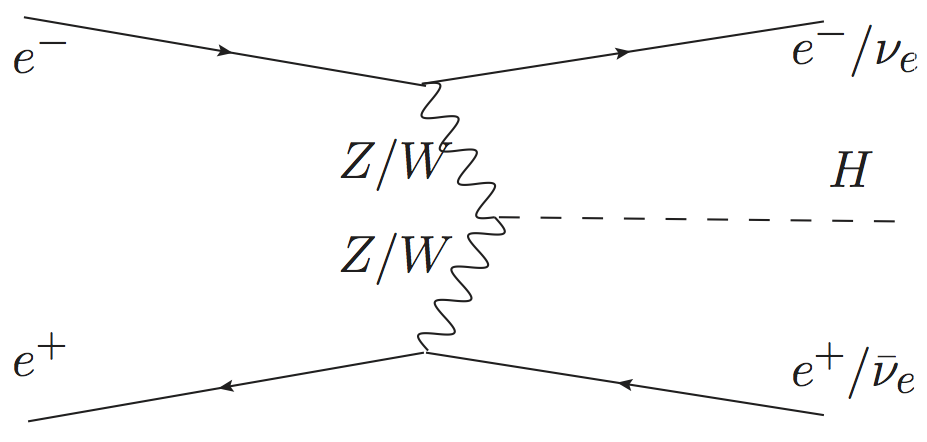}
\end{subfigure}
\begin{subfigure}[c]{0.4\textwidth}
\includegraphics[width=\linewidth]{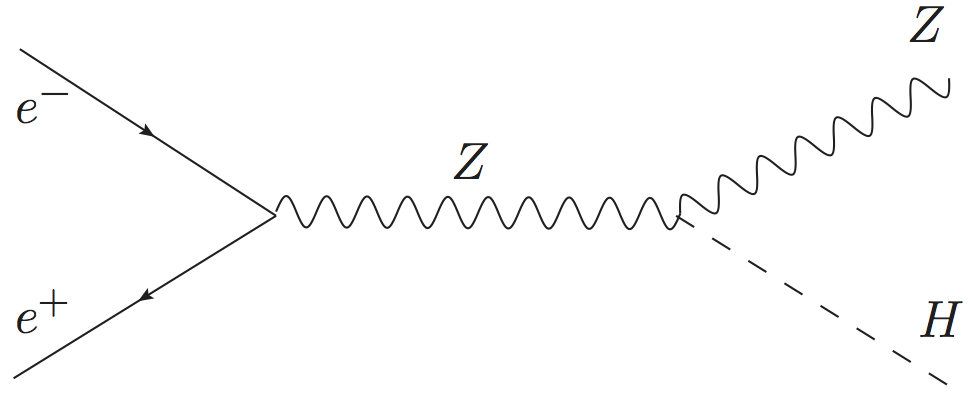}
\end{subfigure}

\bigskip

\begin{subfigure}[c]{0.45\textwidth}
\includegraphics[width=\linewidth]{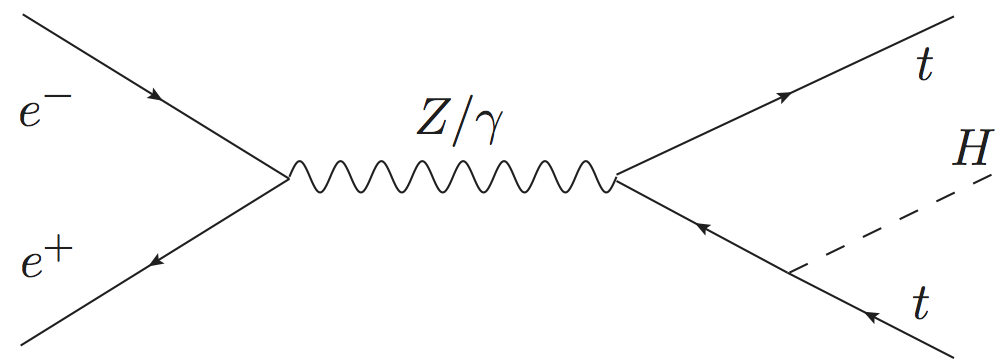}
\end{subfigure}
\end{center}
\caption{Feynman diagrams for leptonic Higgs production. From left to right, and top to bottom,
subprocesses are: vector boson fusion (VBF), Higgs-strahlung, and for higher energies,
associated top production ($t\bar{t} H$).}
\label{fig:leptHiggsprod}
\end{figure}

In lepton collisions, one loses the gluonic contributions that dominate LHC Higgs production, and the remaining
processes are the electroweak modes. The leading modes in leptonic Higgs production are 
the Higgs-strahlung process, and $W$ boson fusion. Additional contributions are present 
in $Z$ boson fusion, and associated heavy quark production at higher energies. 
The corresponding Feynman diagrams are shown in figure \ref{fig:leptHiggsprod}.
Diagrams correspond interchangeably to either $H$ or $S$
production, with the couplings scaled by the appropriate factor of the mixing angle.
Naturally, the Higgs production cross section is larger in hadron collisions than 
in lepton collisions, due to the leading QCD processes; by experimental design the LHC
optimizes Higgs production, with discovery being the priority. Although production cross
sections are smaller at a lepton collider, higher order effects are much less significant
in the case of electroweak processes, offering greater ease of calculability and less
theoretical uncertainty associated with higher order effects. 
\begin{figure}
\begin{center}
	\includegraphics[width=0.75\linewidth]{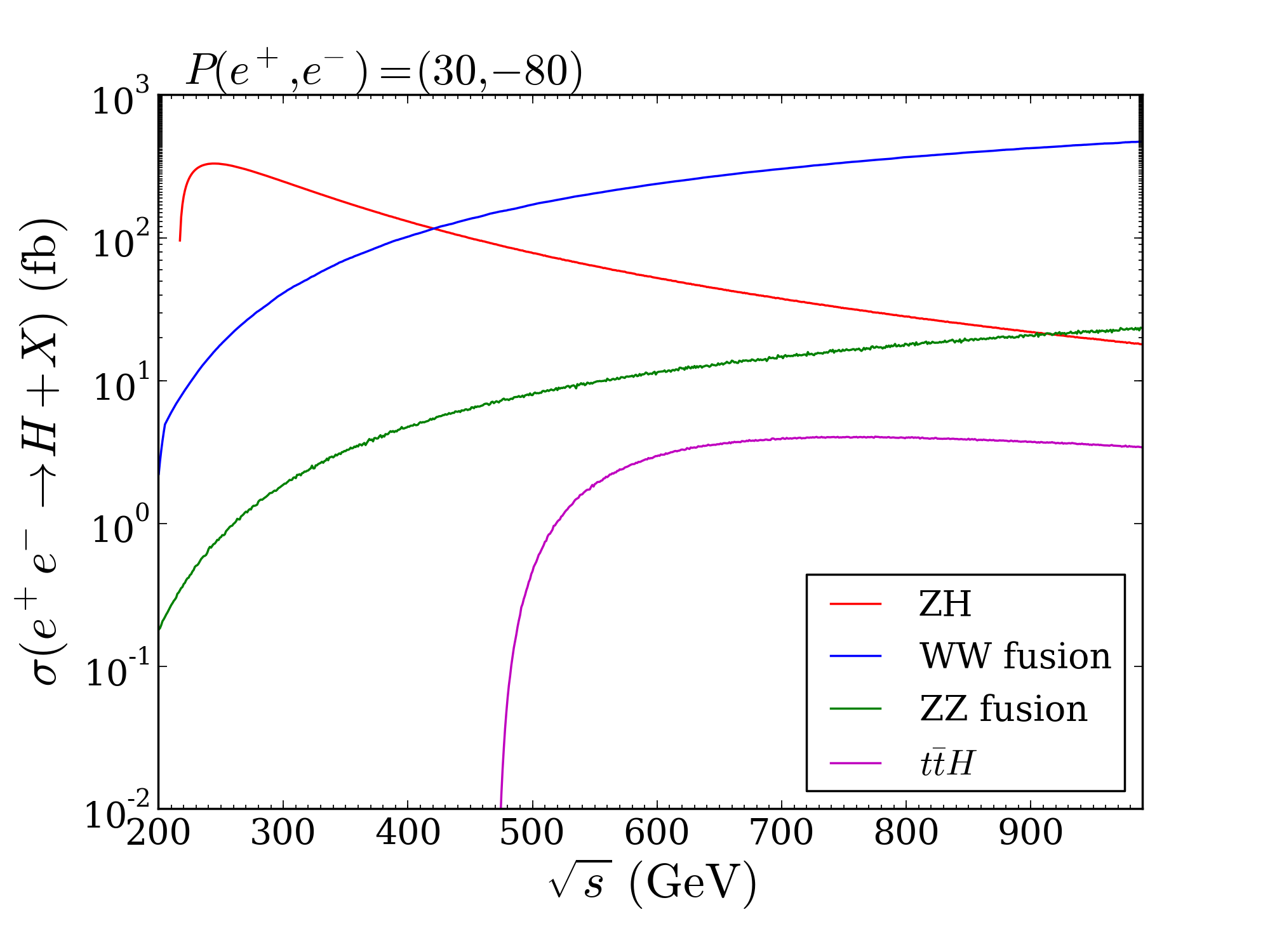}
\end{center}
\caption{Leptonic Higgs production cross section as a function of centre of mass energy,
for beam polarization $P(e^{+}, e^{-}) = (30, -80)$. The Higgs mass is taken to be $125$ GeV.}
\label{fig:leptXSCM}
\end{figure}
Cross sections for Higgs production in $e^- e^+$ collisions are determined at LO,
using \textsc{MadGraph5} \cite{Alwall:2014hca}. The cross sections as a function of
centre of mass energy are shown in figure \ref{fig:leptXSCM}, for beam polarization 
$(P_{e^+}, P_{e^-}) = (30, -80)$. 
More specifically, this denotes an electron beam which is $80\%$ left-polarized, 
and a positron beam that is $30\%$ right-polarized. The formal definition of 
polarization is 
\begin{equation}
P = \frac{N_R - N_L}{N_R + N_L},
\end{equation}
where $N_R$ and $N_L$ are the number of particles with spin parallel or antiparallel
to the direction of motion.
An equivalent interpretation of this number, is the percentage of events
for which the helicity is known, with positive or negative signifying 
right or left.

A possible experimental program includes running at centre of
mass energies $250$ GeV, $500$ GeV, and $1$ TeV, with ability to obtain beam polarizations
of $(P_{e^+}, P_{e^-}) = (30, -80)$ at the lower two energies, and $(20, -80)$ at 
$1$ TeV \cite{TDRvol1}. The displayed cross sections correspond to production of the 
SM-like Higgs, i.e. $m_H=125$ GeV, and omitting the mixing angle factor $\cos^2\theta_h$.
\begin{figure}
\begin{center}
	\begin{subfigure}{0.4\textwidth}
	\includegraphics[width=\linewidth]{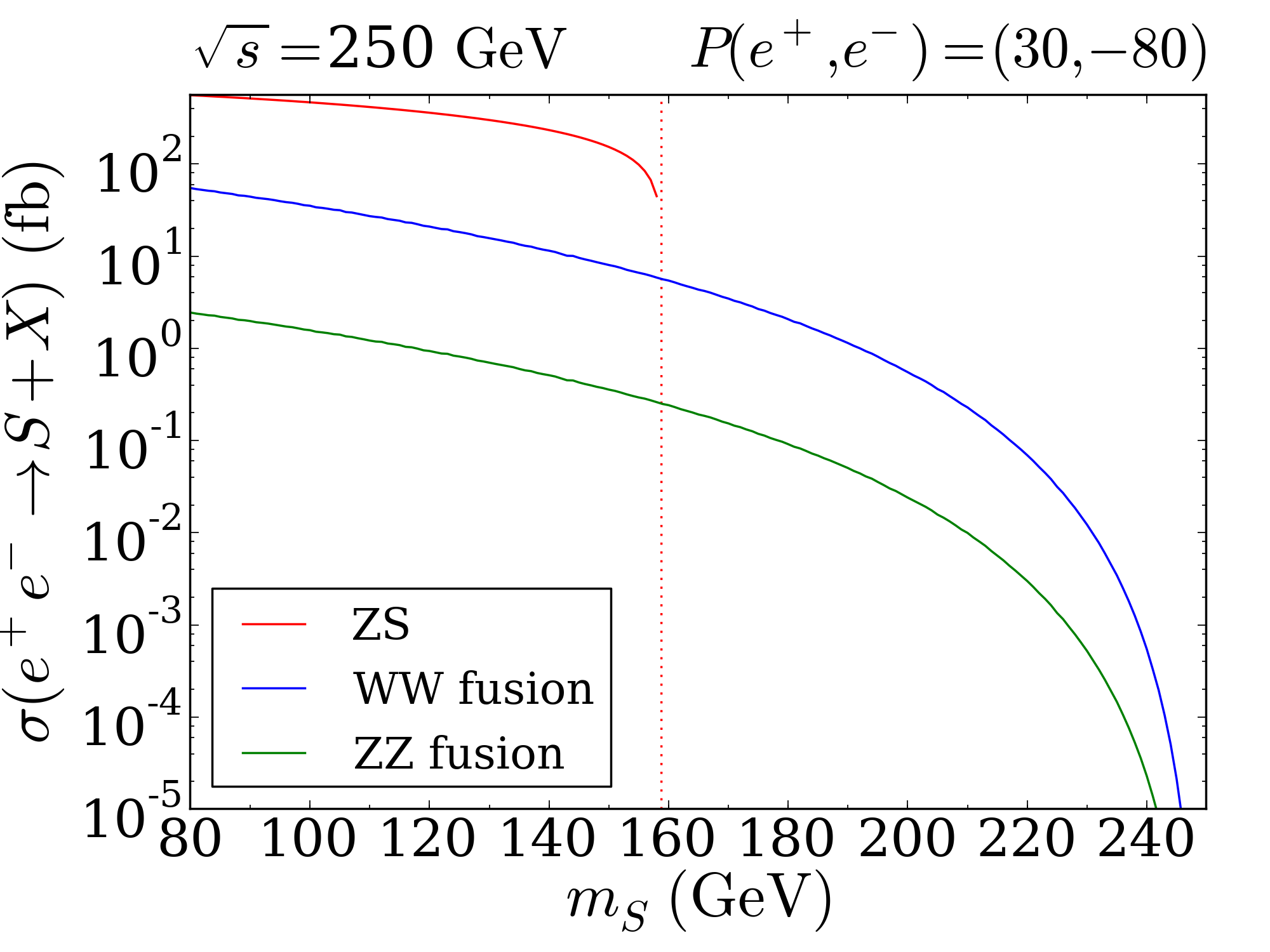}
	\end{subfigure}
	\begin{subfigure}{0.4\textwidth}
	\includegraphics[width=\linewidth]{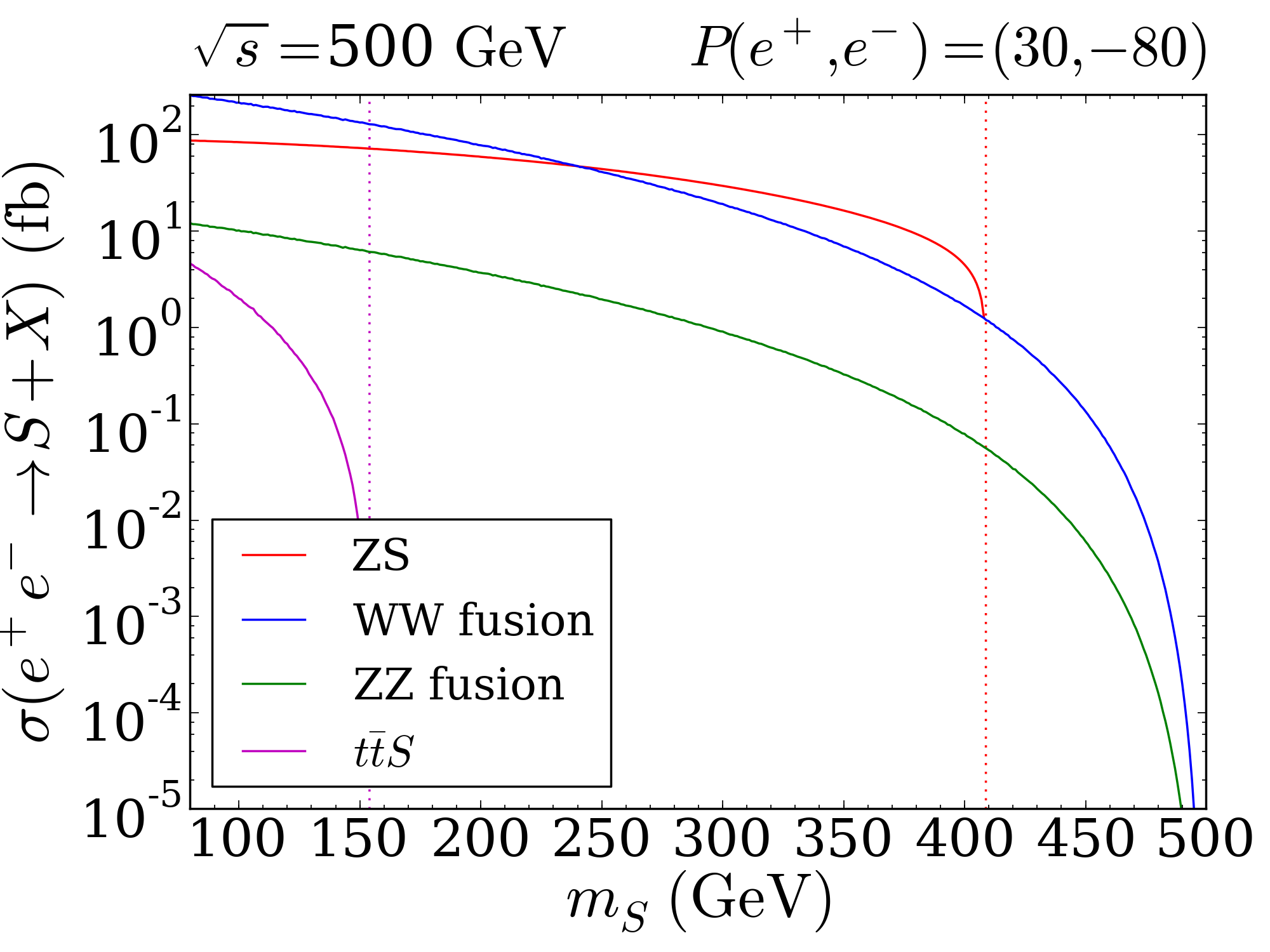}
	\end{subfigure}
	\begin{subfigure}{0.4\textwidth}
	\includegraphics[width=\linewidth]{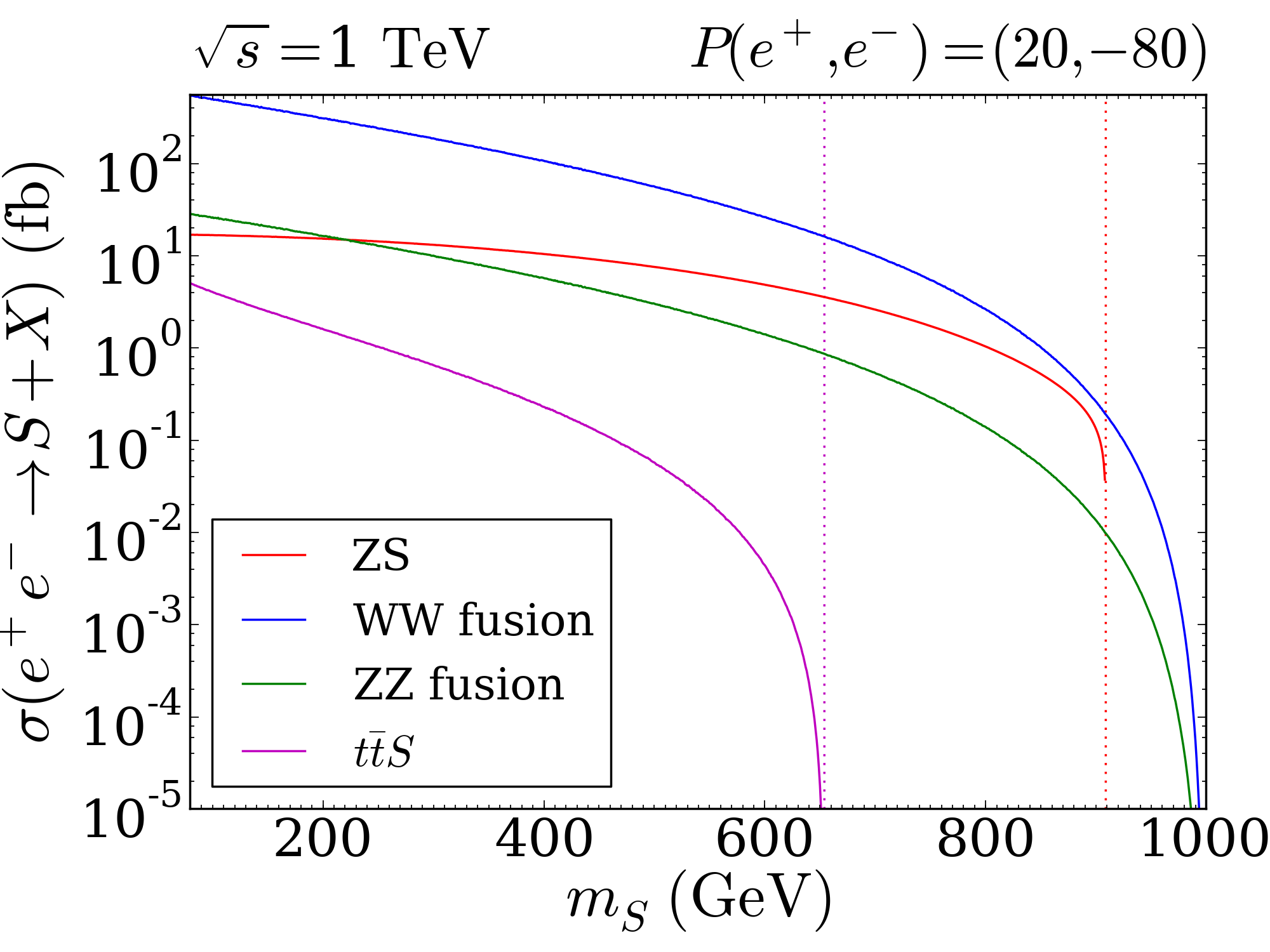}
	\end{subfigure}
\end{center}
\caption{Higgs production cross section in electron-positron collisions, for centre of mass energies, 
$\sqrt{s} = 250, 500$ GeV, and $1$ TeV. Cross sections are given at leading order, for electron 
polarization $-80$, and positron polarization $30$ for $250$ and $500$ GeV, and $20$ for $1$ TeV.}
\label{fig:leptXSmass}
\end{figure}

Depending on the mass of the additional scalar,  it may also be possible to observe
direct production of the new particle. For each centre of mass energy, and the
particular range of $m_S$, there are various possible modes by which the additional
scalar may be observed. In addition to direct $S$ production, the new scalar may also
be observed by Higgs production and subsequent decay to a scalar pair  --- for the light $S$ case 
--- or through double scalar production processes. 
Cross sections for single $S$ production are presented in
Figure \ref{fig:leptXSmass} as a function of  $S$ mass, for proposed operational energies of the 
ILC and respective beam polarizations corresponding to each centre of mass energy. 
Maintaining convention, $S$ production cross sections have been normalized by the mixing factor, omitting the scaling by $\sin^2\theta_h$ in this figure.

An important feature of the ILC is the sensitivity to 
processes in which two Higgs (or two scalars) are produced. Such processes could
be a distinguishing signature of a mixed Higgs model, and would allow the possibility
to probe scalar parameters which are uniquely modified, an effect which is not measurable
at the LHC. In the following section, the potential observation of the di-Higgs processes
is further explored.

\subsection{Di-Higgs Processes}
\label{sscn:diHiggs}

It is interesting to note that the potentially large couplings of the trilinear scalar terms,
and modification to the SM value of the Higgs self-couplings, could result in distinguishing
effects in measurements of the double Higgs processes. Precision measurements required to
probe the Higgs self-couplings and other parameters of the scalar potential, are beyond the
scope of current hadron colliders; such measurements fall under the domain of the ILC.
The scalar interaction vertices give rise to double scalar production processes which may
produce $H$, $S$, or $H$-$S$ pairs. The following will focus on measurement of the double Higgs
($H$ pair) process. 

\begin{figure}
\begin{center}
  \begin{subfigure}{0.425\textwidth}
    \includegraphics[width=\linewidth]{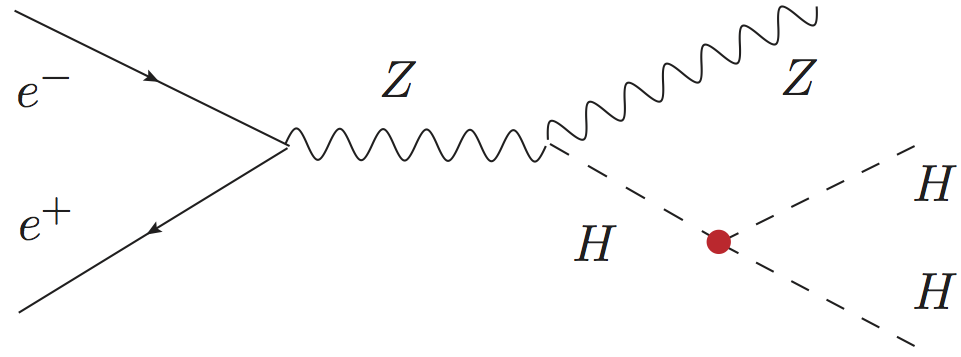}
  \end{subfigure}
  $ \; + \;$
  \begin{subfigure}{0.425\textwidth}
    \includegraphics[width=\linewidth]{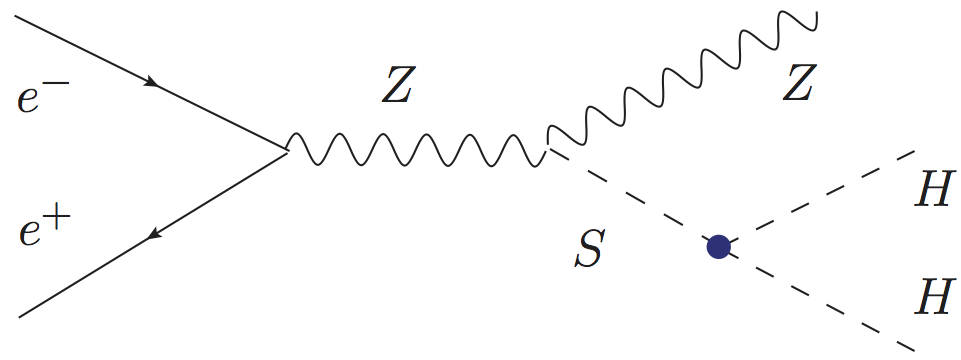}
  \end{subfigure}
\end{center}
\caption{Diagrams showing contribution of Higgs self coupling (left) and mixed scalar coupling
(right) to di-Higgs production.}
\label{fig:diHiggsFeyn}
\end{figure}

An example Feynman diagram for di-Higgs production is given in figure \ref{fig:diHiggsFeyn}, 
showing the contributions of the self-coupling and mixed scalar coupling, to the double
Higgs-strahlung process. 
Additional diagrams contributing to this process are given in appendix \ref{app:diag}. These
are referred to as `background' subprocesses, in the sense that the contributions are not proportional to either the self-coupling, or mixed scalar coupling. These contributions arise from the Higgs-vector boson interactions. Hence, the dependence of the di-Higgs cross section on these scalar couplings 
is not straightforward; translating measurement of the cross section to a measurement of either one of the couplings is therefore more complex than in the SM case. For this reason, the quantitative effect on measurement of the cross section is analyzed, rather than that of the self-coupling. At higher energies,
approaching $1$ TeV, the $W$ fusion mode becomes more significant. Analogous diagrams exist for
double Higgs production in this channel, which are also given in appendix \ref{app:diag}.
\begin{figure}
\begin{center}
  \begin{subfigure}{0.45\textwidth}
    \includegraphics[width=\linewidth]{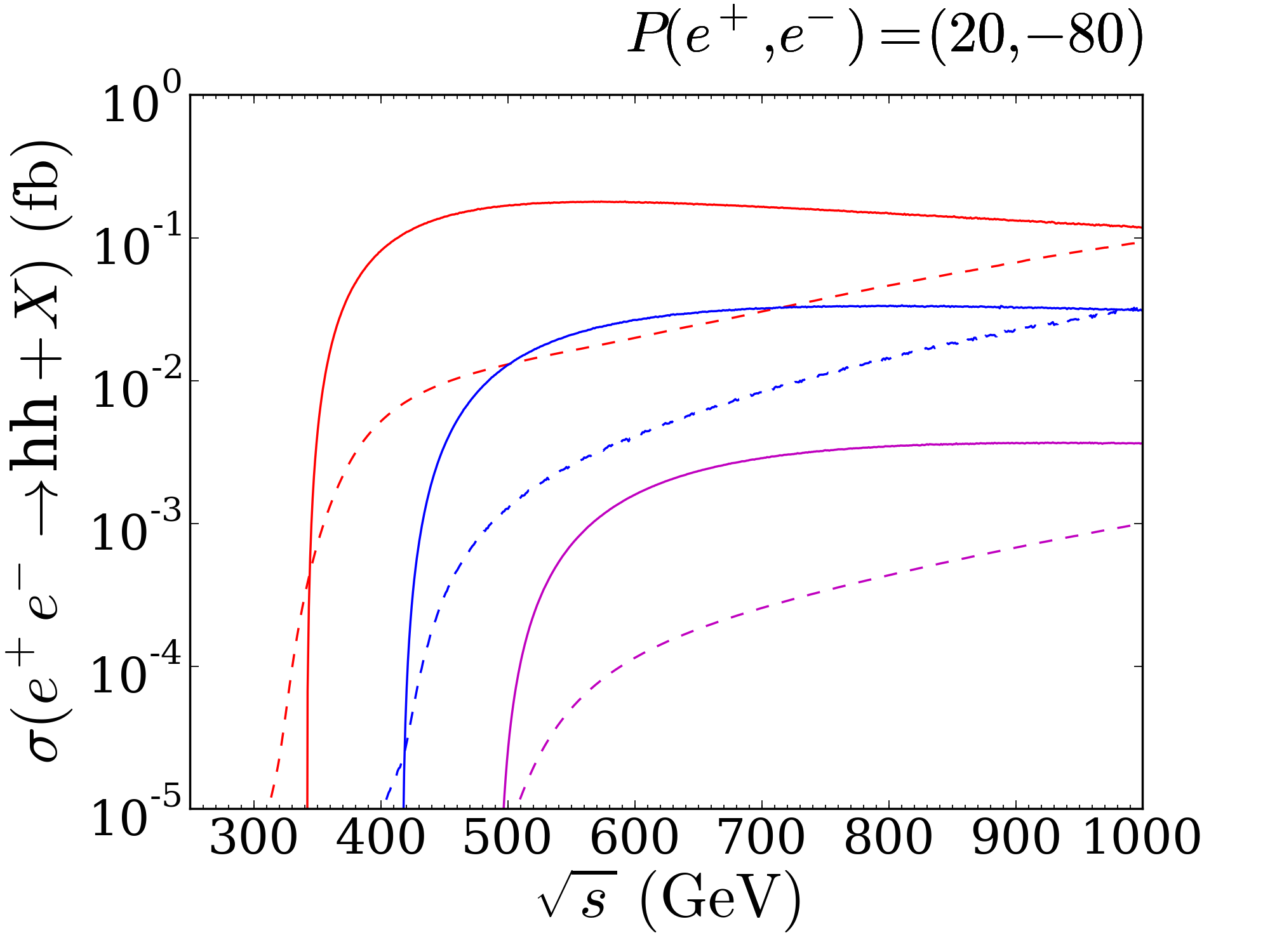}
  \end{subfigure}
  \begin{subfigure}{0.45\textwidth}
    \includegraphics[width=\linewidth]{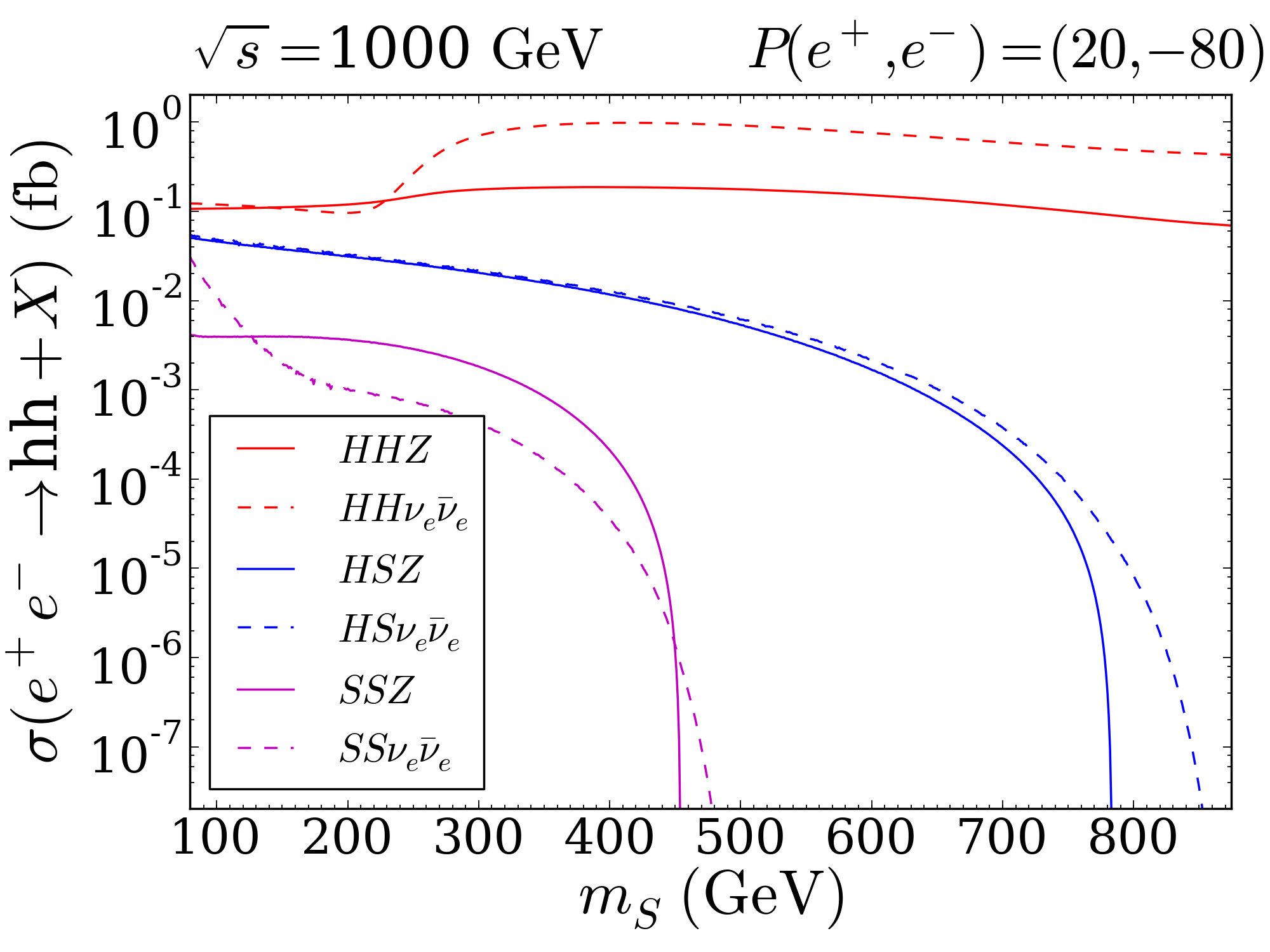}
  \end{subfigure}
\end{center}
\caption{Cross sections for scalar pair production processes, presented both as a function of centre
of mass energy (left) and scalar mass, $m_S$, at $\sqrt{s}=1$ TeV (right). In the former case, the scalar mass and
vev are both taken to be $200$ GeV, and the mixing angle is fixed at $s_{\theta_h}=0.4$.}
\label{fig:diHiggsProdxs}
\end{figure}

Under this model, the cross section for di-Higgs production differs from the SM value in several
aspects:
\begin{itemize}
\item[$\bullet$] the triple Higgs self-coupling is modified, according to
\begin{equation*}
\frac{m_H^2}{2v_1} \rightarrow 
\frac{m_H^2}{2 v_1 v_2}\left( v_2 c_{\theta_h}^3 - v_1 s_{\theta_h}^3 \right)
\end{equation*}
\item[$\bullet$] the additional contribution from the mixed trilinear scalar vertex, proportional
to
\begin{equation*}
\mu = \frac{\sin 2\theta_h }{2 v_1 v_2} \left( v_1 s_{\theta_h}  + 
v_2 c_{\theta_h} \right) \left( m_H^2 + \frac{m_S^2}{2} \right)
\end{equation*}
\item[$\bullet$] scaling by various factors of $\cos \theta_h$ and $\sin \theta_h$ in the Higgs' couplings to vector bosons

\begin{minipage}[c]{0.45\textwidth}
\begin{center}
\includegraphics[width=0.5\linewidth]{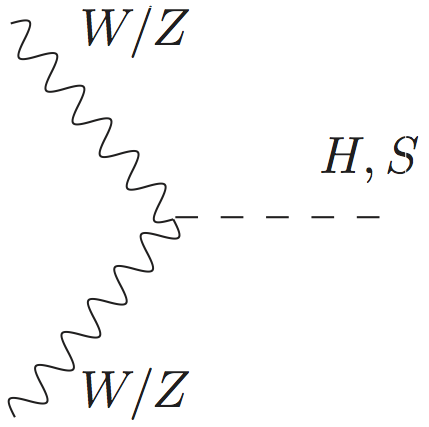} 
\end{center}
\end{minipage}
\begin{minipage}[c]{0.45\textwidth}
$\longrightarrow \quad \cos\theta_h, \: \sin\theta_h$
\end{minipage}

\begin{minipage}[c]{0.45\textwidth}
\begin{center}
\includegraphics[width=0.5\linewidth]{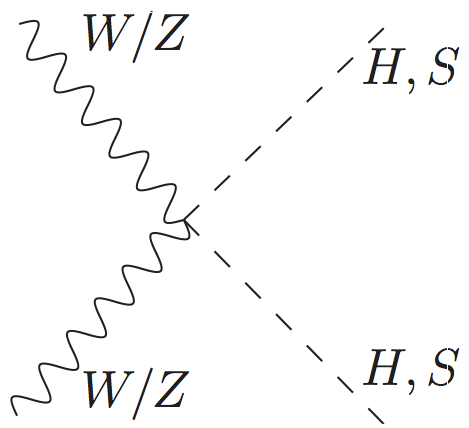} 
\end{center}
\end{minipage}
\begin{minipage}[c]{0.45\textwidth}
$\longrightarrow \quad \cos^2\theta_h, \: \sin^2\theta_h$
\end{minipage}
\end{itemize}

The cross section for the di-Higgs process is given in figure \ref{fig:diHiggsProdxs}, including 
also the $H$-$S$, and double $S$ processes, showing both the $HHZ$ and $HH\nu_e\bar{\nu}_e$ production 
channels. The dependence on both the centre of mass energy, and the $S$ mass is shown. In both cases, the mixing angle is fixed near its maximal allowed value, $\sin\theta_h = 0.4$, and in the latter case the $S$ mass and vev are set to $m_S = 200$ GeV and $v_2 = 200$ GeV. Cross sections are calculated using \textsc{MadGraph5}, and the model as described in section \ref{scn:model} is implemented via \textsc{FeynRulesv2.0} \cite{Alloul:2013bka}.

\begin{figure}
	\begin{minipage}[c]{0.5\textwidth}
	\begin{center}
	 \includegraphics[width=\linewidth]{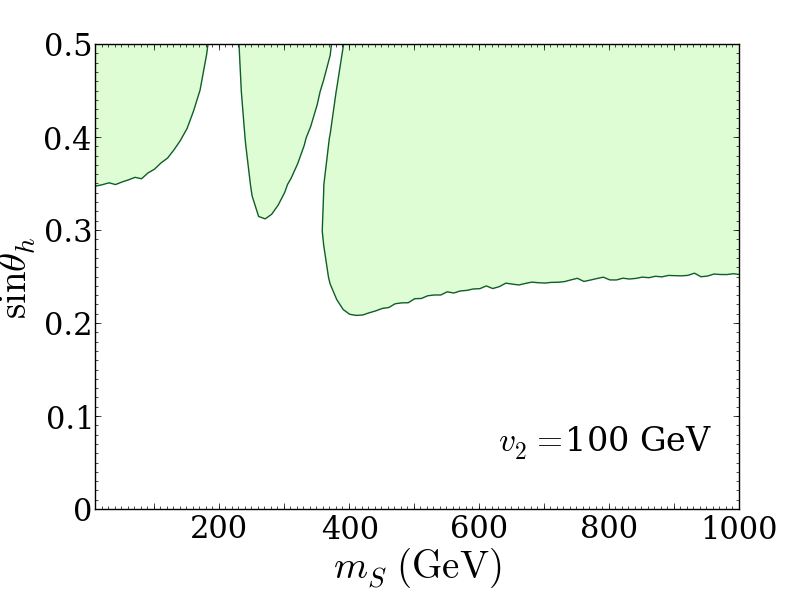}
	 \end{center}
	 \end{minipage}
	 \begin{minipage}[c]{0.5\textwidth}
	\begin{center}
	 \includegraphics[width=\linewidth]{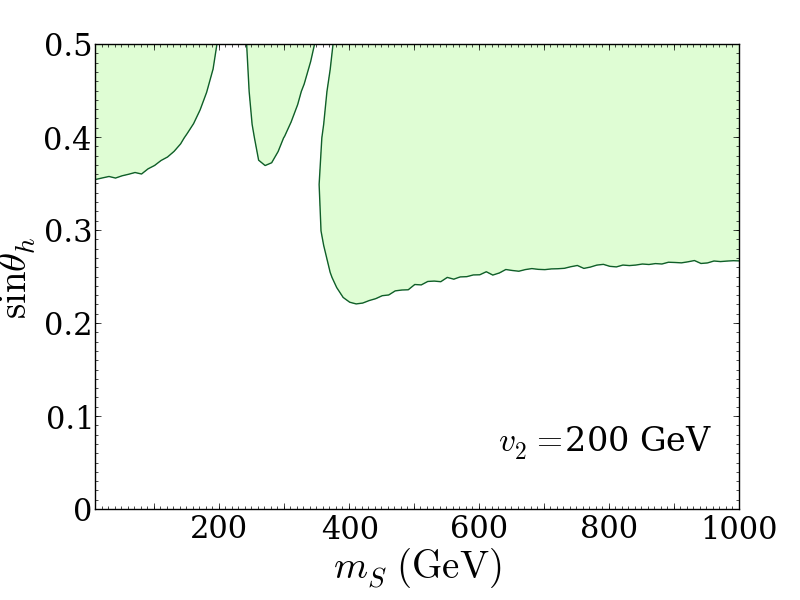}
	 \end{center}
	 \end{minipage}
	 
	 \begin{minipage}[c]{0.5\textwidth}
	\begin{center}
	 \includegraphics[width=\linewidth]{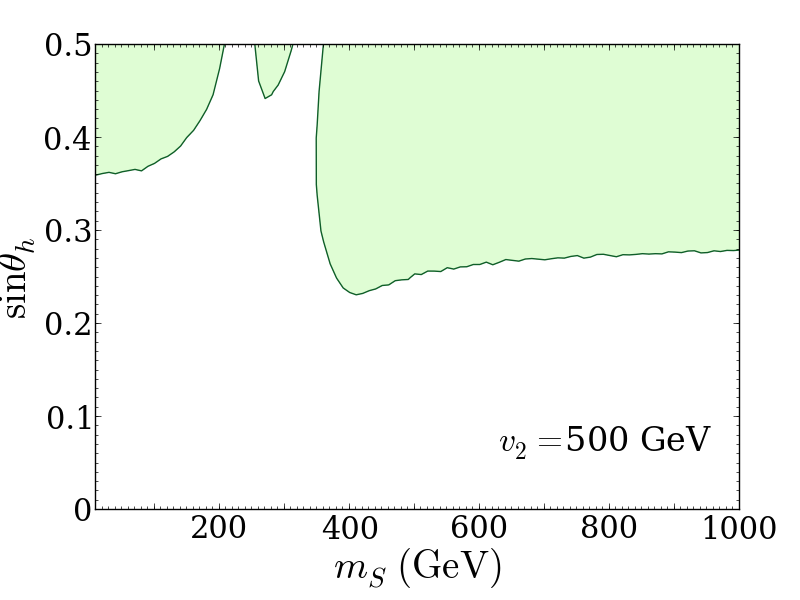}
	 \end{center}
	 \end{minipage}
	 \begin{minipage}[c]{0.5\textwidth}
	\begin{center}
	 \includegraphics[width=\linewidth]{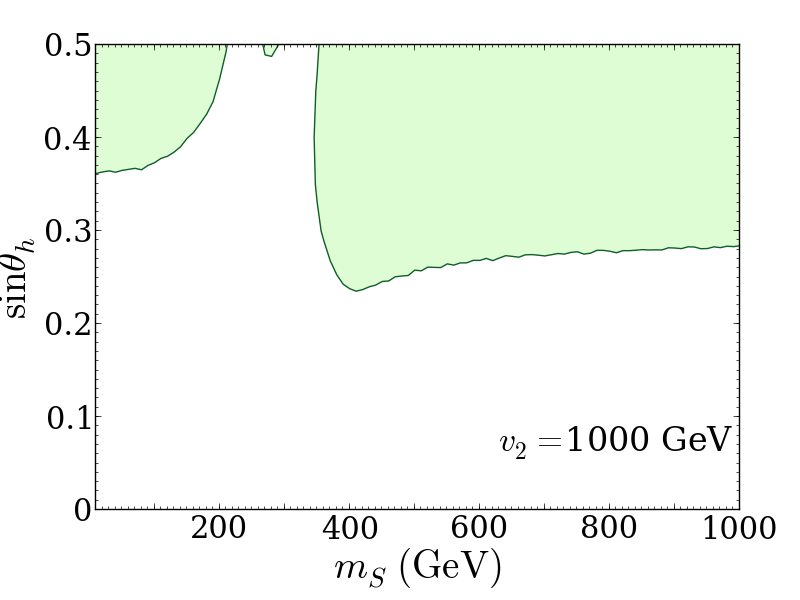}
	 \end{center}
	 \end{minipage}
\caption{Measurable region in $(m_S, \sin\theta_h)$ space, based on fractional change in the
$HHZ$ production cross section, using expected ILC precision in $\Delta\sigma/\sigma$. Value
of $v_2$ is varied discretely, as indicated. Cross sections are calculated at $500$ GeV, and
$(P_{e^-}, P_{e^+}) = (-80, 30)$.}
\label{fig:dsigmaHHZ}
\end{figure}
\begin{figure}
	\begin{minipage}[c]{0.5\textwidth}
	\begin{center}
	 \includegraphics[width=\linewidth]{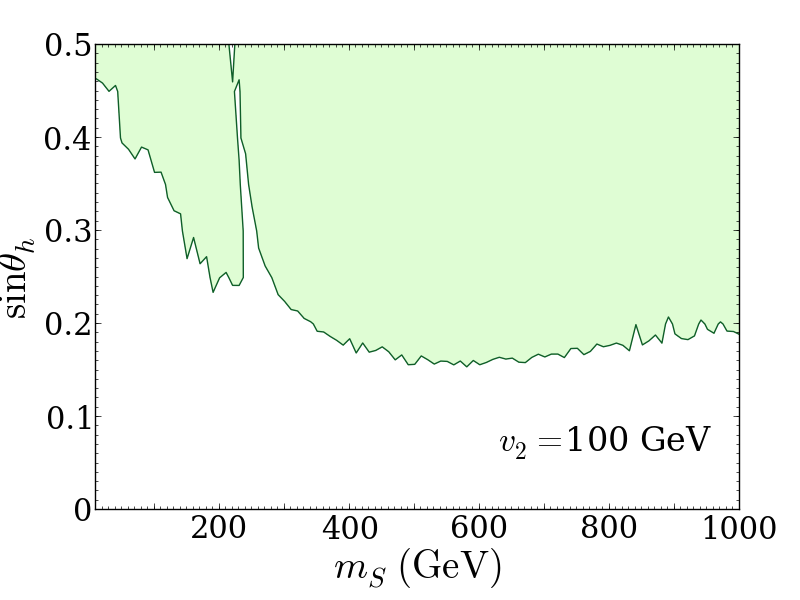}
	 \end{center}
	 \end{minipage}
	 \begin{minipage}[c]{0.5\textwidth}
	\begin{center}
	 \includegraphics[width=\linewidth]{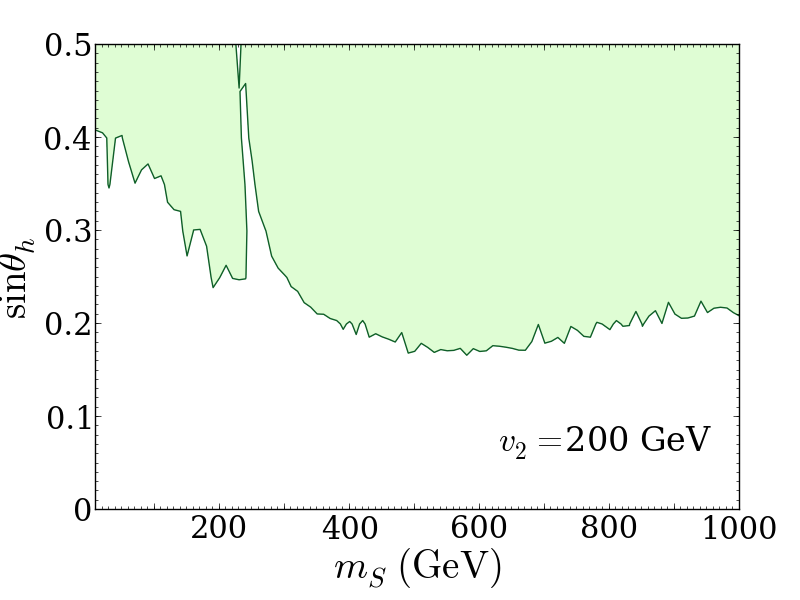}
	 \end{center}
	 \end{minipage}
	 
	 \begin{minipage}[c]{0.5\textwidth}
	\begin{center}
	 \includegraphics[width=\linewidth]{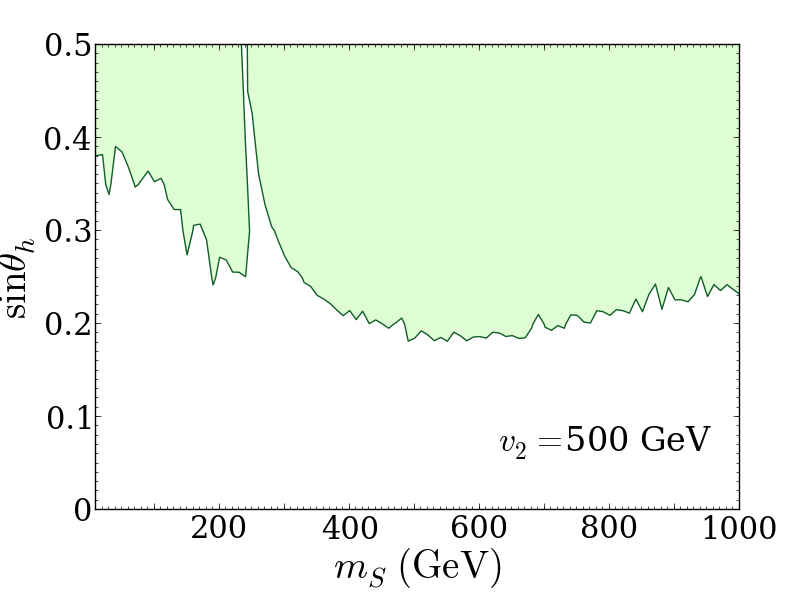}
	 \end{center}
	 \end{minipage}
	 \begin{minipage}[c]{0.5\textwidth}
	\begin{center}
	 \includegraphics[width=\linewidth]{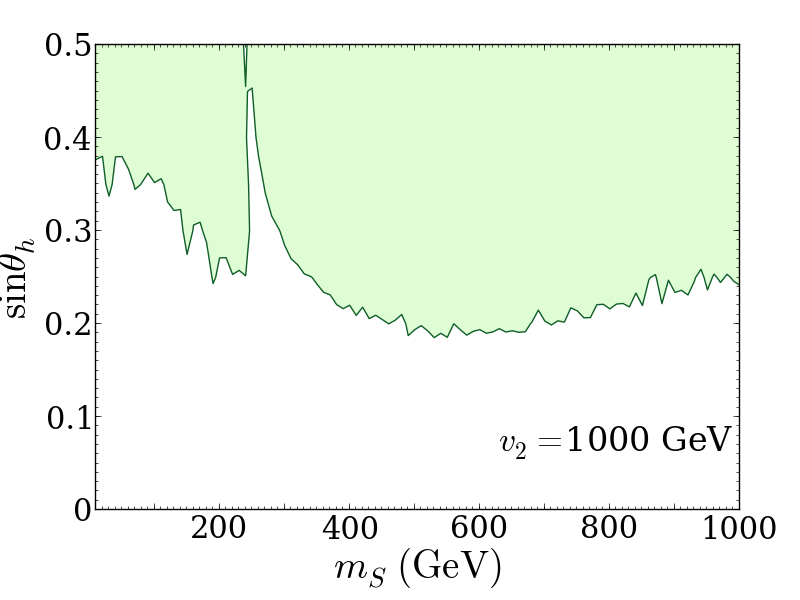}
	 \end{center}
	 \end{minipage}
\caption{Measurable region in $(m_S, \sin\theta_h)$ space, based on fractional change in the
$HH\nu_e\bar{\nu}_e$ production cross section, using expected ILC precision in $\Delta\sigma/\sigma$. Value of $v_2$ is varied discretely, as indicated. Cross sections are
calculated at $1$ TeV, and $(P_{e^-}, P_{e^+}) = (-80, 20)$.}
\label{fig:dsigmaHHvv}
\end{figure}

The expected precision in measurement of the cross section for di-Higgs production has been 
determined by ILC projections obtained by full detector simulations of SM processes. The
technical design report gives an expected precision of $\Delta\sigma/\sigma = 0.27$ for the 
$HHZ$ process \cite{TDRvol2}, and $\Delta\sigma/\sigma = 0.23$ for the $HH\nu_e\bar{\nu}_e$ is obtained
in ref. \cite{selfcoupHvv}. The fractional change in the di-Higgs production cross sections for both the $HHZ$ and $HH\nu_e\bar{\nu}_e$ processes is determined, and the parameter regions over which
the effect of the mixed scalar model would be measurable is established. 
The fractional change in cross section, compared with the SM value, for both the $HHZ$ and 
$HH\nu_e\bar{\nu}_e$ processes is calculated and plotted over the remaining scalar parameter
space, allowed by existing LHC constraints. Specifically, the quantity
\begin{equation}
\frac{\lvert \sigma(m_S, \sin\theta_h, v_2) - \sigma_{SM} \rvert}{\sigma_{SM}}
\end{equation}
is determined.

The region over which the effect of the scalar mixing is within
the expected ILC precision of the measurement is presented, showing the region of scalar 
parameter space which is within the reach of measurement via di-Higgs production. 
Results are presented in figures \ref{fig:dsigmaHHZ} and \ref{fig:dsigmaHHvv}, in the
$HHZ$ and $HH\nu_e\bar{\nu}_e$ channels respectively. The strongest dependence is found to be in 
$(m_S, \sin\theta_h)$, while varying the value of $v_2$ has a minimal effect on the bounds of the measurable region. Cross sections are calculated for $HHZ$ production at $\sqrt{s} = 500$ GeV 
and $(P_{e^-}, P_{e^+}) = (-80, 30)$, and for $HH\nu_e\bar{\nu}_e$ at $\sqrt{s} = 1$ TeV
and $(P_{e^-}, P_{e^+}) = (-80, 20)$ --- consistent with the values corresponding to the
expected precision in the cross section measurements. The greatest change in the cross section
is found to be for larger values of $m_S$, due to the exchanged $S$ going on shell and decaying
to an $HH$ pair, in the contribution $\propto \mu$. For light $S$, there is still a possible
measurable effect for $\sin\theta_h > 0.25-0.4$ that is within the region allowed by existing
LHC constraints.

\section{Direct Detection}
\label{scn:DD}

\begin{figure}
\begin{center}
\begin{minipage}[c]{0.4\textwidth}
\begin{center}
\includegraphics[width=0.8\linewidth]{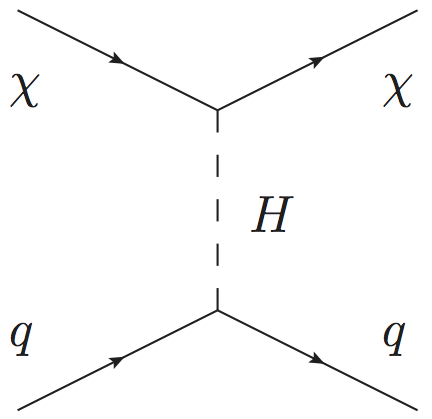}
\end{center}
\end{minipage}
$+$
\begin{minipage}[c]{0.4\textwidth}
\begin{center}
\includegraphics[width=0.8\linewidth]{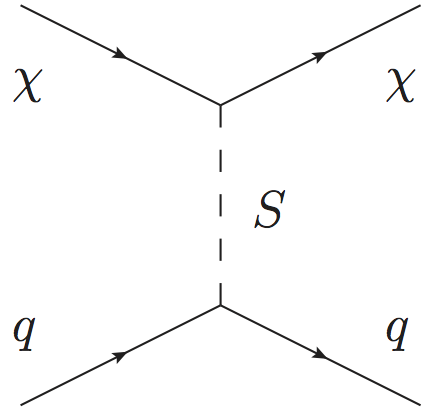}
\end{center}
\end{minipage}
\end{center}
\caption{Feynman diagrams for DM-quark scattering.}
\label{fig:feynDD}
\end{figure}

In any model that includes a generic weakly interacting massive particle (WIMP), limits 
inevitably arise from DM direct detection searches. 
In this case, the spin-independent
cross section for DM scattering by nucleons receives a contribution from the Higgs
and scalar couplings to the DM candidate $\chi$. The scattering process occurs by 
$t$-channel exchange of either $H$ or $S$. The corresponding Feynman
diagrams for the parton subprocesses are shown in figure \ref{fig:feynDD}. 

The spin-independent cross section for $\chi$-nucleon scattering is then given by
\begin{equation}
\sigma_{SI} = \left( \frac{s_{\theta_h} c_{\theta_h}}{v_2}\right)^2 
m_{\chi}^2 \left( \frac{f_N m_N}{v_1} \right)^2 
\frac{\mu^2}{\pi}  \left( \frac{1}{m_H^2} - \frac{1}{m_S^2} \right)^2.
\end{equation}
The Higgs coupling to nucleons (rather than partons) is accounted for in the factor
$f_N m_N/v_1$, with $f_N = 0.303$ \cite{Cline:2013gha}. Here, $\mu$ is the $\chi$-nucleon
reduced mass, $\mu = m_{\chi} m_N/(m_{\chi} + m_N)$ and the difference in proton and
neutron mass is assumed negligible, taking $m_N=0.946$ GeV.

Current results from LUX \cite{LUX2015} give the most stringent limit on direct detection
of WIMP-nucleon scattering. They present an upper limit at $90 \%$ c.l. on the
spin-independent cross section, as a function of the WIMP mass. 
\begin{figure}
\begin{center}
	\begin{minipage}[c]{0.495\textwidth}
	\begin{center}
	\includegraphics[width=\linewidth]{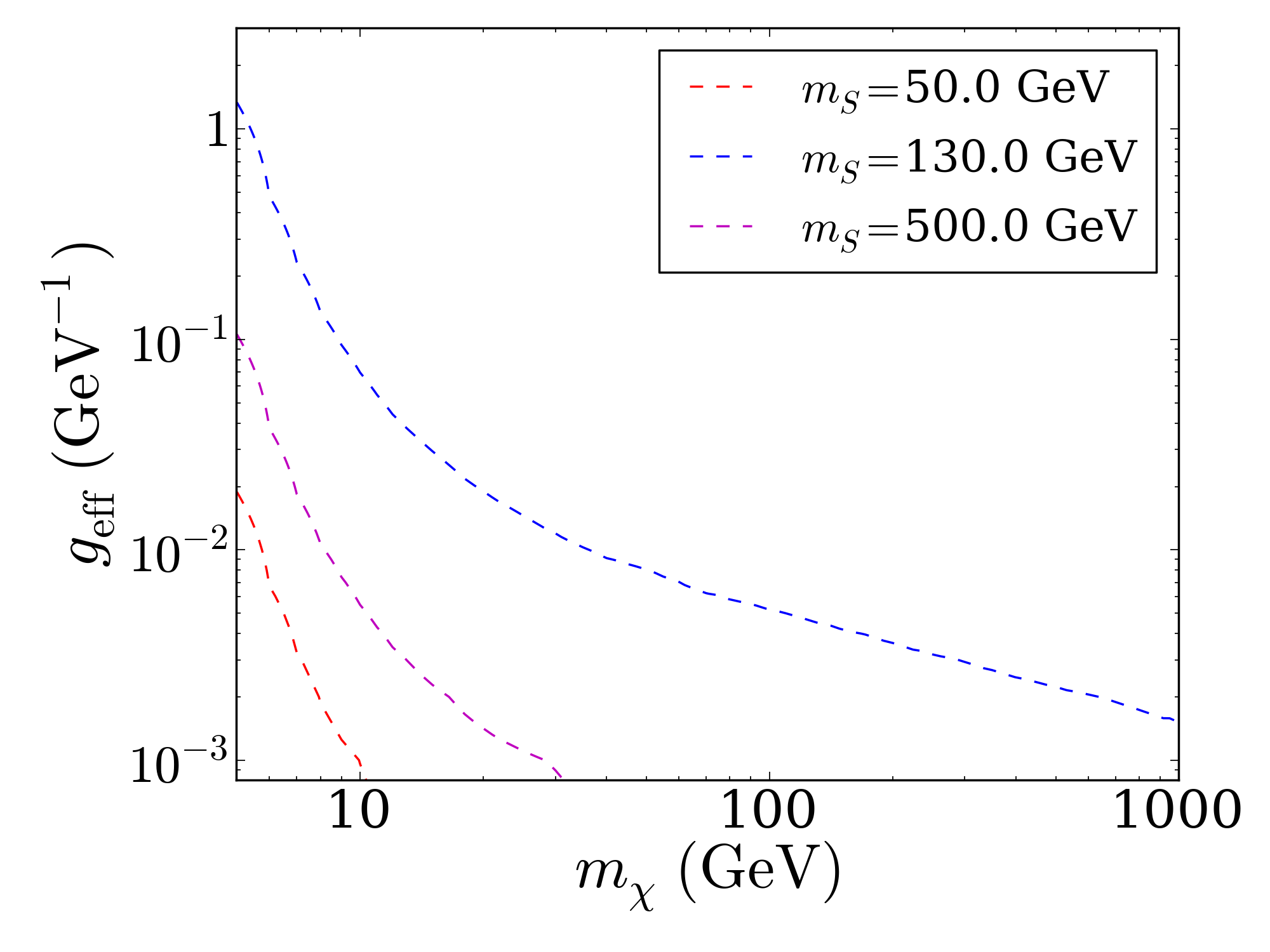}
	\end{center}
	\end{minipage}
	\begin{minipage}[c]{0.495\textwidth}
	\begin{center}
	\includegraphics[width=\linewidth]{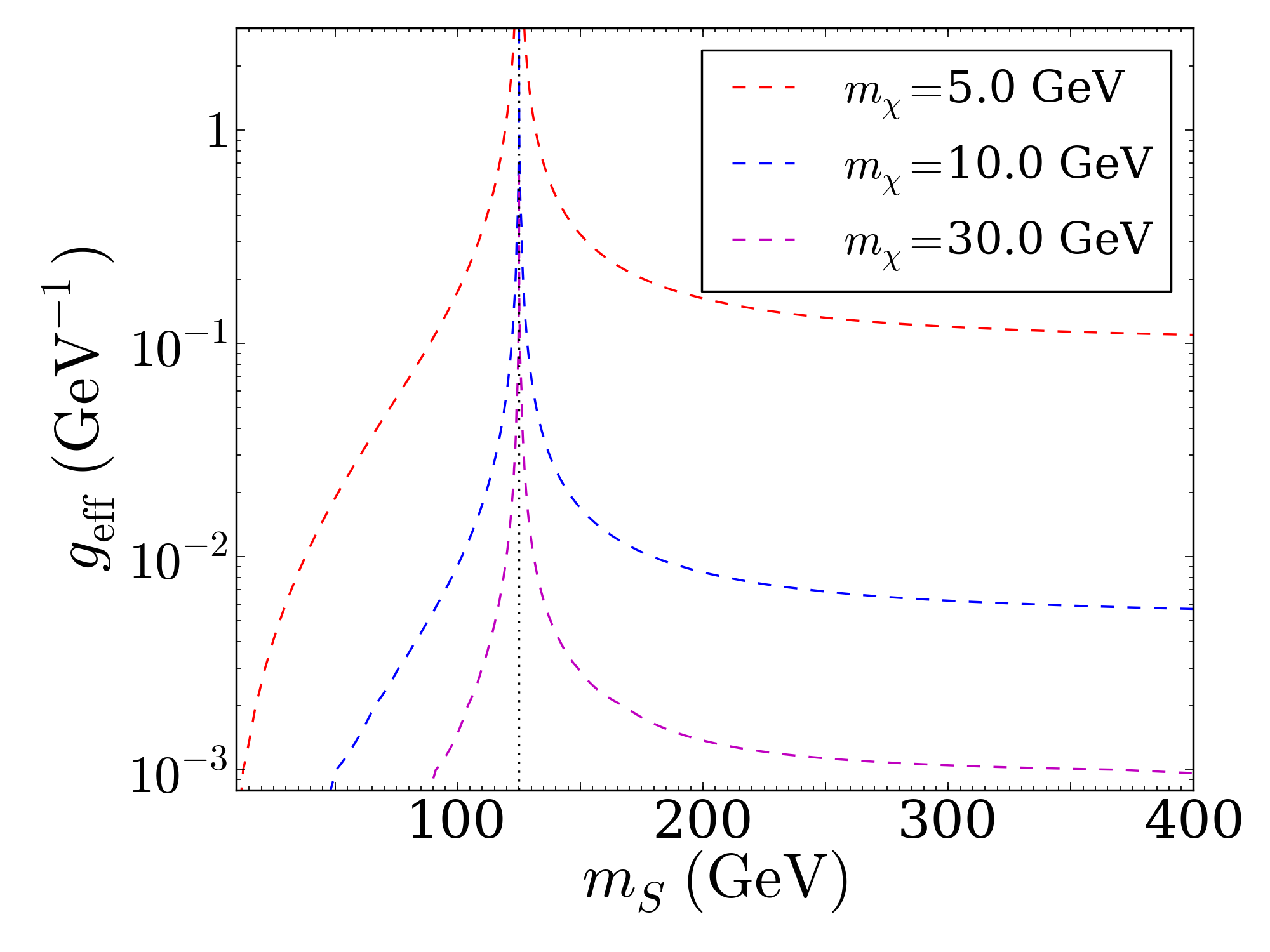}
	\end{center}
	\end{minipage}
\end{center}
\caption{LUX upper limit on the effective coupling $g_{\mathrm{eff}} = s_{\theta_h}
c_{\theta_h}/v_2$. The dependence on both the dark matter mass (left) and scalar mass (right) are
shown.}
\label{fig:LUXlimit}
\end{figure}
The resulting limit on the effective coupling, denoted 
$g_{\mathrm{eff}} = s_{\theta_h} c_{\theta_h}/v_2$, and its dependence on both
the scalar and DM masses is determined, as shown in figure \ref{fig:LUXlimit}.
The upper limit on $g_{\mathrm{eff}}$, based on the LUX $90 \%$ c.l. bound, is determined 
as a function of $m_{\chi}$, while discretely varying $m_{S}$, and vice versa. 
Discrete values of $m_S = \{ 50, 130, 500 \}$ GeV, and $m_{\chi} = \{ 5, 10, 30 \}$ 
GeV are chosen.

\begin{figure}
\begin{center}
\includegraphics[width=0.575\textwidth]{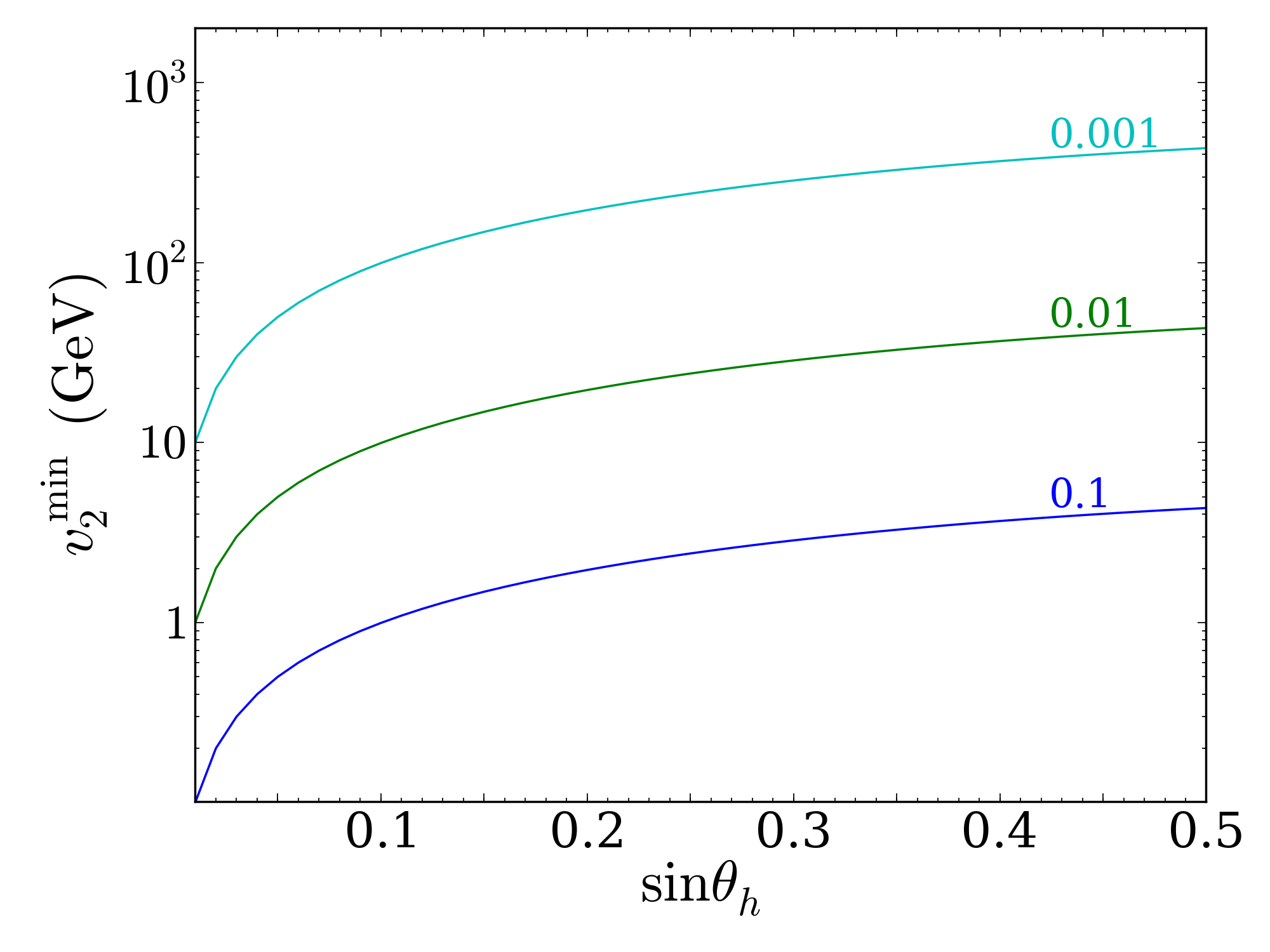}
\end{center}
\caption{Dependence of the minimum allowed value of $v_{2}$ on $\sin \theta_h$,
for several values of the upper bound on $g_{\mathrm{eff}}$.}
\label{fig:v2min}
\end{figure}

Noting first the $m_{\chi}$ dependence of the effective coupling limit, the limit
is most stringent for heavier $\chi$, due to the additional factor of $m_{\chi}^2$ 
in the scalar-dark matter coupling. The limit relaxes for light $\chi$, roughly less
than $10$ GeV. 
Considering the dependence of this limit on the scalar mass, the coupling is more
tightly constrained for $m_S \lesssim 100$ GeV, while the limit weakens slightly for heavier $m_S > m_H$. 
A special case is seen as $S$ approaches the degenerate limit, $m_{S} \sim m_H$, due to 
destructive interference of the $H$ and $S$ contributions. In this narrow region, the upper
bound on the coupling is substantially relaxed. 
For lighter $\chi$, the upper limit on the effective coupling is between $0.01 - 0.1$ GeV$^{-1}$, 
and can be as large as $\sim 1$ GeV$^{-1}$ for values of $m_S$ near $m_H$. 
For heavier $\chi$, this coupling is restricted to be $\mathcal{O}(10^{-3})$ GeV$^{-1}$, or smaller. 
For illustrative purposes, the dependence of the minimum allowed value of 
the scalar vev on the value of $\sin \theta_h$ is shown in figure 
\ref{fig:v2min}, for several maximum values of $g_{\mathrm{eff}}$.

\section{Conclusion}
\label{scn:conc}

Although the current LHC Higgs data are consistent with the Standard Model, there
is still the possibility that the observed Higgs could be part of an extended theory,
in particular, that a Higgs sector with multiple scalars could be avoiding detection
at the current levels of precision. 
A mixed two-Higgs scenario has been considered, under the addition of a real singlet scalar field to the
Standard Model scalar sector. 
Strict limits from the measured signal strength in LHC Higgs production, and exclusion
searches constrain the mixing angle such that  the upper bound on $|\sin\theta_h|$ is
between 0.3 and 0.5 for various ranges of the scalar mass, above $m_S \sim 80$ GeV,
while LEP exclusion place a stricter bound for smaller masses, of $|\sin\theta_h| \lesssim 0.15$.
For a subset of the parameter space, additional limits arise on the scalar parameters, including the vev of the additional
scalar field, from new contributions to the total width of the Higgs, due to decays to a scalar pair.
For the case of a light $S$, the region of scalar parameter space allowed by measurements
of the total Higgs width is determined, giving the allowed values of the mixing angle and 
scalar vev. Smaller values of $v_2$, i.e. $v_2 \sim 10$ GeV tend to be disfavoured in this case.

This analysis primarily considered scalar masses of approximately $100$ GeV or larger, but
masses as low as $1-10$ GeV were briefly considered in some cases --- specifically the contribution
to the Higgs width from decays to a scalar pair, and in the region excluded by LEP data.
Although $\sim 1$ GeV scalar masses were not the focus here, it should be noted that 
in the region of $m_S \lesssim 4$ GeV, there is potential for the mixing angle to be strictly
constrained by recent LHCb data corresponding to a search for hidden-sector bosons in 
$B^0$ decays \cite{Aaij:2015tna}.

The model is also extended to include a possible application to dark matter, via the
addition of a chiral Yukawa coupling of the scalar singlet to a dark fermion. Two Majorana
mass states generally result as the scalar acquires a vev, which are equivalently expressed as a single
Dirac fermion in the degenerate case which is considered here. The initial implications of the simpler
case of a single dark fermion in this extended Higgs portal framework are investigated,
with the extended case of two non-degenerate dark matter species left for future work.
Bounds on the Higgs invisible width place limits on the dark sector of the model.
Specifically, such measurements constrain the scalar vev to be approximately greater than 
$100$ or $500$ GeV, for $\sin\theta_h = 0.1, 0.4$ respectively, for values of $m_{\chi}$
in the region kinematically allowed by the $H \rightarrow \chi \bar{\chi}$ decay. The lower
bound on $v_2$ is relaxed for very light $\chi$, $m_{\chi} \sim 1$ GeV. Alternatively,
the constraint on $v_2$ is avoided for $m_{\chi} > m_H/2$.
Additional limits on the parameters describing the dark sector interaction are obtained by direct detection
results, from $H$ and $S$ mediated contributions to the spin-independent $\chi$-nucleon 
scattering cross section.
The maximally allowed value of the effective coupling,
$g_{\mathrm{eff}} = \sin\theta_h \cos\theta_h/v_2$, is determined as a function of the dark matter
mass, with additional dependence on the mass of the new scalar. The resulting constraint
places a lower bound on the scalar vev, which varies with the values of the mixing angle and 
masses of $S$ and $\chi$, and disfavours heavier $\chi$. 
It is noted than possible tension with the LUX result is avoided under three possibilities:
either lighter dark matter with $m_{\chi} \lesssim 10$ GeV, $v_2$ values of order $1$ TeV,
or values of $m_S$ which are near the SM Higgs mass.  

Under the scenario in which an additional scalar has evaded detection at current hadron
colliders, the remaining parameter space is probed in a Higgs precision environment,
investigating the discovery potential at ILC.
The presence of the additional scalar, and induced mixing, results in unique contributions
to the di-Higgs production process. Based on expected precision in measurement of the
di-Higgs cross section, it is shown that there exists a measurable region in the remaining
allowed parameter space, over which the effect of scalar mixing could be observed.
Future experiments in Higgs precision will offer interesting results, and the opportunity to
further constrain, or perhaps detect, new physics in the Higgs sector.

\appendix
\section{Di-Higgs Production Diagrams}
\label{app:diag}

Additional Feynman diagrams contributing to di-Higgs production processes at ILC
are given here. Included, are the background diagrams for $HHZ$ production, and
the diagrams for $HH\nu_e\bar{\nu}_e$ production, via $WW$ fusion, including
both background, and the scalar trilinear and self coupling contributions.
It should be noted, that $HH\nu_e\bar{\nu}_e$ production also includes the corresponding
Higgs-strahlung diagrams in which the final-state $Z$ decays to $\nu_e \bar{\nu}_e$. 

\begin{figure}[h!]
\begin{center}
	 \begin{minipage}[c]{0.3\textwidth}
	 \begin{center}
	 \includegraphics[width=\linewidth]{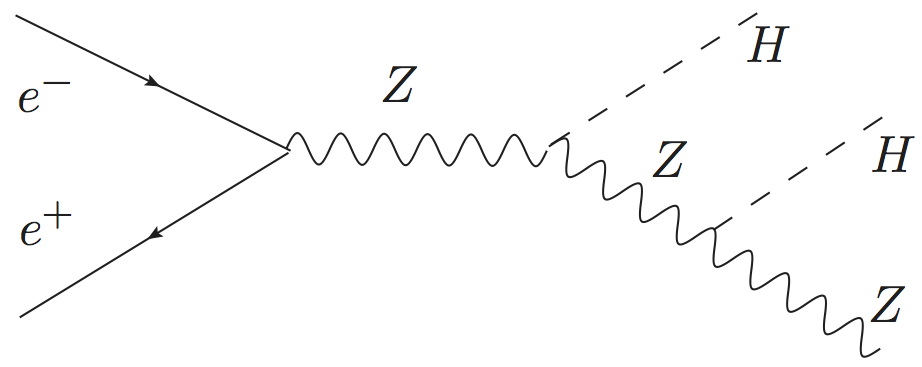}
	 \end{center}
	 \end{minipage}
	 \begin{minipage}[c]{0.3\textwidth}
	 \begin{center}
	 \includegraphics[width=\linewidth]{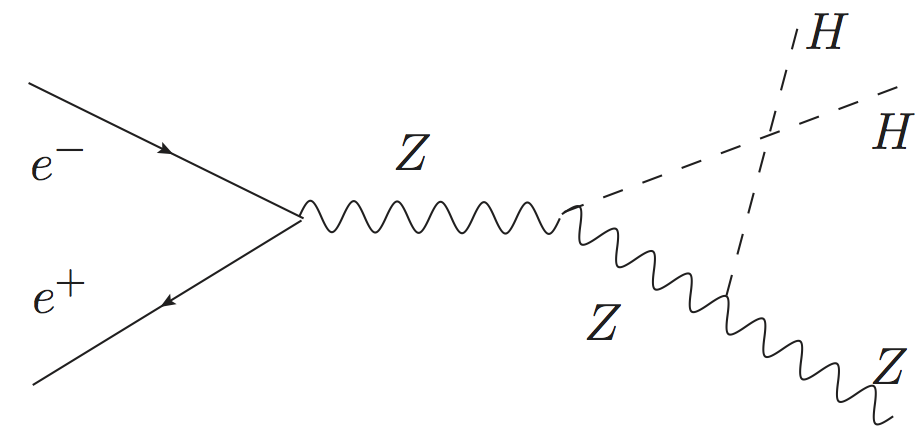}
	 \end{center}
	 \end{minipage}
	 \begin{minipage}[c]{0.3\textwidth}
	 \begin{center}
	 \includegraphics[width=\linewidth]{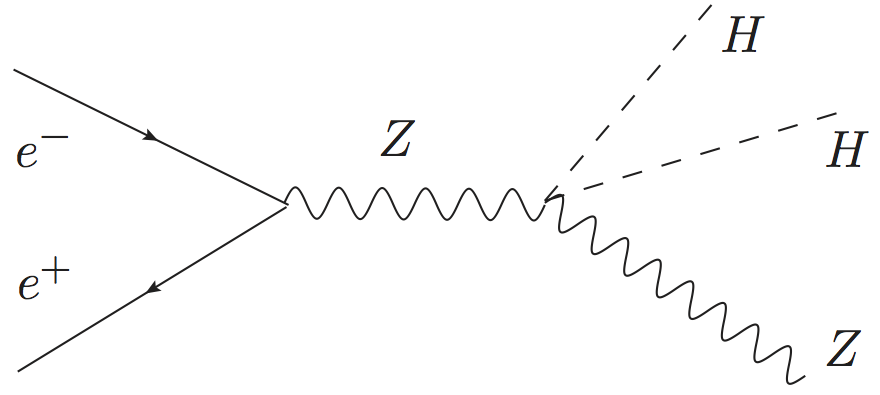}
	 \end{center}
	 \end{minipage}
\end{center}
\caption{Background diagrams for double Higgs-strahlung process.}
\end{figure}
%
\begin{figure}[h!]
\begin{center}
	 \begin{minipage}[c]{0.35\textwidth}
	 \begin{center}
	 \includegraphics[width=\linewidth]{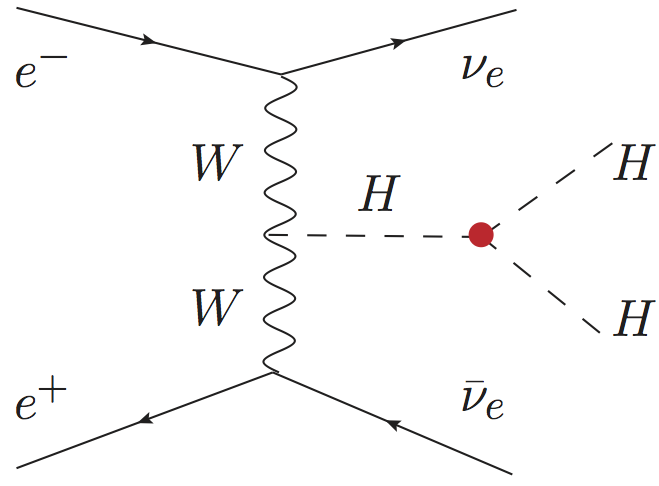}
	 \end{center}
	 \end{minipage}
	 \begin{minipage}[c]{0.35\textwidth}
	 \begin{center}
	 \includegraphics[width=\linewidth]{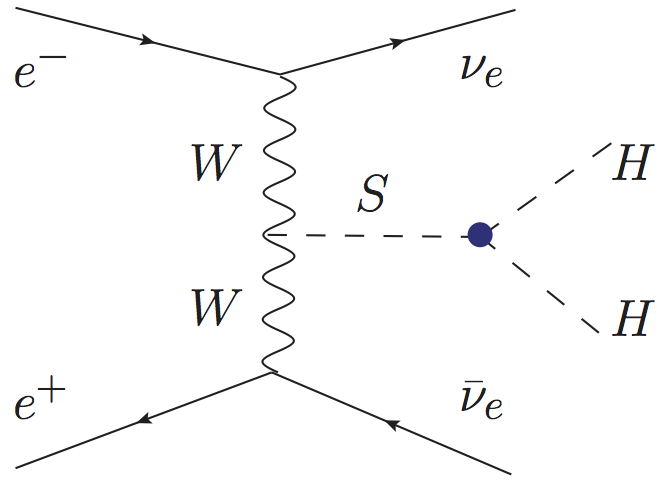}
	 \end{center}
	 \end{minipage}
\end{center}
\caption{Contribution of Higgs self coupling and mixed scalar coupling to di-Higgs production
via $WW$ fusion.}
\end{figure}
%
\begin{figure}[h!]
\begin{center}
	 \begin{minipage}[c]{0.3\textwidth}
	 \begin{center}
	 \includegraphics[width=\linewidth]{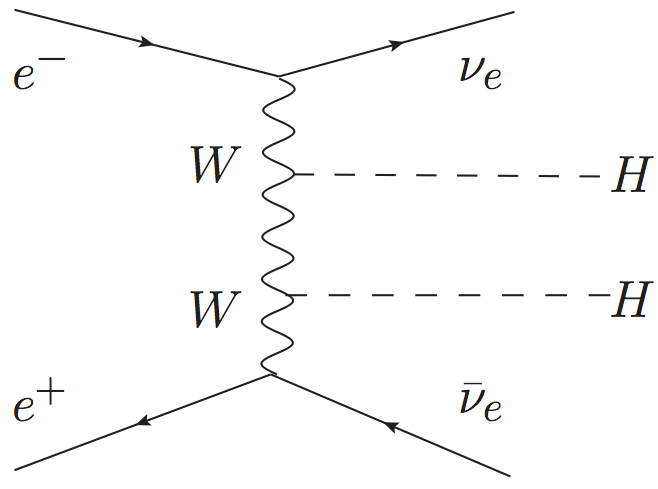}
	 \end{center}
	 \end{minipage}
	 \begin{minipage}[c]{0.3\textwidth}
	 \begin{center}
	 \includegraphics[width=\linewidth]{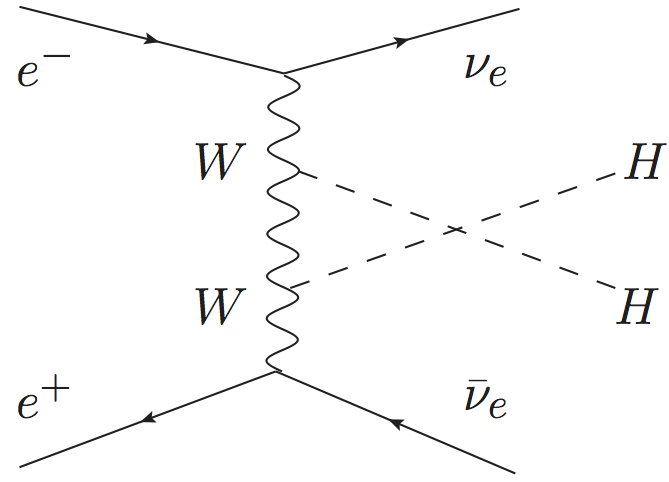}
	 \end{center}
	 \end{minipage}
	 \begin{minipage}[c]{0.3\textwidth}
	 \begin{center}
	 \includegraphics[width=\linewidth]{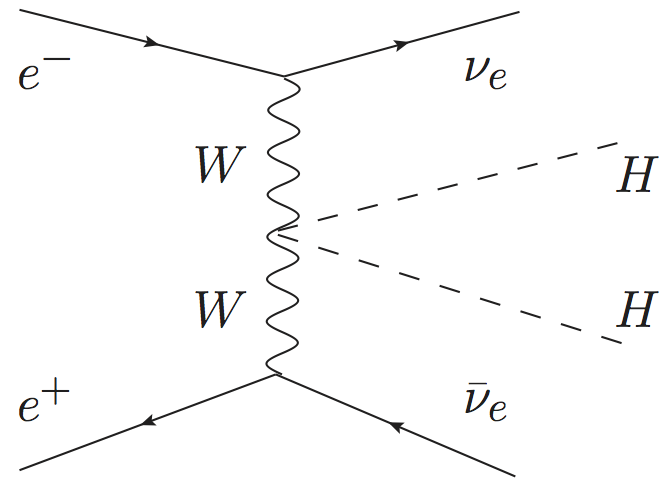}
	 \end{center}
	 \end{minipage}
\end{center}
\caption{Background diagrams contributing to di-Higgs $WW$ fusion process.}
\end{figure}

\acknowledgments
The author would like to thank Jim Cline for valuable comments, guidance, and supervision
of this project, as well as Brigitte Vachon and Guy Moore
for helpful discussions. Research is supported by
the Natural Sciences and Engineering Research Council (NSERC) of Canada. 

\bibliographystyle{JHEP}
\bibliography{hpscx}

\providecommand{\href}[2]{#2}\begingroup\raggedright\begin{thebibliography}{100}

\bibitem{Robens:2015gla}
T.~Robens and T.~Stefaniak, \emph{{Status of the Higgs Singlet Extension of the
  Standard Model after LHC Run 1}},
  \href{http://dx.doi.org/10.1140/epjc/s10052-015-3323-y}{\emph{Eur. Phys. J.}
  {\bf C75} (2015) 104}, [\href{http://arxiv.org/abs/1501.02234}{{\tt
  1501.02234}}].

\bibitem{Pruna:2013bma}
G.~M. Pruna and T.~Robens, \emph{{Higgs singlet extension parameter space in
  the light of the LHC discovery}},
  \href{http://dx.doi.org/10.1103/PhysRevD.88.115012}{\emph{Phys. Rev.} {\bf
  D88} (2013) 115012}, [\href{http://arxiv.org/abs/1303.1150}{{\tt
  1303.1150}}].

\bibitem{Godunov:2015nea}
S.~I. Godunov, A.~N. Rozanov, M.~I. Vysotsky and E.~V. Zhemchugov,
  \emph{{Extending the Higgs sector: an extra singlet}},
  \href{http://dx.doi.org/10.1140/epjc/s10052-015-3826-6}{\emph{Eur. Phys. J.}
  {\bf C76} (2016) 1}, [\href{http://arxiv.org/abs/1503.01618}{{\tt
  1503.01618}}].

\bibitem{Martin-Lozano:2015dja}
V.~Martín~Lozano, J.~M. Moreno and C.~B. Park, \emph{{Resonant Higgs boson
  pair production in the $ hh\to b\overline{b}\ WW\to
  b\overline{b}{\ell}^{+}\nu {\ell}^{-}\overline{\nu} $ decay channel}},
  \href{http://dx.doi.org/10.1007/JHEP08(2015)004}{\emph{JHEP} {\bf 08} (2015)
  004}, [\href{http://arxiv.org/abs/1501.03799}{{\tt 1501.03799}}].

\bibitem{Falkowski:2015iwa}
A.~Falkowski, C.~Gross and O.~Lebedev, \emph{{A second Higgs from the Higgs
  portal}}, \href{http://dx.doi.org/10.1007/JHEP05(2015)057}{\emph{JHEP} {\bf
  05} (2015) 057}, [\href{http://arxiv.org/abs/1502.01361}{{\tt 1502.01361}}].

\bibitem{Berlin:2015wwa}
A.~Berlin, S.~Gori, T.~Lin and L.-T. Wang, \emph{{Pseudoscalar Portal Dark
  Matter}}, \href{http://dx.doi.org/10.1103/PhysRevD.92.015005}{\emph{Phys.
  Rev.} {\bf D92} (2015) 015005}, [\href{http://arxiv.org/abs/1502.06000}{{\tt
  1502.06000}}].

\bibitem{Kouvaris:2014uoa}
C.~Kouvaris, I.~M. Shoemaker and K.~Tuominen, \emph{{Self-Interacting Dark
  Matter through the Higgs Portal}},
  \href{http://dx.doi.org/10.1103/PhysRevD.91.043519}{\emph{Phys. Rev.} {\bf
  D91} (2015) 043519}, [\href{http://arxiv.org/abs/1411.3730}{{\tt
  1411.3730}}].

\bibitem{Buckley:2014fba}
M.~R. Buckley, D.~Feld and D.~Goncalves, \emph{{Scalar Simplified Models for
  Dark Matter}},
  \href{http://dx.doi.org/10.1103/PhysRevD.91.015017}{\emph{Phys. Rev.} {\bf
  D91} (2015) 015017}, [\href{http://arxiv.org/abs/1410.6497}{{\tt
  1410.6497}}].

\bibitem{Ghorbani:2014qpa}
K.~Ghorbani, \emph{{Fermionic dark matter with pseudo-scalar Yukawa
  interaction}},
  \href{http://dx.doi.org/10.1088/1475-7516/2015/01/015}{\emph{JCAP} {\bf 1501}
  (2015) 015}, [\href{http://arxiv.org/abs/1408.4929}{{\tt 1408.4929}}].

\bibitem{Basso:2013nza}
L.~Basso, O.~Fischer and J.~J. van Der~Bij, \emph{{A renormalization group
  analysis of the Hill model and its HEIDI extension}},
  \href{http://dx.doi.org/10.1016/j.physletb.2014.01.064}{\emph{Phys. Lett.}
  {\bf B730} (2014) 326--331}, [\href{http://arxiv.org/abs/1309.6086}{{\tt
  1309.6086}}].

\bibitem{Basso:2012nh}
L.~Basso, O.~Fischer and J.~J. van~der Bij, \emph{{A singlet-triplet extension
  for the Higgs search at LEP and LHC}},
  \href{http://dx.doi.org/10.1209/0295-5075/101/51004}{\emph{Europhys. Lett.}
  {\bf 101} (2013) 51004}, [\href{http://arxiv.org/abs/1212.5560}{{\tt
  1212.5560}}].

\bibitem{Fedderke:2014wda}
M.~A. Fedderke, J.-Y. Chen, E.~W. Kolb and L.-T. Wang, \emph{{The Fermionic
  Dark Matter Higgs Portal: an effective field theory approach}},
  \href{http://dx.doi.org/10.1007/JHEP08(2014)122}{\emph{JHEP} {\bf 08} (2014)
  122}, [\href{http://arxiv.org/abs/1404.2283}{{\tt 1404.2283}}].

\bibitem{Baek:2011aa}
S.~Baek, P.~Ko and W.-I. Park, \emph{{Search for the Higgs portal to a singlet
  fermionic dark matter at the LHC}},
  \href{http://dx.doi.org/10.1007/JHEP02(2012)047}{\emph{JHEP} {\bf 02} (2012)
  047}, [\href{http://arxiv.org/abs/1112.1847}{{\tt 1112.1847}}].

\bibitem{Queiroz:2014pra}
F.~S. Queiroz, K.~Sinha and A.~Strumia, \emph{{Leptoquarks, Dark Matter, and
  Anomalous LHC Events}},
  \href{http://dx.doi.org/10.1103/PhysRevD.91.035006}{\emph{Phys. Rev.} {\bf
  D91} (2015) 035006}, [\href{http://arxiv.org/abs/1409.6301}{{\tt
  1409.6301}}].

\bibitem{Feng:2014vea}
L.~Feng, S.~Profumo and L.~Ubaldi, \emph{{Closing in on singlet scalar dark
  matter: LUX, invisible Higgs decays and gamma-ray lines}},
  \href{http://dx.doi.org/10.1007/JHEP03(2015)045}{\emph{JHEP} {\bf 03} (2015)
  045}, [\href{http://arxiv.org/abs/1412.1105}{{\tt 1412.1105}}].

\bibitem{Martin:2014sxa}
A.~Martin, J.~Shelton and J.~Unwin, \emph{{Fitting the Galactic Center
  Gamma-Ray Excess with Cascade Annihilations}},
  \href{http://dx.doi.org/10.1103/PhysRevD.90.103513}{\emph{Phys. Rev.} {\bf
  D90} (2014) 103513}, [\href{http://arxiv.org/abs/1405.0272}{{\tt
  1405.0272}}].

\bibitem{Cline:2013gha}
J.~M. Cline, K.~Kainulainen, P.~Scott and C.~Weniger, \emph{{Update on scalar
  singlet dark matter}}, \href{http://dx.doi.org/10.1103/PhysRevD.92.039906,
  10.1103/PhysRevD.88.055025}{\emph{Phys. Rev.} {\bf D88} (2013) 055025},
  [\href{http://arxiv.org/abs/1306.4710}{{\tt 1306.4710}}].

\bibitem{Izaguirre:2014vva}
E.~Izaguirre, G.~Krnjaic and B.~Shuve, \emph{{The Galactic Center Excess from
  the Bottom Up}},
  \href{http://dx.doi.org/10.1103/PhysRevD.90.055002}{\emph{Phys. Rev.} {\bf
  D90} (2014) 055002}, [\href{http://arxiv.org/abs/1404.2018}{{\tt
  1404.2018}}].

\bibitem{Varzielas:2015sno}
I.~Medeiros~Varzielas and O.~Fischer, \emph{{Non-Abelian family symmetries as
  portals to dark matter}},
  \href{http://dx.doi.org/10.1007/JHEP01(2016)160}{\emph{JHEP} {\bf 01} (2016)
  160}, [\href{http://arxiv.org/abs/1512.00869}{{\tt 1512.00869}}].

\bibitem{Agashe:2014kda}
{\scshape Particle Data Group} collaboration, K.~A. Olive et~al., \emph{{Review
  of Particle Physics}},
  \href{http://dx.doi.org/10.1088/1674-1137/38/9/090001}{\emph{Chin. Phys.}
  {\bf C38} (2014) 090001}.

\bibitem{Bechtle:2015pma}
P.~Bechtle, S.~Heinemeyer, O.~Stal, T.~Stefaniak and G.~Weiglein,
  \emph{{Applying Exclusion Likelihoods from LHC Searches to Extended Higgs
  Sectors}}, \href{http://dx.doi.org/10.1140/epjc/s10052-015-3650-z}{\emph{Eur.
  Phys. J.} {\bf C75} (2015) 421}, [\href{http://arxiv.org/abs/1507.06706}{{\tt
  1507.06706}}].

\bibitem{arXiv:1311.0055}
P.~Bechtle et~al., \emph{{HiggsBounds-4: Improved Tests of Extended Higgs
  Sectors against Exclusion Bounds from LEP, the Tevatron and the LHC}},
  {\emph{Eur. Phys. J.} {\bf C74} (2014) 2693},
  [\href{http://arxiv.org/abs/1311.0055}{{\tt 1311.0055}}].

\bibitem{arXiv:1301.2345}
P.~Bechtle et~al., \emph{{Recent Developments in HiggsBounds and a Preview of
  HiggsSignals}}, {\emph{PoS} {\bf CHARGED2012} (2012) 024},
  [\href{http://arxiv.org/abs/1301.2345}{{\tt 1301.2345}}].

\bibitem{arXiv:1102.1898}
P.~Bechtle, O.~Brein, S.~Heinemeyer, G.~Weiglein and K.~E. Williams,
  \emph{{HiggsBounds 2.0.0: Confronting Neutral and Charged Higgs Sector
  Predictions with Exclusion Bounds from LEP and the Tevatron}},
  \href{http://dx.doi.org/10.1016/j.cpc.2011.07.015}{\emph{Comput. Phys.
  Commun.} {\bf 182} (2011) 2605--2631},
  [\href{http://arxiv.org/abs/1102.1898}{{\tt 1102.1898}}].

\bibitem{arXiv:0811.4169}
P.~Bechtle, O.~Brein, S.~Heinemeyer, G.~Weiglein and K.~E. Williams,
  \emph{{HiggsBounds: Confronting Arbitrary Higgs Sectors with Exclusion Bounds
  from LEP and the Tevatron}},
  \href{http://dx.doi.org/10.1016/j.cpc.2009.09.003}{\emph{Comput. Phys.
  Commun.} {\bf 181} (2010) 138--167},
  [\href{http://arxiv.org/abs/0811.4169}{{\tt 0811.4169}}].

\bibitem{Searches:2001aa}
{\scshape LEP Higgs Working Group for Higgs boson searches} collaboration,
  \emph{{Flavor independent search for hadronically decaying neutral Higgs
  bosons at LEP}},  in \emph{{Lepton and photon interactions at high energies.
  Proceedings, 20th International Symposium, LP 2001, Rome, Italy, July 23-28,
  2001}}, 2001.
\newblock \href{http://arxiv.org/abs/hep-ex/0107034}{{\tt hep-ex/0107034}}.

\bibitem{Abbiendi:2002qp}
{\scshape OPAL} collaboration, G.~Abbiendi et~al., \emph{{Decay mode
  independent searches for new scalar bosons with the OPAL detector at LEP}},
  \href{http://dx.doi.org/10.1140/epjc/s2002-01115-1}{\emph{Eur. Phys. J.} {\bf
  C27} (2003) 311--329}, [\href{http://arxiv.org/abs/hep-ex/0206022}{{\tt
  hep-ex/0206022}}].

\bibitem{Searches:2001ab}
{\scshape OPAL, DELPHI, LEP Higgs Working for Higgs boson searches, L3 CERN,
  ALEPH} collaboration, \emph{{Searches for invisible Higgs bosons: Preliminary
  combined results using LEP data collected at energies up to 209-GeV}},  in
  \emph{{Lepton and photon interactions at high energies. Proceedings, 20th
  International Symposium, LP 2001, Rome, Italy, July 23-28, 2001}}, 2001.
\newblock \href{http://arxiv.org/abs/hep-ex/0107032}{{\tt hep-ex/0107032}}.

\bibitem{Abdallah:2003ry}
{\scshape DELPHI} collaboration, J.~Abdallah et~al., \emph{{Searches for
  invisibly decaying Higgs bosons with the DELPHI detector at LEP}},
  \href{http://dx.doi.org/10.1140/epjc/s2003-01469-8}{\emph{Eur. Phys. J.} {\bf
  C32} (2004) 475--492}, [\href{http://arxiv.org/abs/hep-ex/0401022}{{\tt
  hep-ex/0401022}}].

\bibitem{Abbiendi:2007ac}
{\scshape OPAL} collaboration, G.~Abbiendi et~al., \emph{{Search for invisibly
  decaying Higgs bosons in $e^{+} e^{-} \to Z_0 h_0$ production at $\sqrt{s}=$
  183 - 209 GeV}},
  \href{http://dx.doi.org/10.1016/j.physletb.2009.09.010}{\emph{Phys. Lett.}
  {\bf B682} (2010) 381--390}, [\href{http://arxiv.org/abs/0707.0373}{{\tt
  0707.0373}}].

\bibitem{Searches:2001ac}
{\scshape OPAL, DELPHI, L3, ALEPH, LEP Higgs Working Group for Higgs boson
  searches} collaboration, \emph{{Search for charged Higgs bosons: Preliminary
  combined results using LEP data collected at energies up to 209-GeV}},  in
  \emph{{Lepton and photon interactions at high energies. Proceedings, 20th
  International Symposium, LP 2001, Rome, Italy, July 23-28, 2001}}, 2001.
\newblock \href{http://arxiv.org/abs/hep-ex/0107031}{{\tt hep-ex/0107031}}.

\bibitem{Abbiendi:2001kp}
{\scshape OPAL} collaboration, G.~Abbiendi et~al., \emph{{Search for Yukawa
  production of a light neutral Higgs boson at LEP}},
  \href{http://dx.doi.org/10.1007/s100520200896}{\emph{Eur. Phys. J.} {\bf C23}
  (2002) 397--407}, [\href{http://arxiv.org/abs/hep-ex/0111010}{{\tt
  hep-ex/0111010}}].

\bibitem{Achard:2004cf}
{\scshape L3} collaboration, P.~Achard et~al., \emph{{Search for an
  invisibly-decaying Higgs boson at LEP}},
  \href{http://dx.doi.org/10.1016/j.physletb.2005.01.030}{\emph{Phys. Lett.}
  {\bf B609} (2005) 35--48}, [\href{http://arxiv.org/abs/hep-ex/0501033}{{\tt
  hep-ex/0501033}}].

\bibitem{Schael:2006cr}
{\scshape DELPHI, OPAL, ALEPH, LEP Working Group for Higgs Boson Searches, L3}
  collaboration, S.~Schael et~al., \emph{{Search for neutral MSSM Higgs bosons
  at LEP}}, \href{http://dx.doi.org/10.1140/epjc/s2006-02569-7}{\emph{Eur.
  Phys. J.} {\bf C47} (2006) 547--587},
  [\href{http://arxiv.org/abs/hep-ex/0602042}{{\tt hep-ex/0602042}}].

\bibitem{Abdallah:2003wd}
{\scshape DELPHI} collaboration, J.~Abdallah et~al., \emph{{Search for charged
  Higgs bosons at LEP in general two Higgs doublet models}},
  \href{http://dx.doi.org/10.1140/epjc/s2004-01732-6}{\emph{Eur. Phys. J.} {\bf
  C34} (2004) 399--418}, [\href{http://arxiv.org/abs/hep-ex/0404012}{{\tt
  hep-ex/0404012}}].

\bibitem{Abdallah:2004wy}
{\scshape DELPHI} collaboration, J.~Abdallah et~al., \emph{{Searches for
  neutral higgs bosons in extended models}},
  \href{http://dx.doi.org/10.1140/epjc/s2004-02011-4}{\emph{Eur. Phys. J.} {\bf
  C38} (2004) 1--28}, [\href{http://arxiv.org/abs/hep-ex/0410017}{{\tt
  hep-ex/0410017}}].

\bibitem{ATLASnotes}
{\scshape ATLAS} collaboration, ``{ATLAS} {CONF} notes 2012-160 2014-049
  2012-135 2012-161 2012-092 2012-018 2012-012 2013-010 2013-013 2012-019
  2012-168 2012-078 2014-050 2012-017 2012-016 2011-094 2013-030 2011-157.''

\bibitem{LHWGnotes}
{LEP Higgs Working Group}, ``{LHWG} notes 2002-02.''

\bibitem{Aad:2014xva}
{\scshape ATLAS} collaboration, G.~Aad et~al., \emph{{Search for the Standard
  Model Higgs boson decay to $\mu^{+}\mu^{-}$ with the ATLAS detector}},
  \href{http://dx.doi.org/10.1016/j.physletb.2014.09.008}{\emph{Phys. Lett.}
  {\bf B738} (2014) 68--86}, [\href{http://arxiv.org/abs/1406.7663}{{\tt
  1406.7663}}].

\bibitem{Chatrchyan:2012sn}
{\scshape CMS} collaboration, S.~Chatrchyan et~al., \emph{{Search for a Higgs
  boson in the decay channel $H \to ZZ^{*} \to q \bar{q} \ell^{-} \ell^{+}$ in
  $pp$ collisions at $\sqrt{s}=7$ TeV}},
  \href{http://dx.doi.org/10.1007/JHEP04(2012)036}{\emph{JHEP} {\bf 04} (2012)
  036}, [\href{http://arxiv.org/abs/1202.1416}{{\tt 1202.1416}}].

\bibitem{Chatrchyan:2014tja}
{\scshape CMS} collaboration, S.~Chatrchyan et~al., \emph{{Search for invisible
  decays of Higgs bosons in the vector boson fusion and associated ZH
  production modes}},
  \href{http://dx.doi.org/10.1140/epjc/s10052-014-2980-6}{\emph{Eur. Phys. J.}
  {\bf C74} (2014) 2980}, [\href{http://arxiv.org/abs/1404.1344}{{\tt
  1404.1344}}].

\bibitem{Aad:2012tfa}
{\scshape ATLAS} collaboration, G.~Aad et~al., \emph{{Observation of a new
  particle in the search for the Standard Model Higgs boson with the ATLAS
  detector at the LHC}},
  \href{http://dx.doi.org/10.1016/j.physletb.2012.08.020}{\emph{Phys. Lett.}
  {\bf B716} (2013) 1--29}, [\href{http://arxiv.org/abs/1207.7214}{{\tt
  1207.7214}}].

\bibitem{Aad:2012tj}
{\scshape ATLAS} collaboration, G.~Aad et~al., \emph{{Search for charged Higgs
  bosons decaying via $H^{+} \to \tau \nu$ in top quark pair events using $pp$
  collision data at $\sqrt{s}=7$ TeV with the ATLAS detector}},
  \href{http://dx.doi.org/10.1007/JHEP06(2012)039}{\emph{JHEP} {\bf 06} (2012)
  039}, [\href{http://arxiv.org/abs/1204.2760}{{\tt 1204.2760}}].

\bibitem{Chatrchyan:2012tx}
{\scshape CMS} collaboration, S.~Chatrchyan et~al., \emph{{Combined results of
  searches for the standard model Higgs boson in $pp$ collisions at
  $\sqrt{s}=7$ TeV}},
  \href{http://dx.doi.org/10.1016/j.physletb.2012.02.064}{\emph{Phys. Lett.}
  {\bf B710} (2013) 26--48}, [\href{http://arxiv.org/abs/1202.1488}{{\tt
  1202.1488}}].

\bibitem{Aad:2011ec}
{\scshape ATLAS} collaboration, G.~Aad et~al., \emph{{Search for a heavy
  Standard Model Higgs boson in the channel $H \to ZZ \to \ell \ell q q$ using
  the ATLAS detector}},
  \href{http://dx.doi.org/10.1016/j.physletb.2011.11.056}{\emph{Phys. Lett.}
  {\bf B707} (2013) 27--45}, [\href{http://arxiv.org/abs/1108.5064}{{\tt
  1108.5064}}].

\bibitem{ATLAS:2012ac}
{\scshape ATLAS} collaboration, G.~Aad et~al., \emph{{Search for the Standard
  Model Higgs boson in the decay channel $H \to$ ZZ(*) $\to 4 \ell$ with 4.8
  fb$^{-1}$ of $pp$ collision data at $\sqrt{s}=7$ TeV with ATLAS}},
  \href{http://dx.doi.org/10.1016/j.physletb.2012.03.005}{\emph{Phys. Lett.}
  {\bf B710} (2013) 383--402}, [\href{http://arxiv.org/abs/1202.1415}{{\tt
  1202.1415}}].

\bibitem{ATLAS:2011af}
{\scshape ATLAS} collaboration, G.~Aad et~al., \emph{{Search for a Standard
  Model Higgs boson in the $H \to ZZ \to \ell^{+} \ell^{-} \nu \bar{\nu}$ decay
  channel with the ATLAS detector}},
  \href{http://dx.doi.org/10.1103/PhysRevLett.107.221802}{\emph{Phys. Rev.
  Lett.} {\bf 107} (2011) 221802}, [\href{http://arxiv.org/abs/1109.3357}{{\tt
  1109.3357}}].

\bibitem{Khachatryan:2014wca}
{\scshape CMS} collaboration, V.~Khachatryan et~al., \emph{{Search for neutral
  MSSM Higgs bosons decaying to a pair of tau leptons in pp collisions}},
  \href{http://dx.doi.org/10.1007/JHEP10(2014)160}{\emph{JHEP} {\bf 10} (2014)
  160}, [\href{http://arxiv.org/abs/1408.3316}{{\tt 1408.3316}}].

\bibitem{ATLAS:2011ae}
{\scshape ATLAS} collaboration, G.~Aad et~al., \emph{{Search for the Higgs
  boson in the $H \to WW \to \ell \nu jj$ decay channel in pp collisions at
  $\sqrt{s} = 7$ TeV with the ATLAS detector}},
  \href{http://dx.doi.org/10.1103/PhysRevLett.107.231801}{\emph{Phys. Rev.
  Lett.} {\bf 107} (2011) 231801}, [\href{http://arxiv.org/abs/1109.3615}{{\tt
  1109.3615}}].

\bibitem{Chatrchyan:2012dg}
{\scshape CMS} collaboration, S.~Chatrchyan et~al., \emph{{Search for the
  standard model Higgs boson in the decay channel $H \to Z Z \to 4 \ell$ in
  $pp$ collisions at $\sqrt{s}=7$ TeV}},
  \href{http://dx.doi.org/10.1103/PhysRevLett.108.111804}{\emph{Phys. Rev.
  Lett.} {\bf 108} (2012) 111804}, [\href{http://arxiv.org/abs/1202.1997}{{\tt
  1202.1997}}].

\bibitem{Aad:2014fia}
{\scshape ATLAS} collaboration, G.~Aad et~al., \emph{{Search for Higgs boson
  decays to a photon and a Z boson in pp collisions at $\sqrt{s}$=7 and 8 TeV
  with the ATLAS detector}},
  \href{http://dx.doi.org/10.1016/j.physletb.2014.03.015}{\emph{Phys. Lett.}
  {\bf B732} (2014) 8--27}, [\href{http://arxiv.org/abs/1402.3051}{{\tt
  1402.3051}}].

\bibitem{Aad:2014yja}
{\scshape ATLAS} collaboration, G.~Aad et~al., \emph{{Search For Higgs Boson
  Pair Production in the $\gamma\gamma b\bar{b}$ Final State using $pp$
  Collision Data at $\sqrt{s}=8$ TeV from the ATLAS Detector}},
  \href{http://dx.doi.org/10.1103/PhysRevLett.114.081802}{\emph{Phys. Rev.
  Lett.} {\bf 114} (2015) 081802}, [\href{http://arxiv.org/abs/1406.5053}{{\tt
  1406.5053}}].

\bibitem{Aad:2014iia}
{\scshape ATLAS} collaboration, G.~Aad et~al., \emph{{Search for Invisible
  Decays of a Higgs Boson Produced in Association with a Z Boson in ATLAS}},
  \href{http://dx.doi.org/10.1103/PhysRevLett.112.201802}{\emph{Phys. Rev.
  Lett.} {\bf 112} (2014) 201802}, [\href{http://arxiv.org/abs/1402.3244}{{\tt
  1402.3244}}].

\bibitem{Aad:2011rv}
{\scshape ATLAS} collaboration, G.~Aad et~al., \emph{{Search for neutral MSSM
  Higgs bosons decaying to $\tau^+ \tau^-$ pairs in proton-proton collisions at
  $\sqrt{s}=7$ TeV with the ATLAS detector}},
  \href{http://dx.doi.org/10.1016/j.physletb.2011.10.001}{\emph{Phys. Lett.}
  {\bf B705} (2013) 174--192}, [\href{http://arxiv.org/abs/1107.5003}{{\tt
  1107.5003}}].

\bibitem{ATLAS:2011aa}
{\scshape ATLAS} collaboration, G.~Aad et~al., \emph{{Search for the Higgs
  boson in the $H \to$ WW(*) $\to \ell_\nu\ell_\nu$ decay channel in $pp$
  collisions at $\sqrt{s}=7$ TeV with the ATLAS detector}},
  \href{http://dx.doi.org/10.1103/PhysRevLett.108.111802}{\emph{Phys. Rev.
  Lett.} {\bf 108} (2012) 111802}, [\href{http://arxiv.org/abs/1112.2577}{{\tt
  1112.2577}}].

\bibitem{ATLAS:2012ad}
{\scshape ATLAS} collaboration, G.~Aad et~al., \emph{{Search for the Standard
  Model Higgs boson in the diphoton decay channel with 4.9 fb$^{-1}$ of $pp$
  collisions at $\sqrt{s}=7$ TeV with ATLAS}},
  \href{http://dx.doi.org/10.1103/PhysRevLett.108.111803}{\emph{Phys. Rev.
  Lett.} {\bf 108} (2012) 111803}, [\href{http://arxiv.org/abs/1202.1414}{{\tt
  1202.1414}}].

\bibitem{Chatrchyan:2013vaa}
{\scshape CMS} collaboration, S.~Chatrchyan et~al., \emph{{Search for a Higgs
  boson decaying into a Z and a photon in $pp$ collisions at $\sqrt{s} =$ 7 and
  8 TeV}}, \href{http://dx.doi.org/10.1016/j.physletb.2013.09.057}{\emph{Phys.
  Lett.} {\bf B726} (2013) 587--609},
  [\href{http://arxiv.org/abs/1307.5515}{{\tt 1307.5515}}].

\bibitem{ATLAS:2012ae}
{\scshape ATLAS} collaboration, G.~Aad et~al., \emph{{Combined search for the
  Standard Model Higgs boson using up to 4.9 fb$^{-1}$ of $pp$ collision data
  at $\sqrt{s}=7$ TeV with the ATLAS detector at the LHC}},
  \href{http://dx.doi.org/10.1016/j.physletb.2012.02.044}{\emph{Phys. Lett.}
  {\bf B710} (2013) 49--66}, [\href{http://arxiv.org/abs/1202.1408}{{\tt
  1202.1408}}].

\bibitem{Aad:2014ioa}
{\scshape ATLAS} collaboration, G.~Aad et~al., \emph{{Search for Scalar
  Diphoton Resonances in the Mass Range $65-600$ GeV with the ATLAS Detector in
  $pp$ Collision Data at $\sqrt{s}$ = 8 $TeV$}},
  \href{http://dx.doi.org/10.1103/PhysRevLett.113.171801}{\emph{Phys. Rev.
  Lett.} {\bf 113} (2014) 171801}, [\href{http://arxiv.org/abs/1407.6583}{{\tt
  1407.6583}}].

\bibitem{Chatrchyan:2012ft}
{\scshape CMS} collaboration, S.~Chatrchyan et~al., \emph{{Search for the
  standard model Higgs boson in the $H$ to $Z Z$ to $2 \ell 2 \nu$ channel in
  $pp$ collisions at $\sqrt{s}=7$ TeV}},
  \href{http://dx.doi.org/10.1007/JHEP03(2012)040}{\emph{JHEP} {\bf 03} (2012)
  040}, [\href{http://arxiv.org/abs/1202.3478}{{\tt 1202.3478}}].

\bibitem{CDFnotes}
{\scshape CDF} collaboration, ``{CDF} notes 10500 7307 10439 10796 9999 10485
  10798 8353 10799 10599 10573 10010 7712 10574.''

\bibitem{D0notes}
{\scshape D0} collaboration, ``{D0} notes 6083 6305 6227 6299 6301 6302 5739
  5845 6286 5757 6296 6183 6295 6171 6309 6276 6304 5873.''

\bibitem{Abazov:2010ci}
{\scshape D0} collaboration, V.~M. Abazov et~al., \emph{{Search for neutral
  Higgs bosons in the multi-$b$-jet topology in 5.2fb$^{-1}$ of $p\bar{p}$
  collisions at $\sqrt{s} = 1.96$ TeV}},
  \href{http://dx.doi.org/10.1016/j.physletb.2011.02.062}{\emph{Phys. Lett.}
  {\bf B698} (2014) 97--104}, [\href{http://arxiv.org/abs/1011.1931}{{\tt
  1011.1931}}].

\bibitem{Aaltonen:2009vf}
{\scshape CDF} collaboration, T.~Aaltonen et~al., \emph{{Search for Higgs
  bosons predicted in two-Higgs-doublet models via decays to tau lepton pairs
  in 1.96-TeV p anti-p collisions}},
  \href{http://dx.doi.org/10.1103/PhysRevLett.103.201801}{\emph{Phys. Rev.
  Lett.} {\bf 103} (2009) 201801}, [\href{http://arxiv.org/abs/0906.1014}{{\tt
  0906.1014}}].

\bibitem{Abazov:2010zk}
{\scshape D0} collaboration, V.~M. Abazov et~al., \emph{{Search for $ZH
  \rightarrow \ell^+\ell^-b\bar{b}$ production in $4.2$~fb$^{-1}$ of $p\bar{p}$
  collisions at $\sqrt{s}=1.96$ TeV}},
  \href{http://dx.doi.org/10.1103/PhysRevLett.105.251801}{\emph{Phys. Rev.
  Lett.} {\bf 105} (2010) 251801}, [\href{http://arxiv.org/abs/1008.3564}{{\tt
  1008.3564}}].

\bibitem{Aaltonen:2011nh}
{\scshape CDF} collaboration, T.~Aaltonen et~al., \emph{{Search for Higgs
  Bosons Produced in Association with $b$-quarks}},
  \href{http://dx.doi.org/10.1103/PhysRevD.85.032005}{\emph{Phys. Rev.} {\bf
  D85} (2012) 032005}, [\href{http://arxiv.org/abs/1106.4782}{{\tt
  1106.4782}}].

\bibitem{Aaltonen:2010cm}
{\scshape CDF} collaboration, T.~Aaltonen et~al., \emph{{Inclusive Search for
  Standard Model Higgs Boson Production in the WW Decay Channel using the CDF
  II Detector}},
  \href{http://dx.doi.org/10.1103/PhysRevLett.104.061803}{\emph{Phys. Rev.
  Lett.} {\bf 104} (2010) 061803}, [\href{http://arxiv.org/abs/1001.4468}{{\tt
  1001.4468}}].

\bibitem{Benjamin:2011sv}
{\scshape TEVNPH Working Group (Tevatron New Phenomena and Higgs Working
  Group), CDF, D0} collaboration, D.~Benjamin, \emph{{Combined CDF and D0 upper
  limits on $gg\to H\to W^+W^-$ and constraints on the Higgs boson mass in
  fourth-generation fermion models with up to 8.2 fb$^{-1}$ of data}},  in
  \emph{{Proceedings, 21st International Europhysics Conference on High energy
  physics (EPS-HEP 2011)}}, 2011.
\newblock \href{http://arxiv.org/abs/1108.3331}{{\tt 1108.3331}}.

\bibitem{Group:2012zca}
{\scshape Tevatron New Physics Higgs Working Group, CDF, D0} collaboration,
  \emph{{Updated Combination of CDF and D0 Searches for Standard Model Higgs
  Boson Production with up to 10.0 fb$^{-1}$ of Data}},  2012.
\newblock \href{http://arxiv.org/abs/1207.0449}{{\tt 1207.0449}}.

\bibitem{Aaltonen:2009ke}
{\scshape CDF} collaboration, T.~Aaltonen et~al., \emph{{Search for charged
  Higgs bosons in decays of top quarks in $p-\bar{p}$ collisions at $\sqrt{s} =
  1.96$ TeV}},
  \href{http://dx.doi.org/10.1103/PhysRevLett.103.101803}{\emph{Phys. Rev.
  Lett.} {\bf 103} (2009) 101803}, [\href{http://arxiv.org/abs/0907.1269}{{\tt
  0907.1269}}].

\bibitem{Abazov:2011qz}
{\scshape D0} collaboration, V.~M. Abazov et~al., \emph{{Search for neutral
  Minimal Supersymmetric Standard Model Higgs bosons decaying to tau pairs
  produced in association with $b$ quarks in $p\bar{p}$ collisions at
  $\sqrt{s}=1.96$ TeV}},
  \href{http://dx.doi.org/10.1103/PhysRevLett.107.121801}{\emph{Phys. Rev.
  Lett.} {\bf 107} (2011) 121801}, [\href{http://arxiv.org/abs/1106.4885}{{\tt
  1106.4885}}].

\bibitem{Abazov:2008wg}
{\scshape D0} collaboration, V.~M. Abazov et~al., \emph{{Search for a scalar or
  vector particle decaying into $Z \gamma$ in $p \bar{p}$ collisions at
  $\sqrt{s}$=1.96 TeV}},
  \href{http://dx.doi.org/10.1016/j.physletb.2008.12.009}{\emph{Phys. Lett.}
  {\bf B671} (2009) 349--355}, [\href{http://arxiv.org/abs/0806.0611}{{\tt
  0806.0611}}].

\bibitem{Aaltonen:2012qt}
{\scshape CDF, D0} collaboration, T.~Aaltonen et~al., \emph{{Evidence for a
  particle produced in association with weak bosons and decaying to a
  bottom-antibottom quark pair in Higgs boson searches at the Tevatron}},
  \href{http://dx.doi.org/10.1103/PhysRevLett.109.071804}{\emph{Phys. Rev.
  Lett.} {\bf 109} (2012) 071804}, [\href{http://arxiv.org/abs/1207.6436}{{\tt
  1207.6436}}].

\bibitem{Abazov:2009yi}
{\scshape D0} collaboration, V.~M. Abazov et~al., \emph{{Search for NMSSM Higgs
  bosons in the $h \to aa \to \mu \mu \mu \mu$, $\mu \mu \tau \tau$ channels
  using $p\bar{p}$ collisions at $\sqrt{s}= 1.96$ TeV}},
  \href{http://dx.doi.org/10.1103/PhysRevLett.103.061801}{\emph{Phys. Rev.
  Lett.} {\bf 103} (2009) 061801}, [\href{http://arxiv.org/abs/0905.3381}{{\tt
  0905.3381}}].

\bibitem{Benjamin:2010xb}
{\scshape Tevatron New Phenomena and Higgs Working Group} collaboration,
  D.~Benjamin et~al., \emph{{Combined CDF and D0 Upper Limits on MSSM Higgs
  Boson Production in tau-tau Final States with up to 2.2 fb$^{-1}$}},
  \href{http://arxiv.org/abs/1003.3363}{{\tt 1003.3363}}.

\bibitem{Abazov:2011jh}
{\scshape D0} collaboration, V.~M. Abazov et~al., \emph{{Search for Higgs
  bosons decaying to $\tau\tau$ pairs in $p\bar {p}$ collisions at $\sqrt{s} =
  1.96$ TeV}},
  \href{http://dx.doi.org/10.1016/j.physletb.2011.12.050}{\emph{Phys. Lett.}
  {\bf B707} (2012) 323--329}, [\href{http://arxiv.org/abs/1106.4555}{{\tt
  1106.4555}}].

\bibitem{Abazov:2011ed}
{\scshape D0} collaboration, V.~M. Abazov et~al., \emph{{Search for associated
  Higgs boson production using like charge dilepton events in $p\bar{p}$
  collisions at $\sqrt{s} = 1.96$ TeV}},
  \href{http://dx.doi.org/10.1103/PhysRevD.84.092002}{\emph{Phys. Rev.} {\bf
  D84} (2011) 092002}, [\href{http://arxiv.org/abs/1107.1268}{{\tt
  1107.1268}}].

\bibitem{TEVNPHWorking:2011aa}
{\scshape CDF, D0} collaboration, T.~W. Group, \emph{{Combined CDF and D0
  Searches for the Standard Model Higgs Boson Decaying to Two Photons with up
  to 8.2 fb$^{-1}$}},  in \emph{{Proceedings, 21st International Europhysics
  Conference on High energy physics (EPS-HEP 2011)}}, 2011.
\newblock \href{http://arxiv.org/abs/1107.4960}{{\tt 1107.4960}}.

\bibitem{Abazov:2010hn}
{\scshape D0} collaboration, V.~M. Abazov et~al., \emph{{Search for $WH$
  associated production in 5.3 fb$^{-1}$ of $p\bar{p}$ collisions at the
  Fermilab Tevatron}},
  \href{http://dx.doi.org/10.1016/j.physletb.2011.02.036}{\emph{Phys. Lett.}
  {\bf B698} (2011) 6--13}, [\href{http://arxiv.org/abs/1012.0874}{{\tt
  1012.0874}}].

\bibitem{Aaltonen:2008ec}
{\scshape CDF} collaboration, T.~Aaltonen et~al., \emph{{Search for a Higgs
  Boson Decaying to Two $W$ Bosons at CDF}},
  \href{http://dx.doi.org/10.1103/PhysRevLett.102.021802}{\emph{Phys. Rev.
  Lett.} {\bf 102} (2009) 021802}, [\href{http://arxiv.org/abs/0809.3930}{{\tt
  0809.3930}}].

\bibitem{Aaltonen:2009dh}
{\scshape CDF} collaboration, T.~Aaltonen et~al., \emph{{Search for a Higgs
  Boson in $W H \to \ell \nu b \bar{b}$ in $p\bar{p}$ Collisions at $\sqrt{s} =
  1.96$ TeV}},
  \href{http://dx.doi.org/10.1103/PhysRevLett.103.101802}{\emph{Phys. Rev.
  Lett.} {\bf 103} (2009) 101802}, [\href{http://arxiv.org/abs/0906.5613}{{\tt
  0906.5613}}].

\bibitem{Abazov:2009aa}
{\scshape D0} collaboration, V.~M. Abazov et~al., \emph{{Search for charged
  Higgs bosons in top quark decays}},
  \href{http://dx.doi.org/10.1016/j.physletb.2009.11.016}{\emph{Phys. Lett.}
  {\bf B682} (2009) 278--286}, [\href{http://arxiv.org/abs/0908.1811}{{\tt
  0908.1811}}].

\bibitem{Abazov:2010ct}
{\scshape D0} collaboration, V.~M. Abazov et~al., \emph{{Search for Higgs boson
  production in dilepton and missing energy final states with 5.4 $fb^{-1}$ of
  $p\bar{p}$ collisions at ${\sqrt s =1.96}$ TeV}},
  \href{http://dx.doi.org/10.1103/PhysRevLett.104.061804}{\emph{Phys. Rev.
  Lett.} {\bf 104} (2010) 061804}, [\href{http://arxiv.org/abs/1001.4481}{{\tt
  1001.4481}}].

\bibitem{Abazov:2009kq}
{\scshape D0} collaboration, V.~M. Abazov et~al., \emph{{Search for Resonant
  Diphoton Production with the D0 Detector}},
  \href{http://dx.doi.org/10.1103/PhysRevLett.102.231801}{\emph{Phys. Rev.
  Lett.} {\bf 102} (2009) 231801}, [\href{http://arxiv.org/abs/0901.1887}{{\tt
  0901.1887}}].

\bibitem{Aad:2014vgg}
{\scshape ATLAS} collaboration, G.~Aad et~al., \emph{{Search for neutral Higgs
  bosons of the minimal supersymmetric standard model in pp collisions at
  $\sqrt{s}$ = 8 TeV with the ATLAS detector}},
  \href{http://dx.doi.org/10.1007/JHEP11(2014)056}{\emph{JHEP} {\bf 11} (2014)
  056}, [\href{http://arxiv.org/abs/1409.6064}{{\tt 1409.6064}}].

\bibitem{arXiv:1509.04670}
{\scshape ATLAS} collaboration, \emph{{Searches for Higgs boson pair production
  in the $hh\to b\bar{b}\tau\tau, \gamma\gamma WW*, \gamma\gamma b\bar{b},
  b\bar{b}b\bar{b}$ channels with the ATLAS detector}},
  \href{http://arxiv.org/abs/1509.04670}{{\tt 1509.04670}}.

\bibitem{arXiv:1506.02301}
{\scshape CMS} collaboration, \emph{{Search for diphoton resonances in the mass
  range from 150 to 850 GeV in $pp$ collisions at $\sqrt{s} = 8$ TeV}},
  {\emph{Phys. Lett.} {\bf B750} (2015) 494},
  [\href{http://arxiv.org/abs/1506.02301}{{\tt 1506.02301}}].

\bibitem{arXiv:1504.00936}
{\scshape CMS} collaboration, \emph{{Search for a Higgs boson in the mass range
  from 145 to 1000 GeV decaying to a pair of W or Z bosons}},
  \href{http://arxiv.org/abs/1504.00936}{{\tt 1504.00936}}.

\bibitem{arXiv:1507.05930}
{\scshape ATLAS} collaboration, \emph{{Search for an additional, heavy Higgs
  boson in the $H\rightarrow ZZ$ decay channel at $\sqrt{s}$ = 8 TeV in $pp$
  collision data with the ATLAS detector}},
  \href{http://arxiv.org/abs/1507.05930}{{\tt 1507.05930}}.

\bibitem{arXiv:1509.00389}
{\scshape ATLAS} collaboration, \emph{{Search for a high-mass Higgs boson
  decaying to a $W$ boson pair in $pp$ collisions at $\sqrt{s} = 8$ TeV with
  the ATLAS detector}},  \href{http://arxiv.org/abs/1509.00389}{{\tt
  1509.00389}}.

\bibitem{Stal:2013hwa}
O.~Stål and T.~Stefaniak, \emph{{Constraining extended Higgs sectors with
  HiggsSignals}}, {\emph{PoS} {\bf EPS-HEP2013} (2013) 314},
  [\href{http://arxiv.org/abs/1310.4039}{{\tt 1310.4039}}].

\bibitem{Bechtle:2013xfa}
P.~Bechtle, S.~Heinemeyer, O.~Stål, T.~Stefaniak and G.~Weiglein,
  \emph{{$HiggsSignals$: Confronting arbitrary Higgs sectors with measurements
  at the Tevatron and the LHC}},
  \href{http://dx.doi.org/10.1140/epjc/s10052-013-2711-4}{\emph{Eur. Phys. J.}
  {\bf C74} (2014) 2711}, [\href{http://arxiv.org/abs/1305.1933}{{\tt
  1305.1933}}].

\bibitem{Aad:2015gba}
{\scshape ATLAS} collaboration, G.~Aad et~al., \emph{{Measurements of the Higgs
  boson production and decay rates and coupling strengths using $pp$ collision
  data at $\sqrt{s}=7$ and $8$ TeV in the ATLAS experiment}},
  \href{http://arxiv.org/abs/1507.04548}{{\tt 1507.04548}}.

\bibitem{Aad:2014eha}
{\scshape ATLAS} collaboration, G.~Aad et~al., \emph{{Measurement of Higgs
  boson production in the diphoton decay channel in pp collisions at
  center-of-mass energies of 7 and 8 TeV with the ATLAS detector}},
  \href{http://dx.doi.org/10.1103/PhysRevD.90.112015}{\emph{Phys. Rev.} {\bf
  D90} (2014) 112015}, [\href{http://arxiv.org/abs/1408.7084}{{\tt
  1408.7084}}].

\bibitem{Aad:2015vsa}
{\scshape ATLAS} collaboration, G.~Aad et~al., \emph{{Evidence for the
  Higgs-boson Yukawa coupling to tau leptons with the ATLAS detector}},
  \href{http://dx.doi.org/10.1007/JHEP04(2015)117}{\emph{JHEP} {\bf 04} (2015)
  117}, [\href{http://arxiv.org/abs/1501.04943}{{\tt 1501.04943}}].

\bibitem{ATLAS:2014aga}
{\scshape ATLAS} collaboration, G.~Aad et~al., \emph{{Observation and
  measurement of Higgs boson decays to WW$^*$ with the ATLAS detector}},
  \href{http://dx.doi.org/10.1103/PhysRevD.92.012006}{\emph{Phys. Rev.} {\bf
  D92} (2015) 012006}, [\href{http://arxiv.org/abs/1412.2641}{{\tt
  1412.2641}}].

\bibitem{ATLAS-CONF-2015-005}
{\scshape ATLAS} collaboration, ATLAS, \emph{{Study of the Higgs boson decaying
  to $WW^*$ produced in association with a weak boson with the ATLAS detector
  at the LHC}},  2015.

\bibitem{Aad:2014eva}
{\scshape ATLAS} collaboration, G.~Aad et~al., \emph{{Measurements of Higgs
  boson production and couplings in the four-lepton channel in pp collisions at
  center-of-mass energies of 7 and 8 TeV with the ATLAS detector}},
  \href{http://dx.doi.org/10.1103/PhysRevD.91.012006}{\emph{Phys. Rev.} {\bf
  D91} (2015) 012006}, [\href{http://arxiv.org/abs/1408.5191}{{\tt
  1408.5191}}].

\bibitem{Aad:2014xzb}
{\scshape ATLAS} collaboration, G.~Aad et~al., \emph{{Search for the $b\bar{b}$
  decay of the Standard Model Higgs boson in associated $(W/Z)H$ production
  with the ATLAS detector}},
  \href{http://dx.doi.org/10.1007/JHEP01(2015)069}{\emph{JHEP} {\bf 01} (2015)
  069}, [\href{http://arxiv.org/abs/1409.6212}{{\tt 1409.6212}}].

\bibitem{Khachatryan:2014ira}
{\scshape CMS} collaboration, V.~Khachatryan et~al., \emph{{Observation of the
  diphoton decay of the Higgs boson and measurement of its properties}},
  \href{http://dx.doi.org/10.1140/epjc/s10052-014-3076-z}{\emph{Eur. Phys. J.}
  {\bf C74} (2014) 3076}, [\href{http://arxiv.org/abs/1407.0558}{{\tt
  1407.0558}}].

\bibitem{Chatrchyan:2014vua}
{\scshape CMS} collaboration, S.~Chatrchyan et~al., \emph{{Evidence for the
  direct decay of the 125 GeV Higgs boson to fermions}},
  \href{http://dx.doi.org/10.1038/nphys3005}{\emph{Nature Phys.} {\bf 10}
  (2014) 557--560}, [\href{http://arxiv.org/abs/1401.6527}{{\tt 1401.6527}}].

\bibitem{Chatrchyan:2013iaa}
{\scshape CMS} collaboration, S.~Chatrchyan et~al., \emph{{Measurement of Higgs
  boson production and properties in the WW decay channel with leptonic final
  states}}, \href{http://dx.doi.org/10.1007/JHEP01(2014)096}{\emph{JHEP} {\bf
  01} (2014) 096}, [\href{http://arxiv.org/abs/1312.1129}{{\tt 1312.1129}}].

\bibitem{Chatrchyan:2013mxa}
{\scshape CMS} collaboration, S.~Chatrchyan et~al., \emph{{Measurement of the
  properties of a Higgs boson in the four-lepton final state}},
  \href{http://dx.doi.org/10.1103/PhysRevD.89.092007}{\emph{Phys. Rev.} {\bf
  D89} (2014) 092007}, [\href{http://arxiv.org/abs/1312.5353}{{\tt
  1312.5353}}].

\bibitem{CMS-PAS-HIG-13-005}
{\scshape CMS Collaboration} collaboration, CMS, \emph{{Combination of standard
  model Higgs boson searches and measurements of the properties of the new
  boson with a mass near 125 GeV}},  2013.

\bibitem{Aad:2015pla}
{\scshape ATLAS} collaboration, G.~Aad et~al., \emph{{Constraints on new
  phenomena via Higgs boson couplings and invisible decays with the ATLAS
  detector}},  \href{http://arxiv.org/abs/1509.00672}{{\tt 1509.00672}}.

\bibitem{Khachatryan:2015mma}
{\scshape CMS} collaboration, V.~Khachatryan et~al., \emph{{Limits on the Higgs
  boson lifetime and width from its decay to four charged leptons}},
  \href{http://arxiv.org/abs/1507.06656}{{\tt 1507.06656}}.

\bibitem{Alwall:2014hca}
J.~Alwall, R.~Frederix, S.~Frixione, V.~Hirschi, F.~Maltoni, O.~Mattelaer
  et~al., \emph{{The automated computation of tree-level and next-to-leading
  order differential cross sections, and their matching to parton shower
  simulations}}, \href{http://dx.doi.org/10.1007/JHEP07(2014)079}{\emph{JHEP}
  {\bf 07} (2014) 079}, [\href{http://arxiv.org/abs/1405.0301}{{\tt
  1405.0301}}].

\bibitem{TDRvol1}
T.~Behnke et~al., \emph{{The International Linear Collider Technical Design
  Report - Volume 1: Executive Summary}},
  \href{http://arxiv.org/abs/1306.6327}{{\tt 1306.6327}}.

\bibitem{Alloul:2013bka}
A.~Alloul, N.~D. Christensen, C.~Degrande, C.~Duhr and B.~Fuks,
  \emph{{FeynRules 2.0 - A complete toolbox for tree-level phenomenology}},
  \href{http://dx.doi.org/10.1016/j.cpc.2014.04.012}{\emph{Comput. Phys.
  Commun.} {\bf 185} (2014) 2250--2300},
  [\href{http://arxiv.org/abs/1310.1921}{{\tt 1310.1921}}].

\bibitem{TDRvol2}
H.~Baer et~al., \emph{{The International Linear Collider Technical Design
  Report - Volume 2: Physics}},  \href{http://arxiv.org/abs/1306.6352}{{\tt
  1306.6352}}.

\bibitem{selfcoupHvv}
J.~Tian, \emph{{Study of the Higgs self-coupling at the ILC based on full
  detector simulation at $\sqrt{s}=500$ GeV and $\sqrt{s}=1$ TeV}},  2013.

\bibitem{LUX2015}
{\scshape LUX} collaboration, D.~S. Akerib et~al., \emph{{Improved WIMP
  scattering limits from the LUX experiment}},
  \href{http://arxiv.org/abs/1512.03506}{{\tt 1512.03506}}.

\bibitem{Aaij:2015tna}
{\scshape LHCb} collaboration, R.~Aaij et~al., \emph{{Search for hidden-sector
  bosons in $B^0 \!\to K^{*0}\mu^+\mu^-$ decays}},
  \href{http://dx.doi.org/10.1103/PhysRevLett.115.161802}{\emph{Phys. Rev.
  Lett.} {\bf 115} (2015) 161802}, [\href{http://arxiv.org/abs/1508.04094}{{\tt
  1508.04094}}].

\end{thebibliography}\endgroup

\end{document}